\documentclass[preprint,12pt,authoryear]{elsarticle}
\usepackage[a4paper, top=2cm, bottom=2cm, left=2cm, right=2cm]{geometry}
\usepackage{framed,multirow}\setcitestyle{square,numbers,sort&compress}
\biboptions{numbers,sort&compress}
\usepackage{amssymb}
\usepackage{amsmath}
\usepackage{latexsym}
\usepackage{bm}
\usepackage{amssymb}
\usepackage{subfigure}
\usepackage{booktabs}
\usepackage{url}
\usepackage[colorlinks=true,linkcolor=black,citecolor=blue,urlcolor=blue,]{hyperref}
\usepackage[doipre={doi:~}]{uri}
\usepackage{xcolor}
\usepackage{listings}

\begin{document}

\begin{frontmatter}

\title{Parametric Gaussian quadratures for Discrete Unified Gas Kinetic Scheme} 

\author{Lu Wang}
\ead{i_wanglu@163.com}
\author{Hong Liang}
\cortext[cor1]{Corresponding author: 
	E-mail: lianghongstefanie@163.com;}
\author{Jiangrong Xu}
\cortext[cor2]{Corresponding author: 
	E-mail: jrxu@hdu.edu.cn;}

\affiliation{organization={Department of Physics},
	addressline={Hangzhou Dianzi University}, 
	city={Hangzhou},
	postcode={310018}, 
	country={China}}

\begin{abstract}
The discrete unified gas kinetic scheme (DUGKS) has emerged as a promising Boltzmann solver capable of effectively capturing flow physics across all Knudsen numbers. However, simulating rarefied flows at high Knudsen numbers remains computationally demanding. This paper introduces a parametric Gaussian quadrature (PGQ) rule designed to improve the computational efficiency of DUGKS. The PGQ rule employs Gaussian functions for weighting and introduces several novel forms of higher-dimensional Gauss-Hermite quadrature. Initially, the velocity space is mapped to polar or spherical coordinates using a parameterized integral transformation method, which converts multiple integrals into repeated parametric integrals. Subsequently, Gaussian points and weight coefficients are computed based on the newly defined parametric weight functions. The parameters in PGQ allow the distribution of Gaussian points to be adjusted according to computational requirements, addressing the limitations of traditional Gaussian quadratures where Gaussian points are difficult to match the distribution of real particles in rarefied flows. To validate the proposed approach, numerical examples across various Knudsen numbers are provided. The simulation results demonstrate that PGQ offers superior computational efficiency and flexibility compared to the traditional Newton-Cotes rule and the half-range Gaussian Hermite rule, achieving computational efficiency that is tens of times higher than that of the Newton-Cotes method. This significantly enhances the computational efficiency of DUGKS and augments its ability to accurately simulate rarefied flow dynamics.
\end{abstract}
\end{frontmatter}

\section{Introduction}
The multiscale fluid flow problem is prevalent in scientific research and engineering applications, such as microelectromechanical systems (MEMS) \citep{FAN2001}, spacecraft technology \citep{LI2009}, and vacuum technology \citep{var2008}. Due to its significance and wide-ranging applications, there has been increasing interest in developing multiscale simulation methods within a unified framework. The Boltzmann equation provides the theoretical foundation necessary for accurately describing flow behavior across different regimes. However, creating a Boltzmann solver that is applicable across all Knudsen numbers (Kn) presents substantial challenges.

The Direct Simulation Monte Carlo (DSMC) technique, a particle-based stochastic methodology that handles particle transport and collision processes separately, has proven effective in simulating rarefied flows \citep{Bird1995, Myong2019}. Despite its strengths, DSMC becomes computationally demanding for continuum and near-continuum flows, requiring smaller time steps and cell sizes relative to particle collision times and mean free paths. Alternatively, the deterministic discrete velocity method (DVM) offers another strategy for solving the Boltzmann equation by discretizing continuous velocity space into a finite set of discrete velocities \citep{MIEUSSENS2000429, YANG2016291}. This approach transforms the Boltzmann equation into a set of discrete velocity equations, which can be numerically solved using various schemes. However, it is important to note that classical DVMs often employ a splitting treatment similar to DSMC, leading to similar constraints on time step and cell size.

Recent advancements have made strides in overcoming these limitations, notably through the work of Xu and Guo et al. Xu and his colleagues introduced the unified gas kinetic scheme (UGKS) \citep{Xu2010, Xu2011, Huang2012, Xu2015}, marking a significant breakthrough in multiscale simulation. The UGKS integrates particle transport and collision processes in updating the discrete distribution function, liberating the time step determination from particle collision constraints. This integration achieves higher computational efficiency without compromising accuracy, showcasing the UGKS’s capability in tackling multiscale problems. Further simplification and efficiency gains are achieved by the discrete unified gas kinetic scheme (DUGKS) \citep{dugks2013, dugks2015, dugks2021}, which combines the advantages of the UGKS and the classic lattice Boltzmann method (LBM) \citep{TK2016, KIM2008}. The DUGKS has become a prominent and widely adopted multiscale simulation technique due to its effectiveness and versatility. Its ability to accurately simulate a wide range of flow regimes has contributed to its increasing popularity within the engineering and scientific research communities \citep{SONG2023, LIU2023}.

The choice of discrete velocities and quadrature formulas in DVM-based solvers is crucial for accurately and efficiently capturing gas flow behavior. Commonly used quadrature rules in this field are the Gauss-Hermite and Newton-Cotes quadrature rules \citep{KIM2008, YANG2017}. Gauss-Hermite quadrature formulas, including the half-range Gauss-Hermite \citep{Ball2002, Ambrus2016, Ambrus2019}, have been extensively studied and proven effective in capturing flows close to equilibrium, particularly for Boltzmann equations with Maxwell equilibrium distribution functions. However, they may not be suitable for simulating high Mach number flows and rarefied flows with pronounced non-equilibrium effects \citep{YANG2016291, dugks2013}. In such cases, the particle distribution function exhibits a concentrated profile with steep slopes in the particle velocity space. Consequently, the abscissas of the Gauss-Hermite rule, which are distributed throughout the entire velocity space, cannot be freely chosen. To address this issue, the Newton-Cotes quadrature rule has been employed, enforcing the use of a finer mesh to discretize the velocity space. Although this method captures non-equilibrium flows, it significantly increases computational cost due to the larger number of discrete velocities. Researchers are actively exploring alternative approaches to mitigate these problems. One avenue involves utilizing alternative Gaussian quadrature formulas such as Gauss-Laguerre \citep{Ambru2012}, Gauss-Legendre \citep{shi2022, shi2023}, and Gauss-Chebyshev \citep{HU2018} quadratures. Additionally, reducing the number of discrete velocities has been investigated to alleviate the computational cost associated with Newton-Cotes quadrature. Techniques such as velocity adaptation \citep{GUTNIC2004, MEHRENBERGER2006}, local refinement methods for velocity grids \citep{chen2019}, and Reduced Order Modeling methods \citep{zhao2020} have been explored.

Despite these efforts, the computational efficiency of DVM-based solvers for rarefied flow remains unsatisfactory. Thus, developing more efficient velocity discretization methods is urgently needed. The computational burden of the Newton-Cotes quadrature rule is substantial due to its inferior accuracy compared to Gaussian quadratures. On the other hand, Gaussian quadratures present their own difficulties, as their predefined abscissas and weights might not closely match the actual velocity distribution of particles. This issue is particularly problematic for distributions exhibiting rapid variation far from the mean, which may not be adequately captured by the predetermined abscissas and weights, resulting in reduced simulation accuracy.

To overcome this limitation, the ability to adjust the abscissas and weights of Gaussian quadratures to meet specific requirements is highly desirable. Two Gaussian quadrature rules stand out for their exceptional characteristics: the generalized Gauss-Laguerre quadrature \citep{perez1996, cassity1965} with the weight function \(w(x) = x^{\alpha} e^{-x}\) \((\alpha>-1)\) and integration interval \([0, \infty)\), and the Gauss-Jacobi quadrature \citep{Gil2020} with the weight function \(w(x) = (1-x)^{\alpha}(1+x)^{\beta}\) \((\alpha, \beta>-1)\) and the interval \([-1, 1]\). The Gauss-Jacobi quadrature rule includes Gauss-Chebyshev quadrature \((|\alpha| = |\beta| = 1/2)\) and Gauss-Legendre quadrature \((\alpha = \beta = 0)\) as particular cases. By adjusting the parameters \(\alpha\) and \(\beta\), the abscissas and weights of these quadratures can be flexibly tailored.

Given the probability density function distribution of the Boltzmann equation over the interval \((- \infty, +\infty)\), direct application of these Gaussian-type quadratures is impractical. Therefore, additional steps such as integration substitution or interval truncation are necessary. Building on these considerations, this paper introduces a novel Gaussian quadrature rule, termed “parametric Gaussian quadrature” (PGQ), specifically tailored for DVM-based solvers within the DUGKS framework. The proposed method utilizes integral transformations to convert the multidimensional Gauss-Hermite weight function \(w(\boldsymbol{x}) = \exp(-\boldsymbol{x}^2)\) into various alternative Gaussian weight forms. This work particularly focuses on two specific forms: 2D PGQ based on polar coordinate transformation and 3D PGQ based on spherical coordinate transformation. These transformations aim to enhance the accuracy and efficiency of rarefied flow simulations, effectively overcoming the limitations of DVM and conventional quadrature rules.

The structure of this paper is arranged as follows: Section \ref{sec2} begins with a concise overview of the governing equation and the DUGKS method. In Section \ref{sec3}, the methodology behind the PGQ is explained in comprehensive detail, highlighting its significance and advantages. To ascertain the efficiency and accuracy of the proposed method, Section \ref{sec4} presents a variety of benchmark tests, including analyses on Couette flow \citep{SUN2002, Huang2013, wl2023}, thermal creep flow \citep{ZHU2019}, Rayleigh flow \citep{SUN2002, Huang2013}, cavity flow \citep{zhao2020, ZHU2019}, and cylinder flow. These tests serve to validate the numerical results and demonstrate the superior performance of the PGQs. Section \ref{sec5} concludes the paper by summarizing the key findings and discussing potential avenues for future research.

\section{Methodology}
\label{sec2}
\subsection{BGK-Shakhov model}
\label{sec2.1}
The construction of the DUGKS is based on the Boltzmann equation, which incorporates the relaxation time assumption. Specifically, for the thermal compressible case, the Boltzmann equation with the BGK–Shakhov collision model is considered. This collision model expands upon the basic BGK approximation by introducing an additional relaxation term designed to account for non-equilibrium effects encountered in rarefied gas dynamics, particularly under different Prandtl ($Pr$) numbers. Considering two-dimensional (2D) flows, the dimensionless Shakhov model \citep{Shakhov1968, Li2021} for monatomic gases is written as follows:
\begin{equation}	
	\frac{\partial f}{\partial t}+\xi _x\frac{\partial f}{\partial x}+\xi _y\frac{\partial f}{\partial y}=\varOmega \equiv \frac{f^s-f}{\tau},
	\label{eq:OSM}
\end{equation}
\begin{equation}	
	f^s=f^{eq}\left[ 1+\frac{2\left( 1-Pr \right) \left( c_xq_x+c_yq_y \right)}{5\rho T^2}\left( 2\frac{c_{x}^{2}+c_{y}^{2}+\xi _{z}^{2}}{T}-5 \right) \right],
	\label{eq:fs}
\end{equation}
\begin{equation}	
	f^{eq}=\frac{\rho}{\left( \pi T \right) ^{3/2}}\exp \left( -\frac{c_{x}^{2}+c_{y}^{2}+\xi _{z}^{2}}{T} \right),
	\label{eq:feq}
\end{equation}
\begin{equation}
	\tau =\frac{5\left( \alpha_0 +1 \right) \left( \alpha_0 +2 \right) \sqrt{\pi}KnT^{\chi -1}}{2\alpha_0 \left( 5-2\omega_0 \right) \left( 7-2\omega_0 \right) \rho}.
	\label{eq:tau}
\end{equation}
Here, $f=f\left( t,x,y,\xi _x,\xi _y,\xi _z \right)$ represents the distribution function for molecules with velocities $\left( \xi _x,\xi _y \right)$ at position $\left( x,y \right)$ and time $t$. $\tau$ denotes the relaxation time, $f^s$ refers to the Shakhov equilibrium distribution function, and $f^{eq}$ represents the Maxwellian equilibrium distribution function. The energy-dependent deflection-angle exponent is denoted as $\alpha_0$, indicating the index of the variable soft sphere (VSS) molecular model. Additionally, $\omega_0$ represents the VSS model index, while $\chi$ represents the temperature exponent of the coefficient of viscosity. The molecular thermal velocity is given as $c_i=\xi _i-u_i$. Furthermore, variables such as $\rho$, $u_i$, $T$ and $q_i$ respectively correspond to density, mean flow velocity, temperature, and heat flux. These quantities can be defined in terms of moments of the distribution function:
\begin{equation}	
	\left[ \begin{array}{c}
		\rho\\
		\rho u_i\\
		\rho T\\
		q_i\\
	\end{array} \right] =\int{\mathcal{M}fd\xi _xd\xi _yd\xi _z},
	\label{eq:fmoments}
\end{equation}
\begin{equation}	
	\mathcal{M}=\left[ 1,\xi _i,\frac{2}{3}\left( c_{x}^{2}+c_{y}^{2}+\xi _{z}^{2} \right) ,c_i\left( c_{x}^{2}+c_{y}^{2}+\xi _{z}^{2} \right) \right] ^T.	
	\label{eq:fusai}
\end{equation}

In the context described, it has been observed that the evolution of the distribution function $f\left( t,x,y,\xi _x,\xi _y,\xi _z \right) $ is independent of an additional variable $\xi_z$. To reduce the number of independent variables, the reduced Shakhov model is commonly introduced \citep{Li2021, YANG1995}:
\begin{equation}	
	\frac{\partial g_i}{\partial t}+\xi _x\frac{\partial g_i}{\partial x}+\xi _y\frac{\partial g_i}{\partial y}=\frac{g_{i}^{s}-g_i}{\tau},
	\label{eq:RSM}
\end{equation}
where $i=0,1$. The reduced distribution functions $g_0$ and $g_1$ are obtained by integrating $f\left( t,x,y,\xi _x,\xi _y,\xi _z \right)$ with respect to $\xi_z$ using weighting factors of $1$ and ${\xi_z}^2$, respectively. Their definitions can be uniformly expressed as follows:
\begin{equation}	
	g_i\left( t,x,y,\xi _x,\xi _y \right) =\int{\left( \xi _{z}^{2} \right) ^if}d\xi _z.
\end{equation}

Additionally, the equilibrium distribution functions in the reduced Shakhov model can be expressed in a unified form:
\begin{equation}	
	g_{i}^{s}=\left( \frac{T}{2} \right) ^ig^{eq}\left[ 1+\frac{4\left( 1-Pr \right) \left( c_xq_x+c_yq_y \right)}{5\rho T^2}\left( \frac{c_{x}^{2}+c_{y}^{2}}{T}-2+i \right) \right],
	\label{eq:GS}
\end{equation}
\begin{equation}	
	g^{eq}=\frac{\rho}{\pi T}\exp \left( -\frac{\left| \xi _x-u_x \right|^2+\left| \xi _y-u_y \right|^2}{T} \right).
	\label{eq:Geq}
\end{equation}

It is essential to clarify that Eq.~(\ref{eq:GS}) corresponds to the formulation found in Ref. \citep{Li2021, YANG1995}, though it is presented herein in a more streamlined version. Within the framework of the reduced Shakhov model, the macroscopic flow variables are expressed as follows:
\begin{equation}	
	\left[ \begin{array}{c}
		\rho\\
		\rho u_i\\
		\rho T\\
		q_i\\
	\end{array} \right] =\int{\mathcal{M}\left[ \begin{array}{c}
			g_0\\
			g_1\\
		\end{array} \right] d\xi _xd\xi _y},
\end{equation}
Here, $\mathcal{M}$ represents a moment matrix that consists of different weightings and velocity components:
\begin{equation}
	\mathcal{M}=\left[ \left( 1,0 \right) ,\left( \xi _i,0 \right) ,\frac{2}{3}\left( c_{x}^{2}+c_{y}^{2},1 \right) ,c_i\left( c_{x}^{2}+c_{y}^{2},1 \right) \right] ^T.
\end{equation}

Simulating flows in 2D physical space using the Shakhov model requires performing calculations in a three-dimensional (3D) velocity space, leading to significant computer storage requirements and computational costs. Conversely, the reduced Shakhov model effectively addresses this challenge by solving a slightly more complex set of governing equations, making it the preferred choice among researchers. However, it is worthwhile to consider the Shakhov model from a different perspective. The Shakhov model provides a means to examine the discretization method for 3D velocity space within a 2D physical space. In this paper, we will develop discretization methods for both 2D and 3D velocity spaces. Consequently, both the Shakhov model and the reduced Shakhov model will be employed to facilitate this investigation.

\subsection{Discrete Unified Gas Kinetic Scheme}
\label{sec2.2}
\begin{figure*}[!t]
	\centering
	\includegraphics[width=7cm,height=7cm]{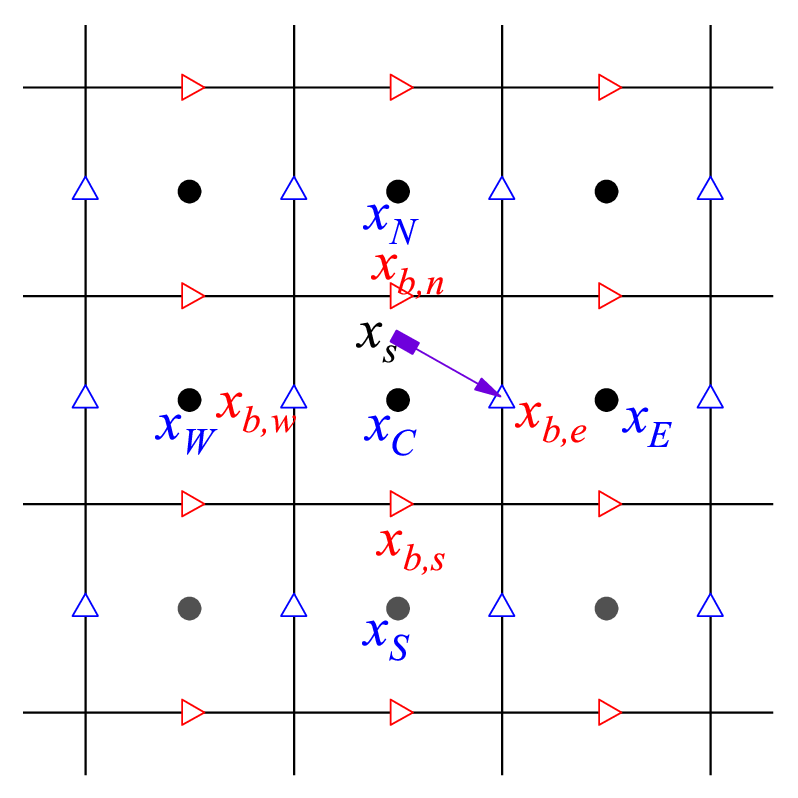}
	\caption{\label{fig:geometry} \centering Schematic of 2D cell geometry.}
\end{figure*}

The application of the DUGKS in solving both the Shakhov model and the reduced Shakhov model demonstrates minimal variation. In this section, we present the solution process for the Shakhov model (from Eq.~(\ref{eq:OSM}) to Eq.~(\ref{eq:fusai})) as a representative example to briefly introduce DUGKS. For comprehensive insights into DUGKS, we refer interested readers to the seminal contributions by Guo et al.\citep{dugks2013, dugks2015, dugks2021}. DUGKS is a two-step Boltzmann solver that merges UGKS and LBM, grounded in the finite volume method. The process begins by partitioning the computational domain into control volumes, as depicted in Fig.~\ref{fig:geometry}. Following this, the calculation of the distribution function is divided into two stages: first at the cell interface \( \boldsymbol{x}_b \) for half a time step, and then at the cell center \( \boldsymbol{x}_C \) for the ensuing full time step. The evaluation of the distribution function at $\boldsymbol{x}_b$ involves integration along characteristic lines, while the formulation of the discrete equation at $\boldsymbol{x}_C$ leverages the finite volume method. This methodology is concisely articulated through the following steps:
\begin{subequations}
	\label{eq:two-step}
	\begin{equation}
		f\left( t^{n+1/2},\boldsymbol{x}_b,\boldsymbol{\xi } \right) =f\left( t^n,\boldsymbol{x}_b-\boldsymbol{\xi }\varDelta t/2,\boldsymbol{\xi } \right) +\frac{\varDelta t}{4}\left[ \varOmega \left( t^{n+1/2},\boldsymbol{x}_b,\boldsymbol{\xi } \right) +\varOmega \left( t^n,\boldsymbol{x}_b-\boldsymbol{\xi }\varDelta t/2,\boldsymbol{\xi } \right) \right] ,
		\label{eq:fb}
	\end{equation}
	\begin{equation}
		f\left( t^{n+1},\boldsymbol{x}_c,\boldsymbol{\xi } \right) =f\left( t^n,\boldsymbol{x}_c,\boldsymbol{\xi } \right) -\varDelta t F^{n+\frac{1}{2}}+\frac{\varDelta t}{2}\left[ \varOmega \left( t^{n+1},\boldsymbol{x}_c,\boldsymbol{\xi } \right) +\varOmega \left( t^n,\boldsymbol{x}_c,\boldsymbol{\xi } \right) \right].
		\label{eq:fC}
	\end{equation}
\end{subequations}
where $F^{n+\frac{1}{2}}$ is the micro-flux per unit volume, which can be discretized into the following form
\begin{equation}
	F^{n+\frac{1}{2}}=\frac{1}{|V_C|}\oint_{\partial V_c}
	{\left( \boldsymbol{\xi}\cdot \boldsymbol{n} \right) f
		\left( t^{n+\frac{1}{2}},\boldsymbol{x}, \boldsymbol{\xi }\right) d\boldsymbol{S}}
	= \frac{\xi _x}{\varDelta x}\left( f_{b,e}^{n+\frac{1}{2}}
	-f_{b,w}^{n+\frac{1}{2}} \right) +\frac{\xi _y}{\varDelta y}\left( f_{b,n}^{n+\frac{1}{2}}
	-f_{b,s}^{n+\frac{1}{2}} \right).
	\label{eq:microFlux}
\end{equation}

As observed, Eq.~(\ref{eq:two-step}) is derived employing a semi-implicit scheme. To transform it into a fully explicit form, Guo et al. \citep{dugks2013, dugks2015} have incorporated four auxiliary distribution functions. These auxiliary functions are delineated as follows:
\begin{subequations}
	\label{eq:fdf}
	\begin{equation}
		\bar{f}_{b}^{n+\frac{1}{2}}=f\left( t^{n+1/2},\boldsymbol{x}_b,\boldsymbol{\xi } \right) -\frac{\varDelta t}{4}\varOmega \left( t^{n+1/2},\boldsymbol{x}_b,\boldsymbol{\xi } \right),
		\label{subeq:fb}
	\end{equation}
	\begin{eqnarray}
		\bar{f}_{s}^{n,+}=f\left( t^n,\boldsymbol{x}_b-\boldsymbol{\xi }\varDelta t/2,\boldsymbol{\xi } \right) +\frac{\varDelta t}{4}\varOmega \left( t^n,\boldsymbol{x}_b-\boldsymbol{\xi }\varDelta t/2,\boldsymbol{\xi } \right),
		\label{subeq:fbplus}
	\end{eqnarray}
	\begin{equation}
		\widetilde{f}_{C}^{n+1}=f\left( t^{n+1},\boldsymbol{x}_c,\boldsymbol{\xi } \right) -\frac{\varDelta t}{2}\varOmega \left( t^{n+1},\boldsymbol{x}_c,\boldsymbol{\xi } \right),
		\label{subeq:fC}
	\end{equation}
	\begin{eqnarray}
		\widetilde{f}_{C}^{n,+}=\frac{4\bar{f}_{C}^{n,+}-\widetilde{f}_{C}^{n}}{3}=f\left( t^n,\boldsymbol{x}_c,\boldsymbol{\xi } \right) +\frac{\varDelta t}{2}\varOmega \left( t^n,\boldsymbol{x}_c,\boldsymbol{\xi } \right).
		\label{subeq:fCplus}
	\end{eqnarray}
\end{subequations}
By introducing four distribution functions $\bar{f}_{b}^{n+\frac{1}{2}}$, $\bar{f}_{s}^{n,+}$, $\widetilde{f}_{C}^{n+1}$, and $\widetilde{f}_{C}^{n,+}$, as described in Eq.~(\ref{eq:fdf}), the two-step scheme for solving the Boltzmann equation, as represented by Eq.~(\ref{eq:two-step}), can be expressed as follows:
\begin{subequations}
	\label{eq:tssi}
	\begin{eqnarray}
		\bar{f}_{b}^{n+\frac{1}{2}}=\bar{f}_{s}^{n,+},
		\label{subeq:fB2}
	\end{eqnarray}
	\begin{equation}
		\widetilde{f}_{C}^{n+1}=\frac{4\bar{f}_{C}^{n,+}-\widetilde{f}_{C}^{n}}{3}-\varDelta tF^{n+\frac{1}{2}}.
		\label{subeq:fC2}
	\end{equation}
\end{subequations}
Here, the distribution function at the starting point of the characteristic line, denoted as $\bar{f}_{s}^{n,+}$, can be approximated using the Taylor expansion around the cell interface or cell centers. For the purpose of illustration, we will now present the approximation around $\boldsymbol{x}_C$ :
\begin{equation}
	\bar{f}_{s}^{n,+}=\bar{f}_{C}^{n,+}+(\boldsymbol{x}_s-\boldsymbol{x}_C)
	\cdot \boldsymbol{\nabla}\bar{f}_{C}^{n,+}.
	\label{subeq:fsTaylor}
\end{equation}

Despite the apparent simplicity of the evolution Eq.~(\ref{eq:tssi}), it entails intricate transformations between the original distribution function and multiple auxiliary distribution functions throughout the calculation process. As a result, the relationship between distribution functions assumes a crucial role in the solving process. Furthermore, it is essential to establish the connection between the auxiliary distribution function and macroscopic quantities. These relationships are summarized as follows:
\begin{equation}
	\bar{f}_{C}^{n,+}=\frac{4\tau -\varDelta t}{4\tau +2\varDelta t}\widetilde{f}_{C}^{n}
	+\frac{3\varDelta t}{4\tau +2\varDelta t}f_{C}^{eq,n},
	\label{eq:fC4}
\end{equation}
\begin{equation}
	f_{b}^{n+\frac{1}{2}}=\frac{4\tau}{4\tau +\varDelta t}\bar{f}_{b}^{n+\frac{1}{2}}
	+\frac{\varDelta t}{4\tau +\varDelta t}f_{b}^{eq,n+\frac{1}{2}},
	\label{eq:fb4}
\end{equation}

\begin{subequations}
	\begin{equation}
		\left[ \begin{array}{c}
			\rho\\
			\rho u_i\\
			\rho T\\
		\end{array} \right] =\int{\mathcal{M}\widetilde{f}d\boldsymbol{\xi }}=\int{\mathcal{M}\bar{f}d\boldsymbol{\xi }},
		\label{subeq:moment1}
	\end{equation}
	\begin{eqnarray}
		q_i=\left\{ \begin{array}{l}
			\frac{2\tau}{2\tau +\varDelta tPr}\int{c_i\left( c_{x}^{2}+c_{y}^{2}+\xi _{z}^{2} \right) \widetilde{f}d\boldsymbol{\xi }},\\
			\frac{4\tau}{4\tau +\varDelta tPr}\int{c_i\left( c_{x}^{2}+c_{y}^{2}+\xi _{z}^{2} \right) \bar{f}d\boldsymbol{\xi }}.\\
		\end{array} \right. 
		\label{subeq:momentq}
	\end{eqnarray}
\end{subequations}
where $\mathcal{M}=\left[ 1,\xi _i,\frac{2}{3}\left( c_{x}^{2}+c_{y}^{2}+\xi _{z}^{2} \right) \right] ^T$. It is evident that the computation of macroscopic variables $\rho$, $u_i$, and $T$ shares similarities with Eq.~(\ref{eq:fmoments}). However, particular emphasis must be placed on the determination of the heat flux $q_i$ calculation.

In DUGKS, the evolution function is represented by $\widetilde{f}$, rather than the original distribution function $f$. To enhance the intuitive expression of the calculation process in DUGKS, we succinctly represent the update of the distribution function $\widetilde{f}$ from time $t^n$ to $t^{n+1}$ as follows:
\begin{eqnarray}
	\widetilde{f_{C}^{n}}\xrightarrow{(\ref{eq:fC4})}\bar{f}_{C}^{n,+}
	\xrightarrow[(\ref{subeq:fsTaylor})]{(\ref{subeq:fB2})}\bar{f}_{b}^{n+\frac{1}{2}}\xrightarrow{(\ref{eq:fb4})}f_{b}^{n+\frac{1}{2}}\xrightarrow[(\ref{eq:microFlux})]{(\ref{subeq:fC2})}\widetilde{f_{C}^{n+1}}.
\end{eqnarray}
\section{Parametric Gaussian quadratures}
\label{sec3}
The choice of an appropriate weight function is essential in utilizing Gaussian quadrature formulas for numerical integration across extensive intervals. Empirical analysis has demonstrated that Gaussian functions are excellent weight functions for numerical integration over the interval \( \left(-\infty,+\infty \right) \), offering exceptional numerical stability and computational efficiency. Building upon this observation and taking into account the structure of the equilibrium distribution functions Eq.~(\ref{eq:feq}) and Eq.~(\ref{eq:Geq}), the weight function employed in this study is defined as
\begin{equation}
	w\left( \boldsymbol{\xi } \right) =\frac{1}{\left( \pi T_m \right) ^{D/2}}\exp \left( -\frac{\boldsymbol{\xi }^2}{T_m} \right),
	\label{eq:wight}
\end{equation}
where $T_m$ is referred to as the median temperature. Additionally, the resulting quadrature formula can be expressed as:
\begin{equation}
	I\left( \boldsymbol{x},t \right) =\int\limits_{-\infty}^{+\infty}{w\left( \boldsymbol{\xi } \right) \mathcal{F}\left( \boldsymbol{x},\boldsymbol{\xi },t \right) d\boldsymbol{\xi }}=\sum_{n=1}^{\mathcal{N}}{\omega_n\mathcal{F}\left( \boldsymbol{x},\boldsymbol{\xi }_n,t \right)}.
	\label{eq:inte}
\end{equation}

For 1D numerical integration, Eq.~(\ref{eq:inte}) corresponds to the standard Gauss-Hermite quadrature formula. When extending the approach to 2D and 3D integrals, the complexity increases.A straightforward extrapolation from 1D to higher dimensions often involves using the tensor product method. This method systematically constructs multidimensional quadrature rules by combining the 1D abscissas and weights along each dimension. However, in this study, we will use the substitution integration method to map the integration region to an alternative space and adjust the weight function accordingly. Specifically, we utilize the generalized polar coordinate transformation (PCT) for double integrals and the generalized spherical coordinate transformation (SCT) for calculating triple integrals.

\subsection{2D Parametric Gaussian quadratures}
\label{sec3.1}

\begin{figure*}[!t]
	\centering
	\includegraphics[width=7cm,height=7cm]{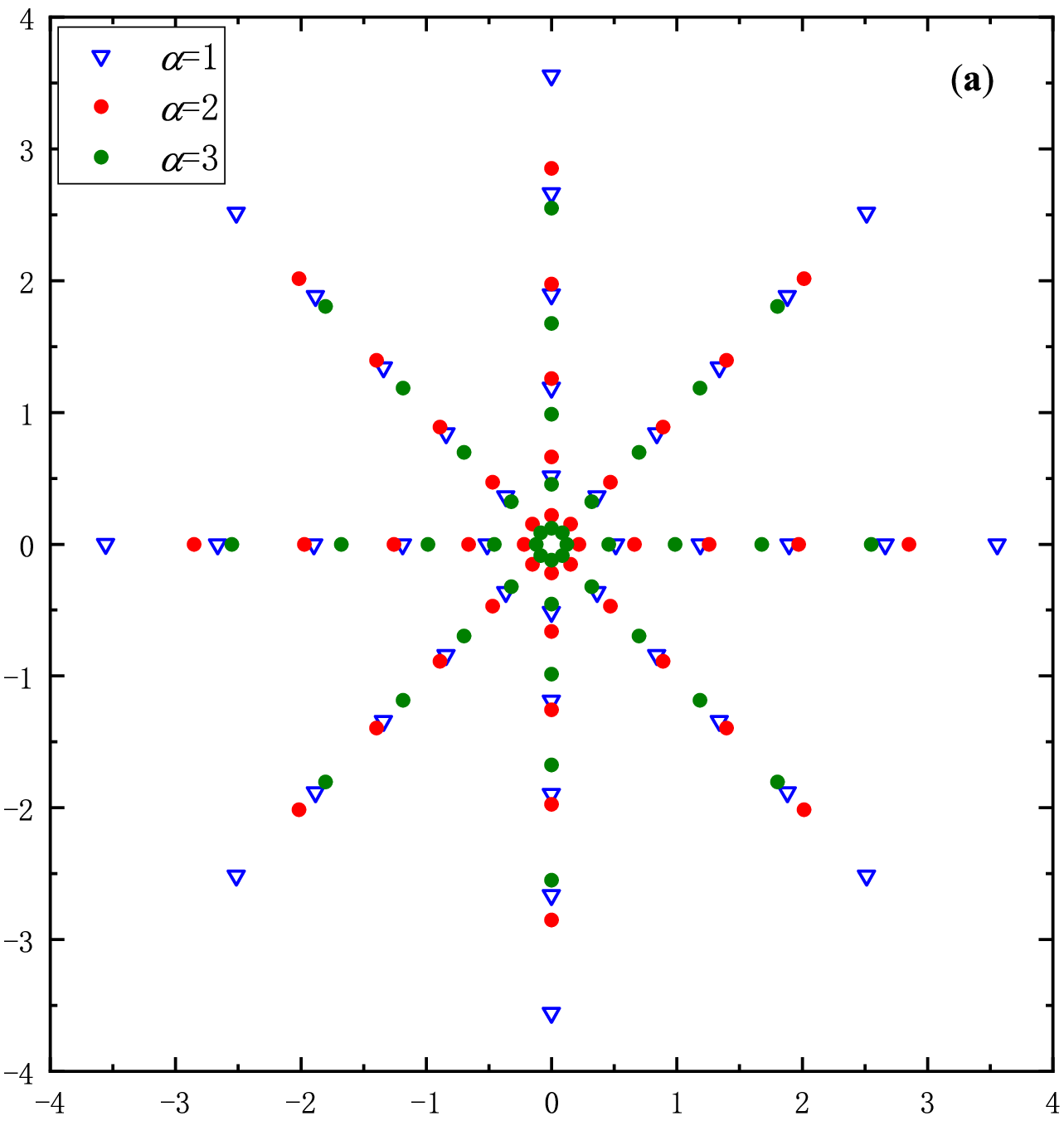}
	\includegraphics[width=7cm,height=7cm]{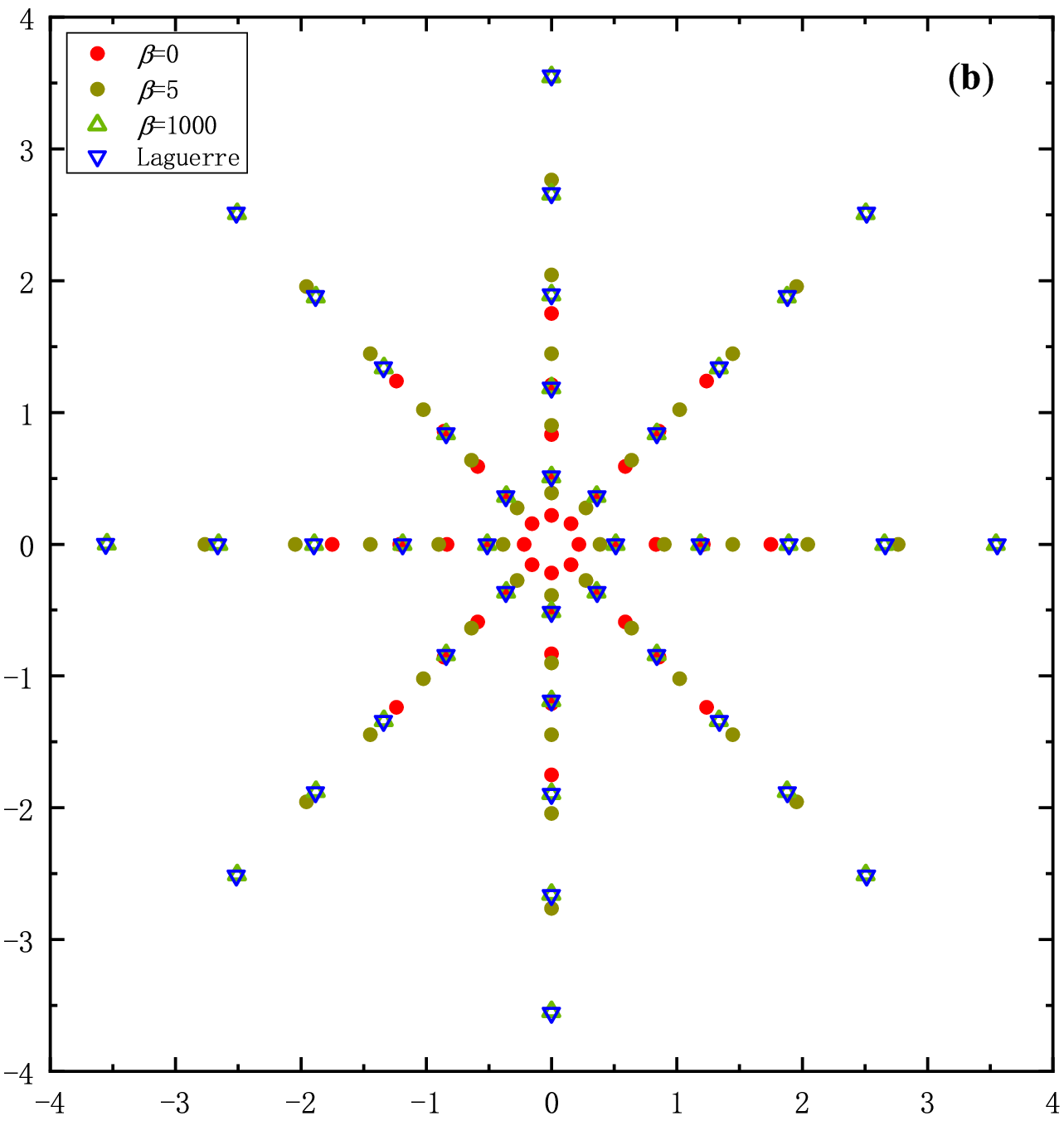}
	\caption{\label{fig:RadDis} \centering Different radial distributions of the PGQ for $N_r=5, N_{\theta}=8$; (a) P1; (b) P2.}
\end{figure*}

\begin{figure*}[!t]
	\centering
	\includegraphics[width=7cm,height=7cm]{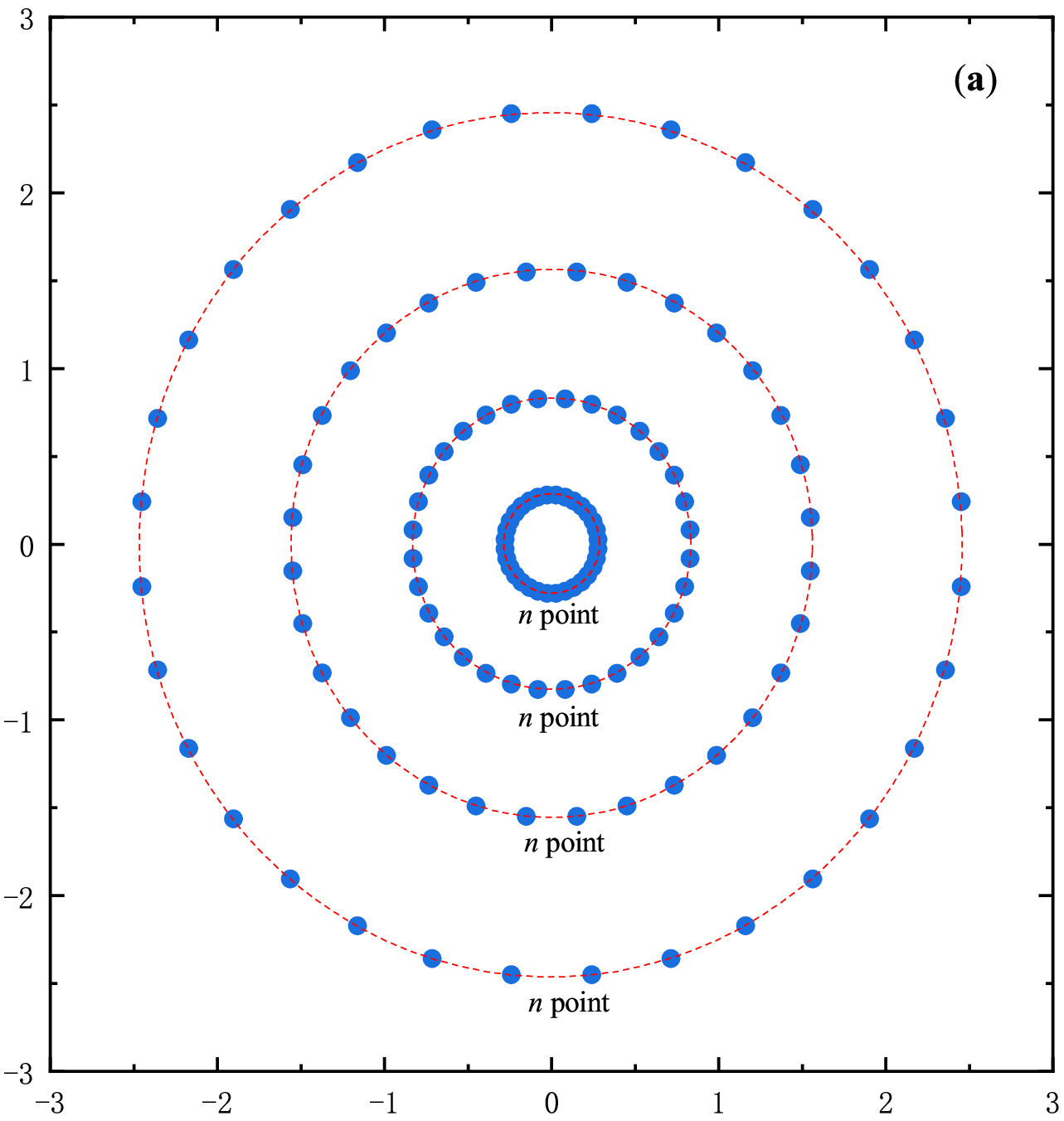}
	\includegraphics[width=7cm,height=7cm]{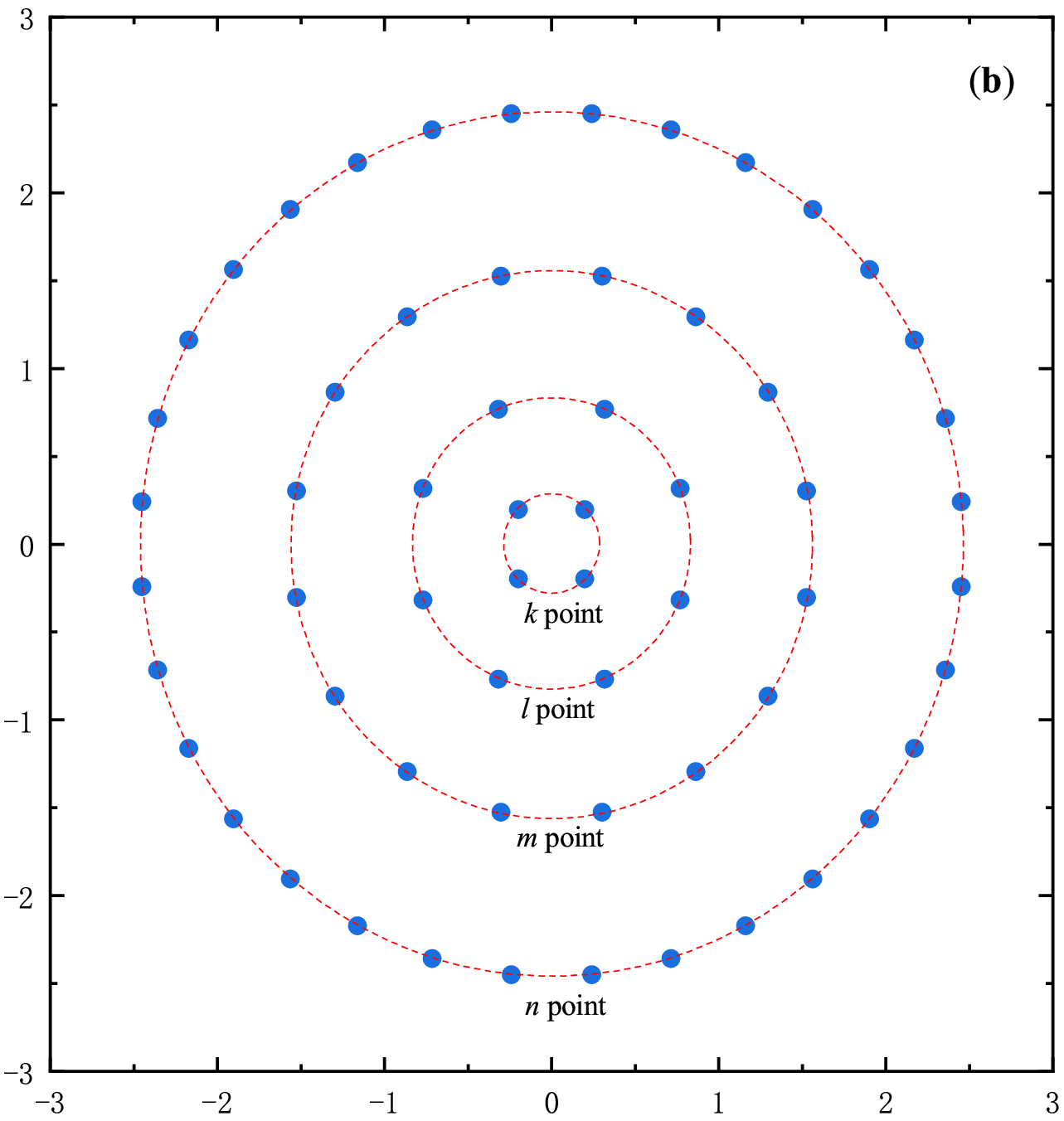}
	\caption{\label{fig:NorDis} \centering Schematic of (a) A-type and (b) B-type discrete velocity distribution.}
\end{figure*}

To calculate double integrals, we adopt the generalized PCT that incorporates a radial function $R\left( r \right)$ and a trigonometric function, specified as:
\begin{equation}	
	\xi _x=\sqrt{T_m}R\left( r \right) \cos \theta,~~ \xi _y=\sqrt{T_m}R\left( r \right) \sin \theta,
	\label{eq:pct}
\end{equation}
where $\theta \in \left[ 0,2\pi \right]$, $R\left( r \right) \in \left( 0,+\infty \right)$. Utilizing this transformation, the weight function and the Jacobian determinant associated with the transformation are determined by:
\begin{equation}	
	e^{-\frac{\xi _{x}^{2}+\xi _{y}^{2}}{T_m}}=e^{-R^2\left( r \right)}, ~~ J=\left| \frac{\partial \left( \xi _x,\xi _y\right)}{\partial \left( r,\theta \right)} \right|=T_mR\left( r \right) R^{\prime}\left( r \right).
	\label{eq:Jac2}
\end{equation}
The double integral is thus reformulated as:
\begin{equation}	
	\mathcal{I}_2=\int\limits_{-\infty}^{+\infty}{\int\limits_{-\infty}^{+\infty}{\left( \frac{1}{\pi T_m}e^{-\frac{\xi _{x}^{2}+\xi _{y}^{2}}{T_m}} \right) \mathcal{F}\left( \xi _x,\xi _y \right) d\xi _xd\xi _y}}=\frac{1}{2\pi}\int\limits_{r_{min}}^{r_{max}}{w^P(r)}\left[ \int\limits_0^{2\pi}{\mathcal{F} \left( \xi _x\left( r,\theta \right) ,\xi _y\left( r,\theta \right) \right)}d\theta \right] dr.
	\label{eq:inte2}
\end{equation}
Here, $w^P(r)=2R\left( r \right) R^{\prime}\left( r \right) e^{-R^2\left( r \right)}$ denotes the weight function in the polar coordinate system. The variability of radial functions significantly influences the diversity of the weight function, facilitating the application of various Gaussian quadrature methods for the integrative computation of Eq.~(\ref{eq:inte2}). Additionally, numerous numerical quadrature techniques are available for integrating \(\theta\) over a finite interval \([0, 2\pi]\), including Gauss-Legendre, Gauss-Chebyshev, and Newton-Cotes methods. In this paper, we utilize the $N_{\theta}$-point periodic trapezoid rule with the following abscissas and weights \citep{kovvali2022}:
\begin{equation}
	\theta_j=\theta_0+\frac{2j\pi}{N_{\theta}},~~
	w_j=\frac{2\pi}{N_{\theta}}, ~~for j=1,2,...,N_{\theta}.
	\label{eq:nptr}
\end{equation}

The selection of the numerical integration method in the radial direction depends on the specific radial function $R(r)$. For instance, consider $R(r) = \sqrt{r^{\alpha}}$, where $\alpha > 0$ and $r \in (0, \infty)$. Under these conditions, the corresponding weight function is expressed as $w^P(r) = \alpha r^{\alpha-1}e^{-r^{\alpha}}$. Notably, when $\alpha = 1$, the weight function simplifies to $w^P(r) = e^{-r}$, permitting the use of the standard Gauss-Laguerre quadrature formula. Conversely, for $\alpha = 2$, the weight function becomes $w^P(r) = 2re^{-r^2}$, and in this instance, Shizgal’s \citep{shizgal1981} methodology offers a reliable framework for calculating the associated Gauss points and weight coefficients.
Moreover, when $R(r) = \sqrt{\ln r^{-(\beta+1)}}$ for $r \in (0, 1)$ and $\beta > -1$, the resultant weight function transforms into $w^P(r) = (\beta + 1)r^{\beta}$. This condition facilitates the use of the Gauss-Jacobi quadrature formula characterized by $w^P(r) = r^{\beta}(1-r)^{\alpha}$ within the interval (0,1), optimally configured to establish precise abscissas and weights. 
To encapsulate these insights, our research delineates two sets of PGQ strategies for 2D cases, as elaborated subsequently.

(P1) Generalized Laguerre rule with weight function $w^P(r)=\alpha r^{\alpha-1}e^{-r^{\alpha}}$
\begin{equation}
	\boldsymbol{\xi}_{n}=\sqrt{T_m r^{\alpha}_{_{\mathcal{L},i}}}\left( \cos \theta_j,~\sin \theta_j \right),~~
	\omega_{n}=\frac{\omega_{_{\mathcal{L},i}}}{N_{\theta}} 
	\label{eq:GL2}
\end{equation}

(P2) Gauss-Jacobi rule with  weight function $w^P(r)=(\beta +1)r^{\beta}$
\begin{equation}
	\boldsymbol{\xi }_{n}=\sqrt{T_m  \ln r^{-(\beta+1)}_{_{\mathcal{J},i}}}\left( \cos \theta_j,~\sin \theta_j \right),~~
	\omega_{n}=\frac{\omega_{_{\mathcal{J},i}}}{N_{\theta}} 
	\label{eq:GJ2}
\end{equation}
where $1\leqslant i\leqslant N_r$, $1\leqslant j\leqslant N_{\theta}$. The values of the abscissas $r_{_{\mathcal{G},i}}$ ($_\mathcal{G}=_\mathcal{L},_\mathcal{J}$) along with their respective weights $\omega_{_{\mathcal{G},i}}$ are outlined in \ref{APPA} and \ref{APPB}. 

It is evident from the description that the Gauss points generated from Eq.~(\ref{eq:GL2}) and Eq.~(\ref{eq:GJ2}) can accommodate an adjustable parameter, either \(\alpha\) or \(\beta\). This capacity for adjustment endows Gaussian quadratures with the flexibility to extend beyond the limitations associated with static Gauss points. Fig.~\ref{fig:RadDis} presents a comparative assessment of Gauss point distributions for varying parameters. Specifically, Fig.~\ref{fig:RadDis}(a) displays the distribution of Gauss points according to the P1 rule with differing $\alpha$ parameters. In this context, $\alpha = 1$ aligns with the outcome of the standard Gauss-Laguerre quadrature, whereas $\alpha = 2$ corresponds to Gauss points delineated by Shizgal (1981). Notably, an increased parameter $\alpha$ results in a more concentrated Gauss point distribution, as evidenced in Fig.~\ref{fig:RadDis}(a). Conversely, Fig.~\ref{fig:RadDis}(b) exhibits the Gauss point distribution pursuant to the P2 rule across various $\beta$ parameters. Here, a lower value of $\beta$ is associated with a denser distribution of Gauss points. Interestingly, as the parameter $\beta$ increases, the Jacobi rule's distribution demonstrates a convergence towards that of the Gauss-Laguerre rule.

Discrete velocities, as determined by Eq.~(\ref{eq:GL2}) and Eq.~(\ref{eq:GJ2}), are dispersed across orbits of circles with varying polar diameters. The principle of integral interval additivity permits a non-uniform distribution of discrete points among various orbits, thereby significantly enhancing the method's adaptability in velocity discretization. This attribute enables the precise allocation of discrete velocity points on selected orbits to meet specific computational demands. Moreover, the weight coefficient, defined by Eq.~(\ref{eq:nptr}), is solely contingent upon the number of discrete points, ensuring that the non-uniform velocity discretization does not add to the algorithm's complexity. Fig.~\ref{fig:NorDis} illustrates the classification of velocity distributions on different orbits into uniform (A-type) and non-uniform (B-type) distributions.

\subsection{3D Parametric Gaussian quadratures}
\label{sec3.2}
For the numerical calculation of triple integrals, the present work aims to  establish a Gaussian quadrature rule with the following form:
\begin{eqnarray}
	\mathcal{I}_3=\int\limits_{-\infty}^{+\infty}{\int\limits_{-\infty}^{+\infty}{\int\limits_{-\infty}^{+\infty}{\left( \frac{1}{\pi T_m} \right) ^{\frac{3}{2}}e^{-\frac{\xi _{x}^{2}+\xi _{y}^{2}+\xi _{z}^{2}}{T_m}}\mathcal{F}\left( \xi _x,\xi _y,\xi _z \right) d\xi _xd\xi _yd\xi _z}}}=\sum_{n=1}^{\mathcal{N}}{\omega_n\mathcal{F}\left( \boldsymbol{x},\boldsymbol{\xi }_n,t \right)}.
	\label{eq:I3}
\end{eqnarray}

There are two common methods for triple integral transformation. A straightforward approach is to add an axial transformation to Eq.~(\ref{eq:pct}), specifically through the implementation of a cylindrical coordinate system. Another effective triple integral transformation is the SCT. In contrast, the spherical distribution may be more consistent with the characteristics of the velocity distribution. Therefore, the generalized SCT is considered in this paper:
\begin{equation}	
	\xi _x=\sqrt{T_m}R\left( r \right)\widehat{\varPhi} \left( \phi \right)\cos \theta, ~~
	\xi _y=\sqrt{T_m}R\left( r \right)\widehat{\varPhi} \left( \phi \right)\sin \theta, ~~
	\xi _z=\sqrt{T_m}R\left( r \right) \varPhi \left( \phi \right) .
	\label{eq:sct}
\end{equation}
Here, $R\left( r \right) \in \left( 0,+\infty \right)$, $\theta \in \left[ 0,2\pi \right]$, $\varPhi \left( \phi \right)  \in \left[ -1,1 \right]$, and $\widehat{\varPhi} \left( \phi \right)=\sqrt{1-\varPhi^{2} \left( \phi\right) }$. Accordingly, the transformed weight function and the Jacobian determinant are

\begin{equation}	
	e^{-\frac{\xi _{x}^{2}+\xi _{y}^{2}+\xi _{z}^{2}}{T_m}}=e^{-R^2\left( r \right)}, ~~ J=\left| \frac{\partial \left( \xi _x,\xi _y,\xi _z \right)}{\partial \left( r,\theta ,\phi \right)} \right|=R^2\left( r \right) R^{\prime}\left( r \right) \varPhi ^{\prime}\left( \phi \right).
	\label{eq:Jac3}
\end{equation}

By the generalized SCT, Eq.~(\ref{eq:I3}) is transformed into the following form:
\begin{eqnarray}
	\mathcal{I}_3=\frac{1}{2\pi ^{\frac{3}{2}}}\int\limits_{r_{min}}^{r_{max}}{w^S\left( r \right) \left\{ \int\limits_0^{2\pi}{\left[ \int\limits_{\phi_{min}}^{\phi_{max}}{w^S\left( \phi \right)\mathcal{F} \left( \xi _x\left( r,\theta ,\phi \right) ,\xi _y\left( r,\theta ,\phi \right) ,\xi _z\left( r,\theta ,\phi \right) \right) d\phi} \right] d\theta} \right\} dr}
	\label{eq:sct3}
\end{eqnarray}
The two weight functions in Eq.~(\ref{eq:sct3}) are $w^S\left( r \right) =R^2\left( r \right) R^\prime\left( r \right) e^{-R^2\left( r \right)}$, and $w^S\left( \phi \right)=\varPhi ^{\prime}\left( \phi \right) $. In this study, the radial functions established in Section \ref{sec3.1}, namely $R\left( r \right)=\sqrt{r^{\alpha}}$ ($\alpha>0,~r\in \left( 0,+\infty \right)$) and $R\left( r \right) =\sqrt{\ln r^{-(\beta+1)}}$ ($\beta>-1,~r\in \left(0,1 \right)$) are further utilized. Moreover, the azimuth function takes the form of a power function, that is $\varPhi \left( \phi \right) = \phi ^{\gamma}(\phi\in \left[ -1,1 \right])$. Assuming $\gamma$ to be an odd number without loss of generality, we formulate the following 3D Gaussian quadratures:

(S1) Generalized Laguerre-Jacobi rule with weight function $w^S\left( r \right)=\alpha r^{3\alpha/2-1}e^{-r^{\alpha}}$ and $w^S\left( \phi \right)=\gamma \phi ^{\gamma -1}$
\begin{equation}
	\boldsymbol{\xi }_n=\sqrt{T_m r^{\alpha}_{_{\mathcal{L},i}}}\left( \widehat{\varPhi}_{_{\mathcal{J},k}}\cos \theta_j,~
	\widehat{\varPhi}_{_{\mathcal{J},k}}\sin \theta_j,~{\varPhi}_{_{\mathcal{J},k}}  \right), ~~
	\omega_n=\frac{\omega_{_{\mathcal{L},i}}\omega_{_{J,k}}}{N_{\theta}} 
\end{equation}

(S2) Logarithmic Jacobi-Jacobi rule with  weight function $w^S\left( r \right)=\sqrt{-\left( \beta +1 \right) ^3\ln r}r^{\beta}$ and $w^S\left( \phi \right)=\gamma \phi ^{\gamma -1}$
\begin{equation}
	\boldsymbol{\xi }_n=\sqrt{T_m\ln r_{_{\mathcal{J}_L,i}}^{-(\beta +1)}}\left(  \widehat{\varPhi}_{_{\mathcal{J},k}}\cos \theta_j,~
	\widehat{\varPhi}_{_{\mathcal{J},k}}\sin \theta_j,~ {\varPhi_{_{\mathcal{J},k}}} \right), ~~
	\omega_n= \frac{\omega_{_{\mathcal{J}_L,i}}\omega_{_{\mathcal{J},k}}}{N_{\theta}}
\end{equation}

In strategy S1, existing literature on Gaussian quadratures offers methods for acquiring the radial distribution points $r^{\alpha}_{_{\mathcal{L},i}}$ and their corresponding weights $\omega_{_{\mathcal{L},i}}$, specifically tailored to particular parameters. For instance, for $\alpha=1,2$, the weight functions are $\sqrt{r}e^{-r}$ and $2r^{2}e^{-r^{2} }$, respectively. These abscissas and weights can be efficiently obtained through the utilization of techniques like the generalized Gauss-Laguerre quadrature and Shizgal’s methodology. Conversely, strategy S2 calls for the development of a new Gaussian quadrature formula to accommodate a specific weight function. The specific abscissas and weights referred to in S2 are detailed in \ref{APPB} for application in this study.

\section{Numerical tests}
\label{sec4}

In this section, we evaluate the effectiveness of the PGQ method through a series of benchmark examples. All simulations are conducted on an Intel$^\circledR$ Core$^\text{TM}$ i7-9700k CPU with a 64-bit operating system.We compare the computational results with those obtained using traditional Newton-Cotes or half-range Gauss-Hermite quadrature rules within the DUGKS framework. The first three numerical examples utilize the 2D PGQ proposed in this study. Subsequently, the performance of the 3D PGQ is demonstrated through the simulation of temperature-discontinuity-induced (TDI) cavity flow in Section \ref{sec4.4}. For simulation of steady flow in this paper, the convergent criterion is given by
\begin{equation}
	\frac{\sqrt{\sum\left|\boldsymbol{u}(\boldsymbol{x},t) -\boldsymbol{u}(\boldsymbol{x},t-1000\Delta t)\right|^2}}{\sqrt{\sum\left|\boldsymbol{u}(\boldsymbol{x},t) \right|^2 }}<10^{-6}
\end{equation}

\subsection{Couette flow}
\label{sec4.1}

\begin{figure*}[!t]
	\centering
	\includegraphics[width=7cm,height=6.5cm]{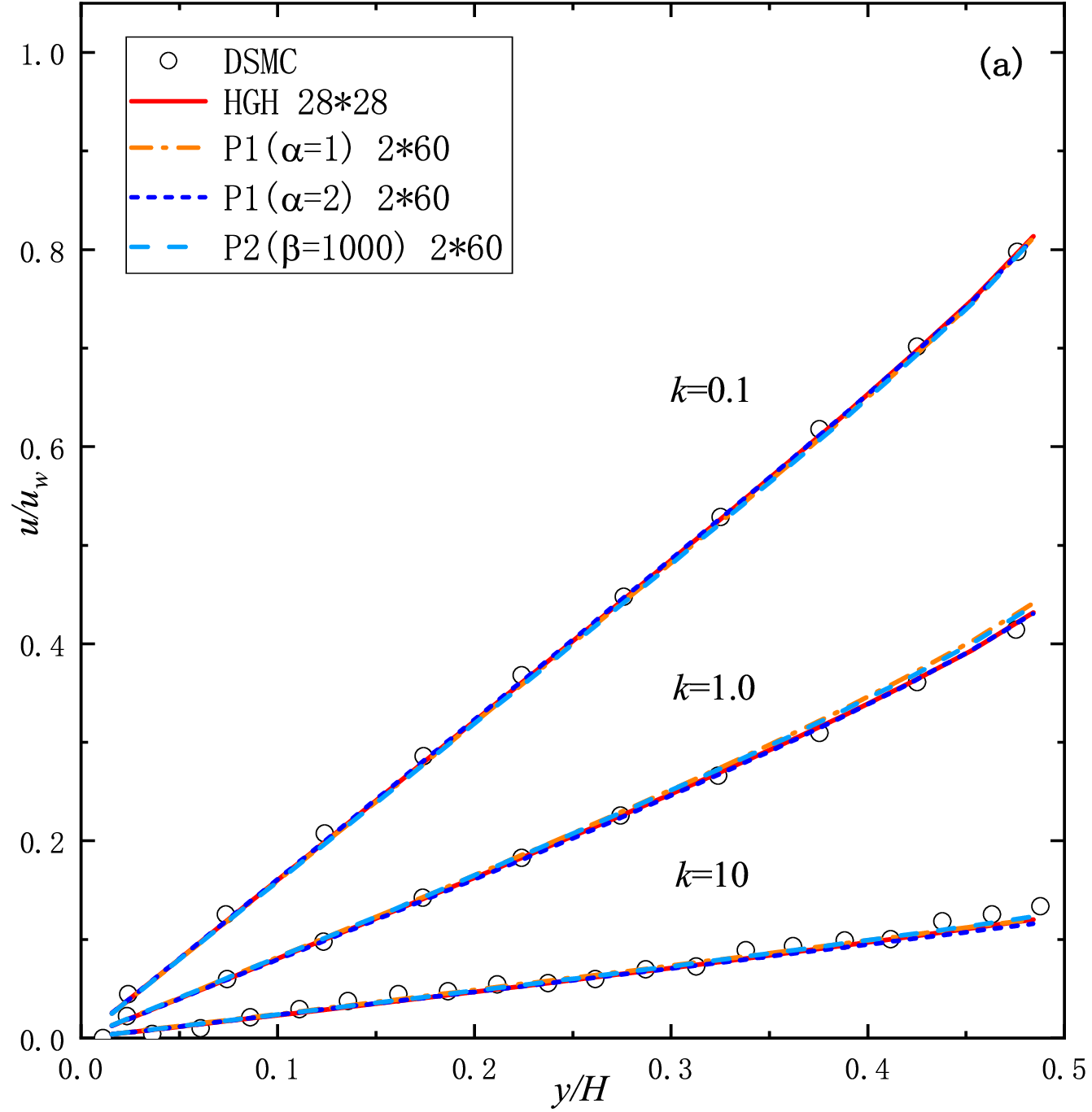}
	\includegraphics[width=7cm,height=6.5cm]{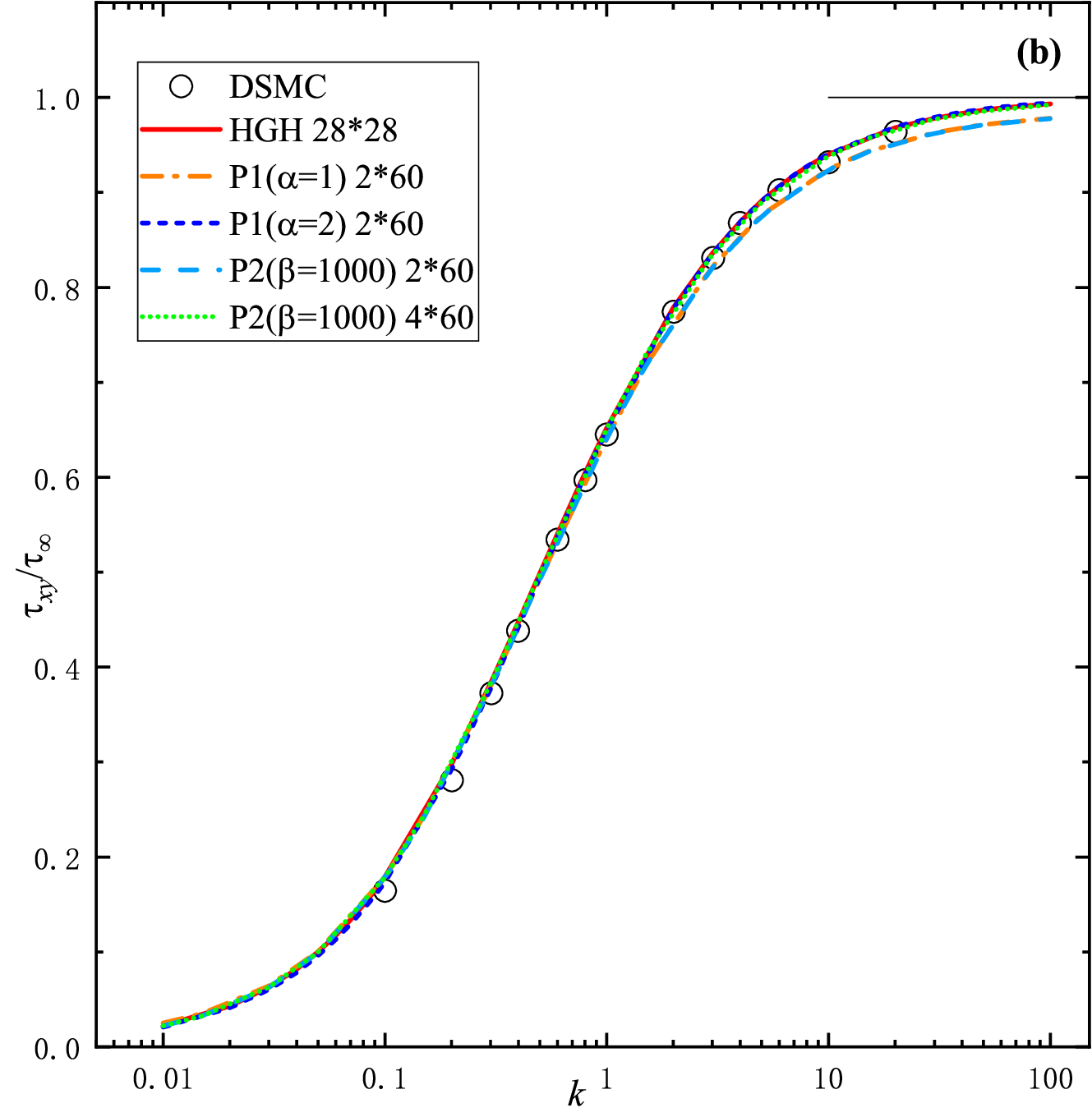}
	\caption{\label{fig:isoCouette} \centering Velocity (a) and stress (b) profiles of the isothermal Couette flow calculated by different Gaussian quadratures.}
\end{figure*}
\begin{figure*}[!t]
	\centering
	\includegraphics[width=7cm,height=6.5cm]{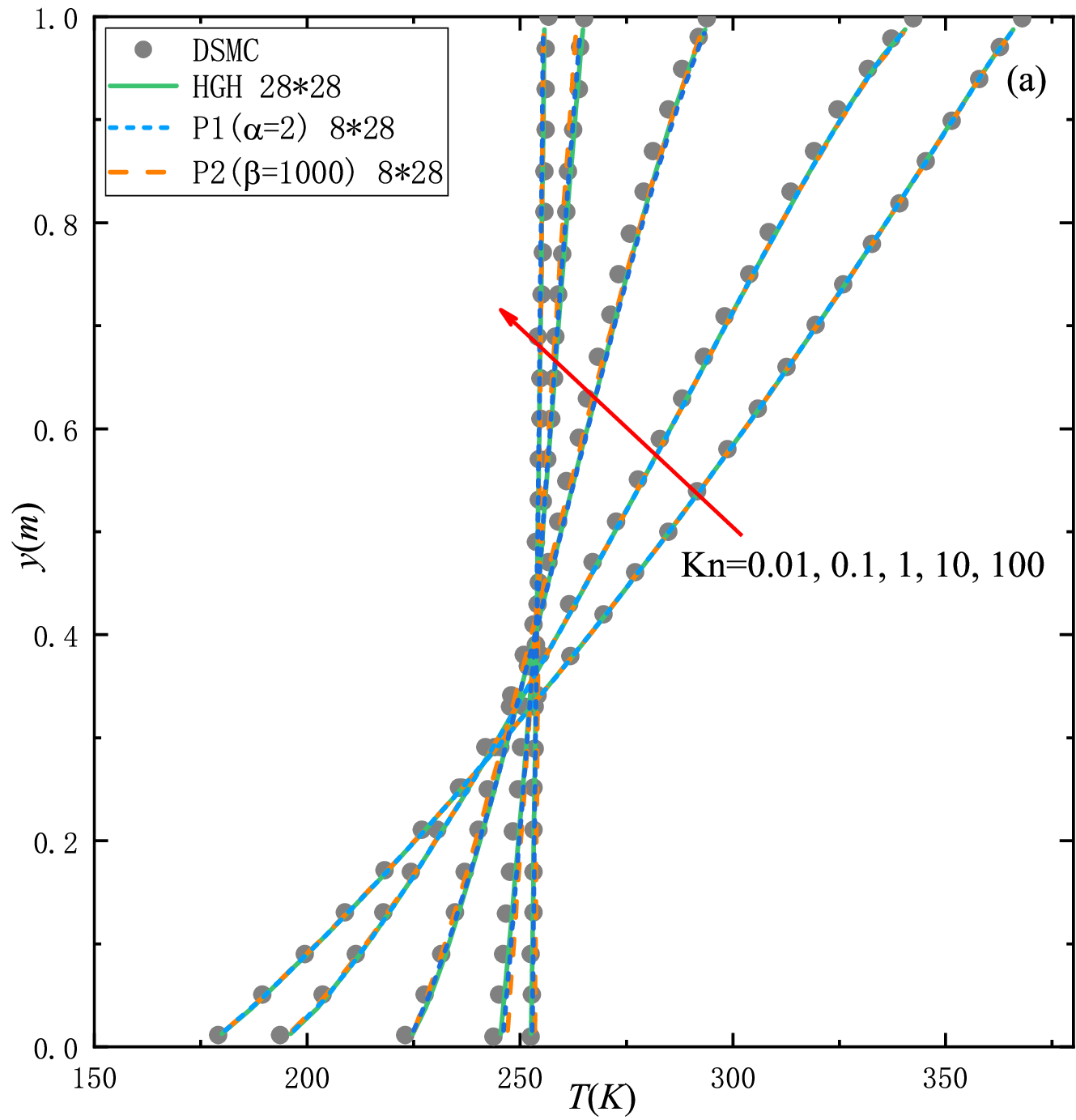}
	\includegraphics[width=7cm,height=6.5cm]{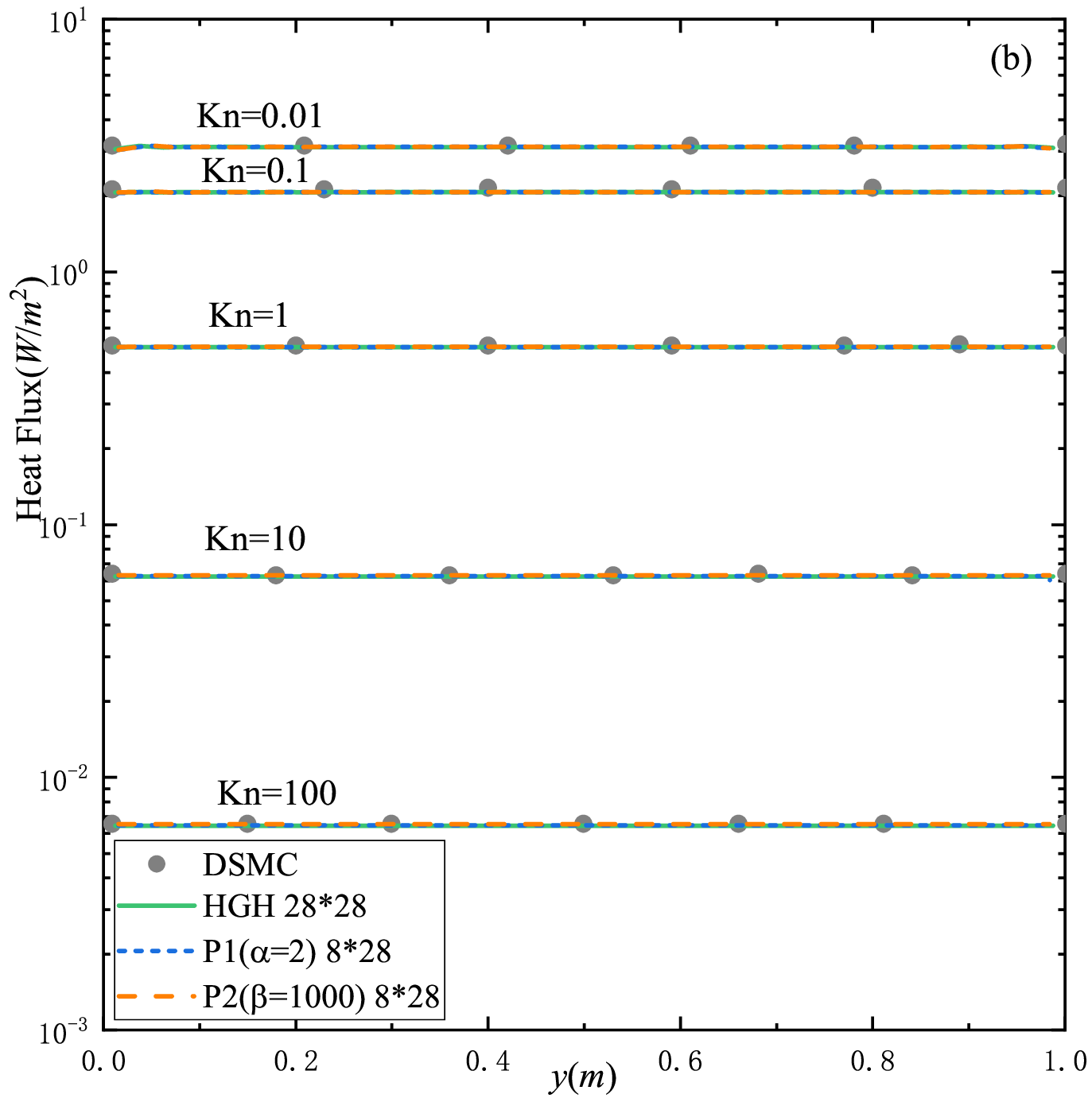}
	\caption{\label{fig:temCouette} \centering Temperature (a) and Heat flux (b) profiles of the thermal Couette flow calculated by different Gaussian quadratures.}
\end{figure*}

The first case considered is the Couette flow of rarefied gas between two infinitely extended parallel plates, encompassing both isothermal and thermal Couette flows. Isothermal Couette flow arises from the relative movement of the two parallel plates, while thermal Couette flow is driven by an imposed temperature differential across the plates. Considering the flow’s periodic nature in both cases, we discretize the computational domain into a \(2 \times 32\) uniform grid. Previous works \citep{Huang2013, wl2023} have confirmed the effectiveness of the half-range Gaussian Hermite (HGH) quadrature rule with a discrete velocity of \(28 \times 28\), closely aligning with the DSMC method throughout the entire flow field. Consequently, these validated results serve as benchmark solutions for evaluating the effectiveness of our proposed method.

The reduced Shakhov model expressed in Eq.~(\ref{eq:RSM}) requires solving two distribution function equations. For isothermal Couette flow, the absence of a thermal field simplifies the procedure, requiring only the solution of the distribution function \(g_0\). Furthermore, we set \(Pr=1\) here to streamline the calculation for the isothermal case. Fig.~\ref{fig:isoCouette} illustrates a comparison of three different Gaussian quadrature rules under various Knudsen numbers, adjusted by \(k=\frac{\sqrt{\pi}}{2}Kn\). These methods include the HGH with \(28 \times 28\) nodes and two PGQ rules with \(2 \times 60\) nodes introduced in Eq.~(\ref{eq:GL2}) and Eq.~(\ref{eq:GJ2}). Generally, for problems effectively addressed by the HGH quadrature rule, the velocity distribution tends to be more dispersed. Hence, a large exponent (\(\beta=1000\)) is employed for the Gauss-Jacobi rule P2. As shown in Fig.~\ref{fig:isoCouette}, for a large value of \(\beta\) (\(\beta=1000\)), the computation results of P2 align with P1 (\(\alpha=1\)), confirming our conclusion from Fig.~\ref{fig:RadDis}(b). Consequently, in subsequent examples, we will no longer present calculation results for both P1 (\(\alpha=1\)) and P2 with a large value of \(\beta\) simultaneously. Regarding the velocity distribution, as shown in Fig.~\ref{fig:isoCouette}(a), the results obtained using PGQ are highly consistent with those of HGH, particularly when considering P1 (\(\alpha=2\)). When examining the stress distribution, calculations using P1 (\(\alpha=2\)) with \(2 \times 60\) discrete velocities exhibit strong agreement with HGH using \(28 \times 28\) discrete velocities. However, for larger Knudsen numbers, the prediction errors of P1 (\(\alpha=1\)) and P2 (\(\beta=1000\)) become more prominent, necessitating more discrete velocities. Overall, the proposed PGQ in this work predicts isothermal Couette flow effectively, with P1 (\(\alpha=2\)) exhibiting the highest computational efficiency.

The thermal Couette flow has been utilized by Sun et al. \citep{SUN2002} and Huang et al. \citep{Huang2013} to assess the capability of information-reservation DSMC and the UGKS in capturing thermal effects. Following their simulation setup, the distance between the two plates is maintained at 1 meter, with temperatures of 373 K and 173 K assigned to the upper and lower plates, respectively. Across a range of Knudsen numbers (Kn = 0.01, 0.1, 1, 10, 100), we scrutinize the performance of PGQ across various flow regimes. Accurate assessment of thermal phenomena necessitates the computation of high-order velocity moments, particularly temperature (T) and heat flux (q). To facilitate this, a denser radial node distribution is employed. Consequently, we opt for an \(8 \times 28\) discrete velocity mesh for both the P1 (\(\alpha=2\)) and P2 (\(\beta=1000\)) methods. Additionally, we set \(\theta_0=\pi/N_{\theta}\) and the median temperature \(T_m=323K\) to optimize computational efficiency. Fig.~\ref{fig:temCouette} demonstrates the excellent agreement between the calculation results obtained using PGQ and HGH. This comparison underscores PGQ’s ability to maintain high computational efficiency in addressing thermal flow problems.

\subsection{Thermal creep flow}
\label{sec4.2}

\begin{table*}[!h]
	\caption{\label{tab:tab1} Velocity discretization settings for thermal creep flow under different Kn numbers}
	\centering
	\begin{tabular}{ccccc}
		\hline
		-- & Kn=0.01 & Kn=0.1 & Kn=1.0 & Kn=10.0 \\
		\hline
		Zhu et al. \citep{ZHU2019} & $16 \times 16$ HGH & $28 \times 28$ HGH & $161 \times 161$ NC & $201 \times 201$ NC \\
		Present & $8 \times 18$ PGQ & $8 \times 36$ PGQ & $8 \times 90$ PGQ & $8 \times 120 $ PGQ \\		
		\hline
	\end{tabular}
\end{table*}

\begin{figure*}[!t]
	\centering
	\includegraphics[width=15cm,height=3.5cm]{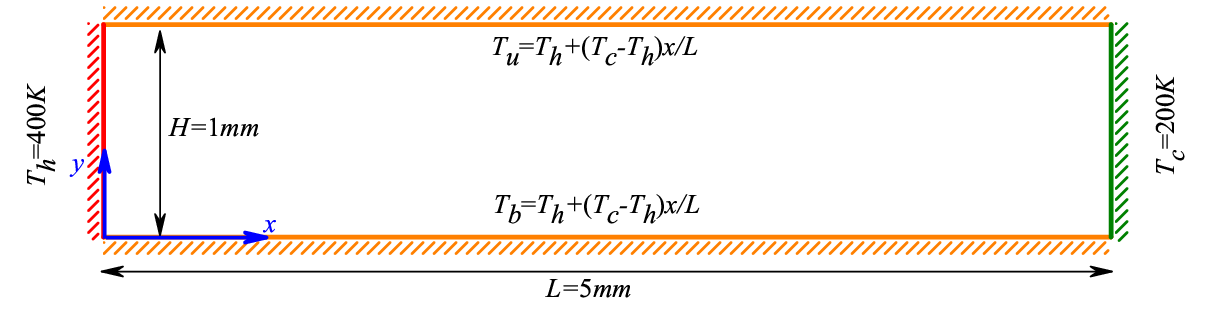}
	\caption{\label{fig:GeoCreep} \centering Schematic diagram of the thermal creep flow.}
\end{figure*}

\begin{figure*}[!t]
	\centering
	\subfigure[Kn=0.01]{
		\label{creep0.01}
		\includegraphics[width=15cm,height=3.3cm]{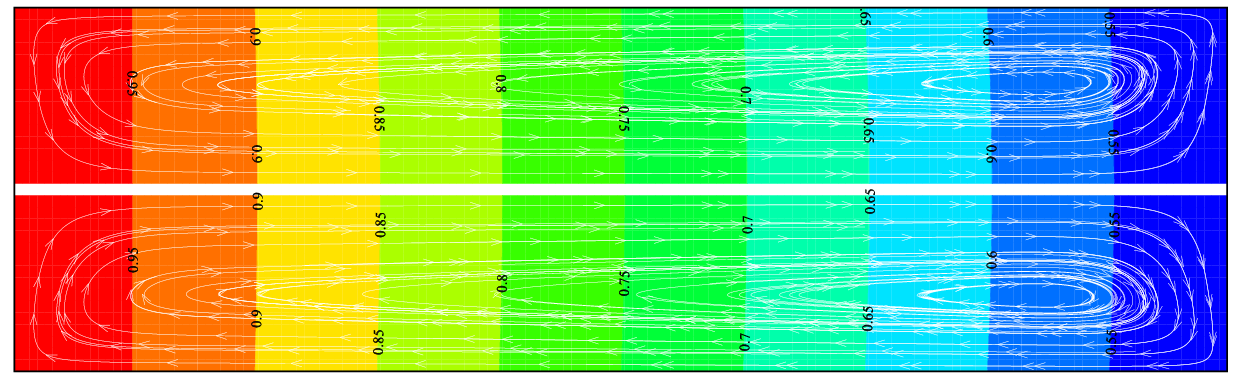}}
	\subfigure[Kn=0.1]{
		\label{creep0.1}
		\includegraphics[width=15cm,height=3.3cm]{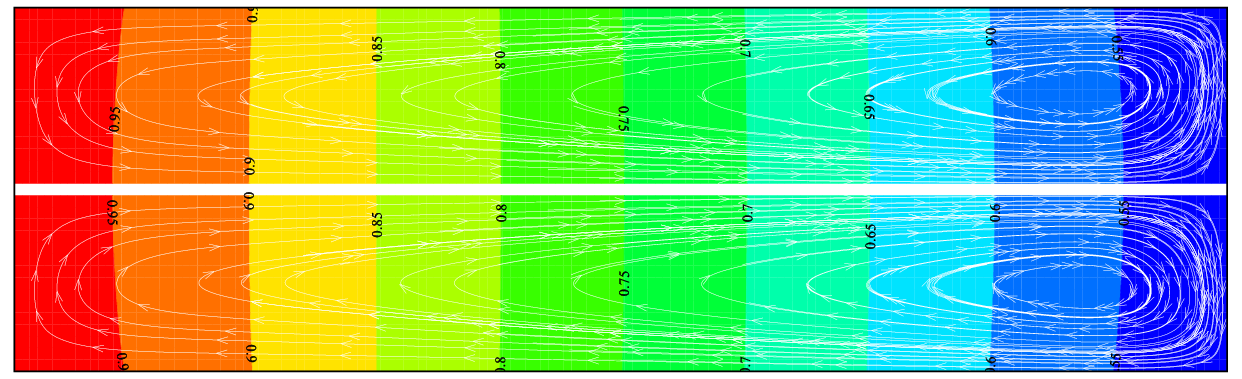}}
	\subfigure[Kn=1.0]{
		\label{creep1.0}
		\includegraphics[width=15cm,height=3.3cm]{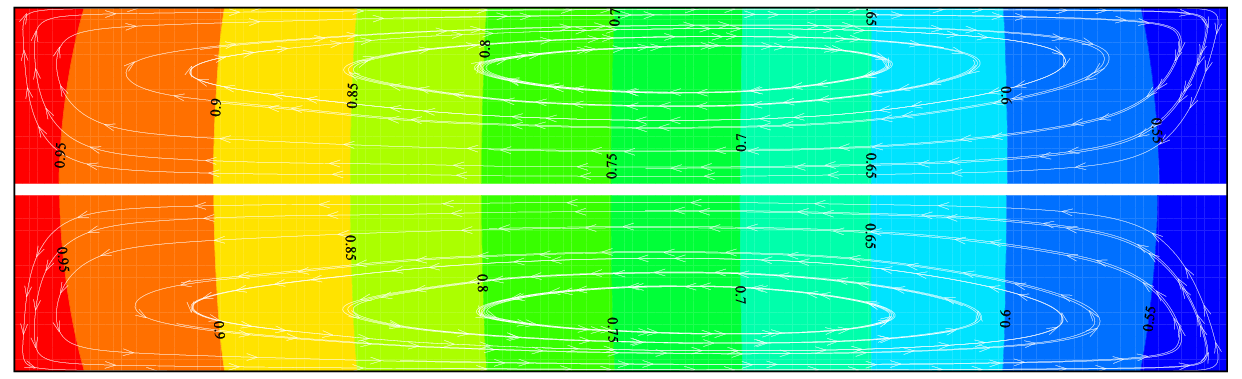}}
	\subfigure[Kn=10.0]{
		\label{creep10.0}
		\includegraphics[width=15cm,height=3cm]{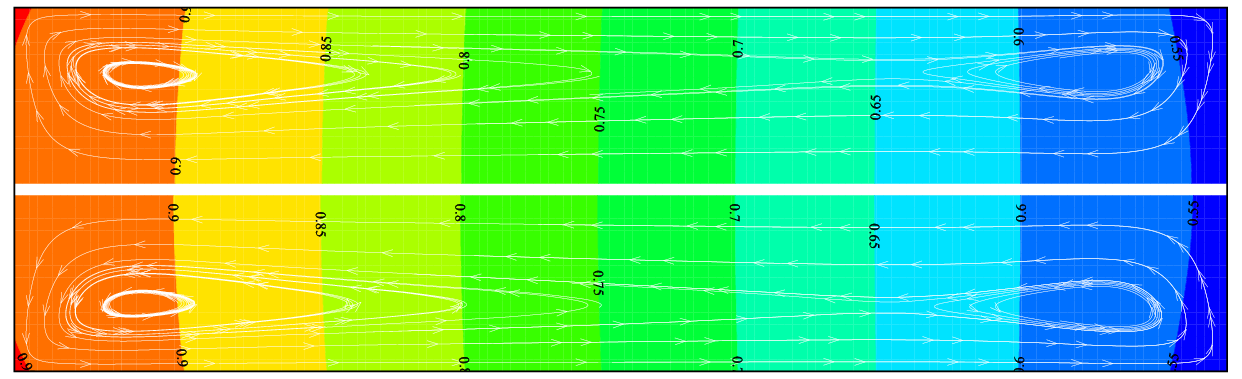}}
	\caption{\label{fig:CreepCount} \centering Temperature contours and streamlines for the thermal creep flow case. Up half: P1 ($\alpha=2$). Down half: P2 ($\beta=1000$).}
\end{figure*}

\begin{figure*}[!t]
	\centering
	\includegraphics[width=7cm,height=6cm]{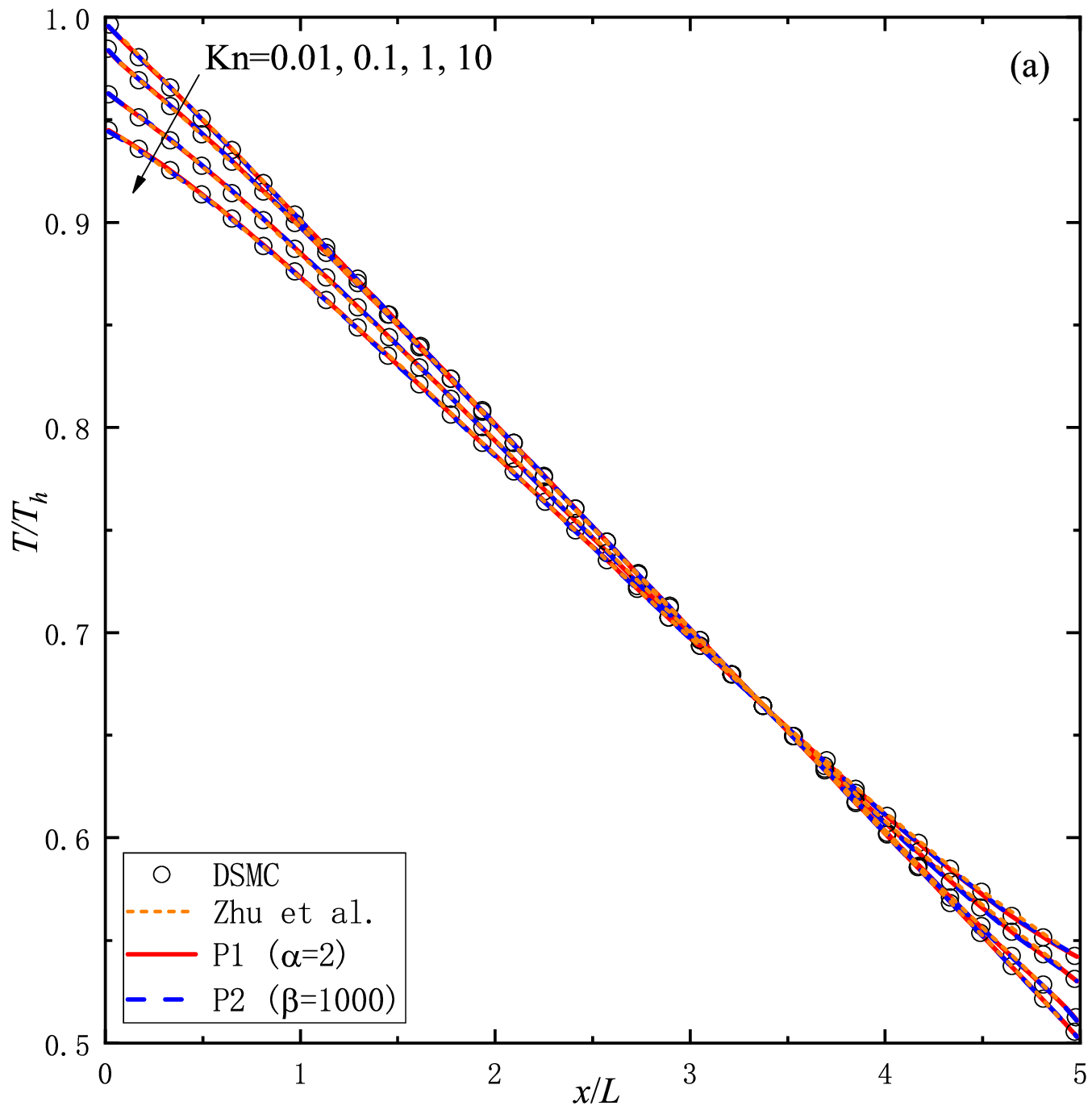}
	\includegraphics[width=7cm,height=6cm]{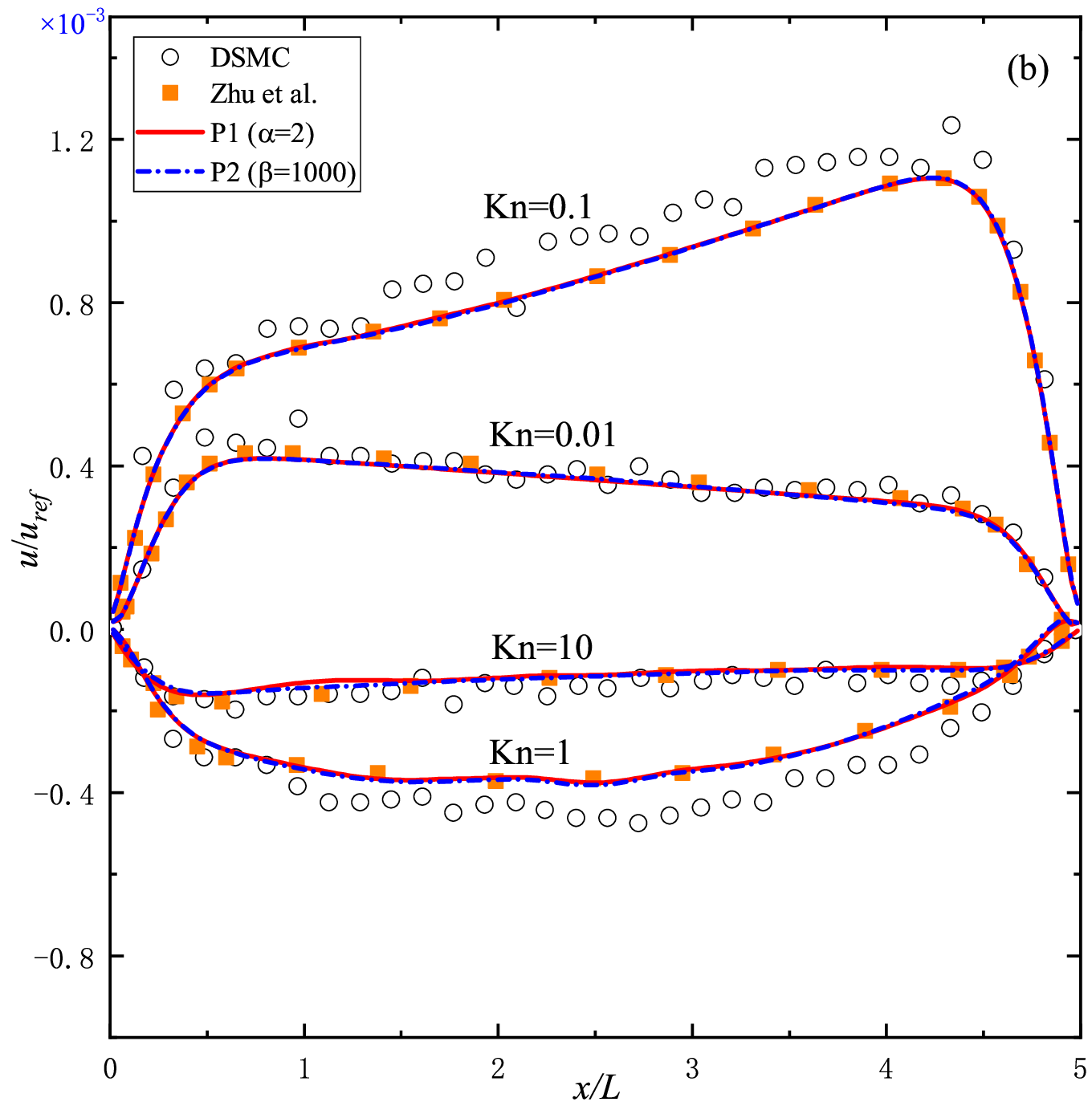}
	\caption{\label{fig:CreepTU} \centering Temperature (a) and U-velocity (b) profiles along the horizontal center line of the thermal creep flow case.}
\end{figure*}

In this section, the PGQ rule is applied to simulate thermal creep flow, focusing on an example used by Zhu et al. \citep{ZHU2019} to validate the DUGKS method. The schematic diagram is depicted in Fig.~\ref{fig:GeoCreep}. Temperatures at the left and right ends are $T_h=400\,K$ and $T_c=200\,K$, respectively. The temperatures on the upper and lower wall surfaces decrease linearly from $T_h$ to $T_c$ along the length of the domain. The reference temperature $T_{\text{ref}}$ is set as $T_h$, whereas the median temperature $T_{\text{m}}=\frac{T_c+T_h}{2}$. Zhu et al. compared the computational results of DUGKS with the DSMC across various Knudsen numbers ($Kn=0.01, ~0.1,~1.0,~10$). The computational domain was discretized into a uniform grid of $50\times250$, with velocity space discretization determined by the half-range Gauss-Hermite quadrature (HGH) and Newton-Cotes (NC) rules. The HGH rule is suitable for low Kn flows, while the NC rule, with finely discretized velocity grids, is employed for high Kn thermal creep flows to capture their strong non-equilibrium effects. To reduce computational cost, a relatively sparse discretization grid is used, consisting of a $32\times160$ uniform grid in physical space, along with two PGQs proposed in this study for velocity space discretization. The settings for discretized velocity are summarized in Table~\ref{tab:tab1}.

Fig.~\ref{fig:CreepCount} compares two forms of PGQ introduced in this study. The temperature contours and streamlines for the thermal creep flow computed by both methods exhibit remarkable consistency, aligning well with existing literature. Fig.~\ref{fig:CreepTU} further quantitatively compares the computational results of P1 ($\alpha=2$) and P2 ($\beta=1000$) with the benchmark solution proposed by Zhu et al. \citep{ZHU2019}. Although there exists a slight disparity in velocity distribution compared to DSMC, the computational results of both PGQ rules closely match those of DUGKS based on HGH and NC rules. It is evident from Table~\ref{tab:tab1} and Fig.~\ref{fig:CreepCount} that the two PGQ rules proposed in this study demonstrate robust performance across simulations involving varying Knudsen numbers. Moreover, they exhibit significantly lower computational overhead compared to the NC rule and even surpass the efficiency of HGH. 

\subsection{Rayleigh flow}
\label{sec4.3}

\begin{table*}[!h]
	\caption{\label{tab:tab2} Comparison between computation time for Rayleigh flow}
	\centering
	\begin{tabular}{c|ccccc}
		\hline
		\multirow{2}{*}{Method}& NC & HGH & P1$(\alpha=1)$ & P1$(\alpha=2)$ & P2$(\beta=5)$\\
		& $(101\times 101)$ & $(28\times 28)$ & $(16\times 32)$ & $(16\times 32)$ & $(16\times 32)$\\
		\hline
		$\xi_{max}^{\ast}$ & 4.0 & 5.2 & 7.2 & 5.8 & 4.2\\
		$\Delta t~(10^{-3})$ & 1.95 & 1.49 & 1.09 & 1.35 & 1.86\\
		Total steps & 121 & 159 & 217 & 175 & 127\\
		CPU time (s) & 133.8 & 12.5 & 11.9 & 8.9 & 6.8\\		
		\hline
	\end{tabular}
\end{table*}

\begin{figure*}[!t]
	\centering
	\includegraphics[width=14cm,height=4cm]{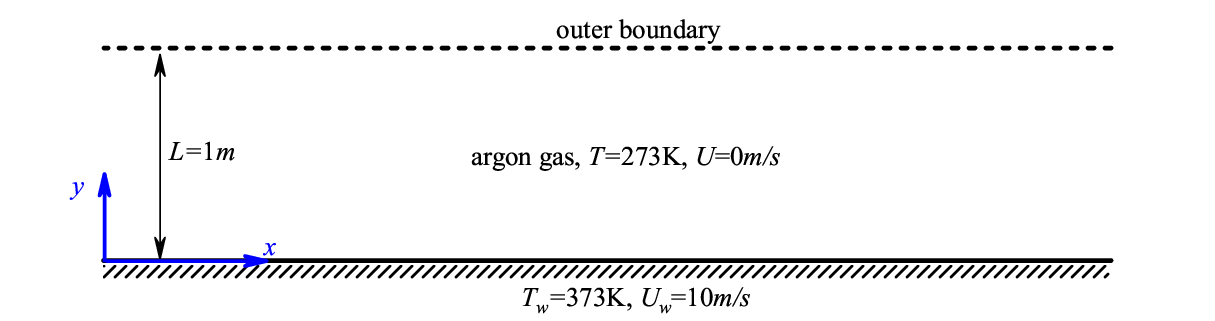}
	\caption{\label{fig:GeoRay} \centering Schematic diagram of the Rayleigh flow.}
\end{figure*}

\begin{figure*}[!t]
	\centering
	\includegraphics[width=5.4cm,height=4.5cm]{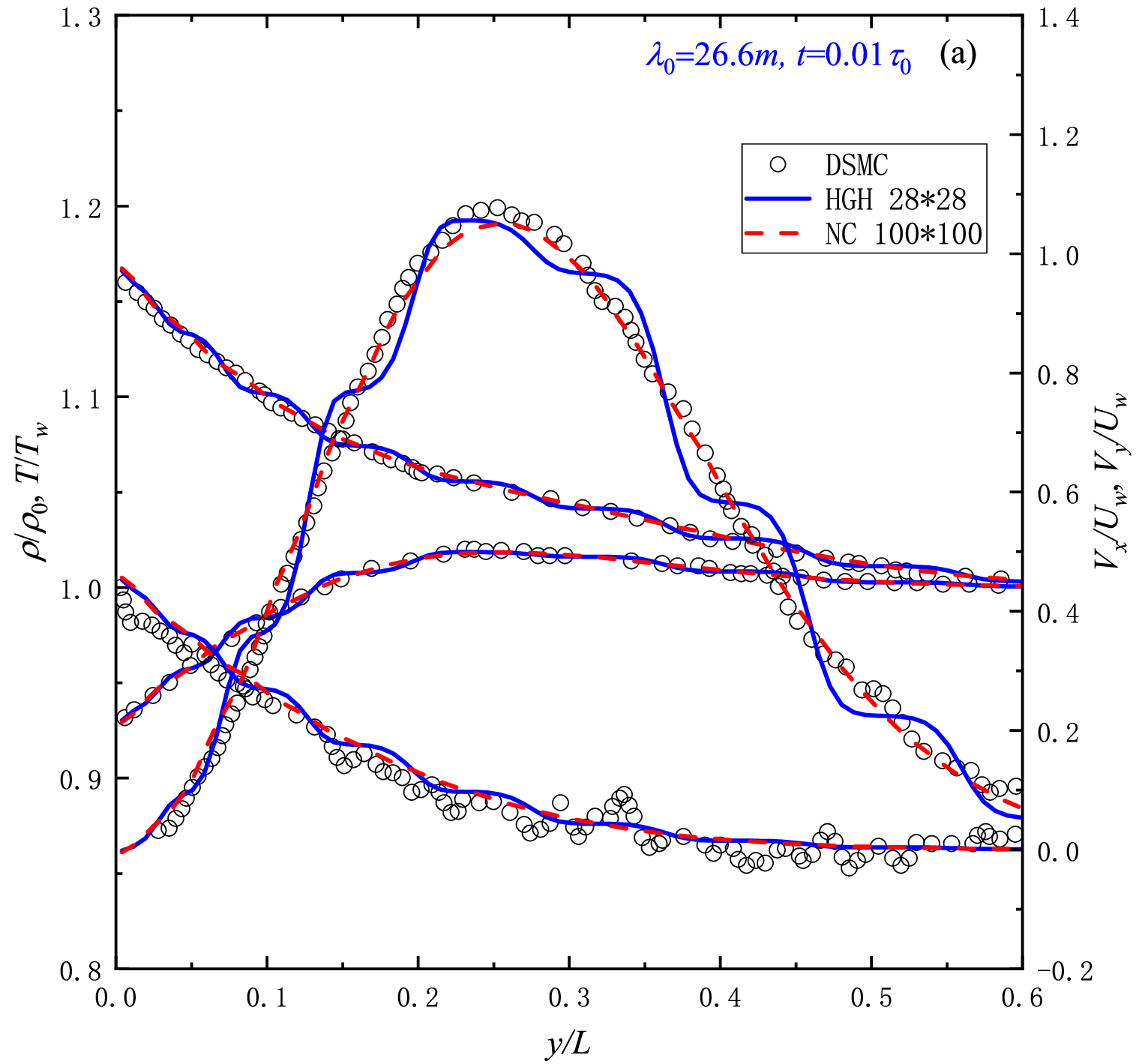}
	\includegraphics[width=5.4cm,height=4.5cm]{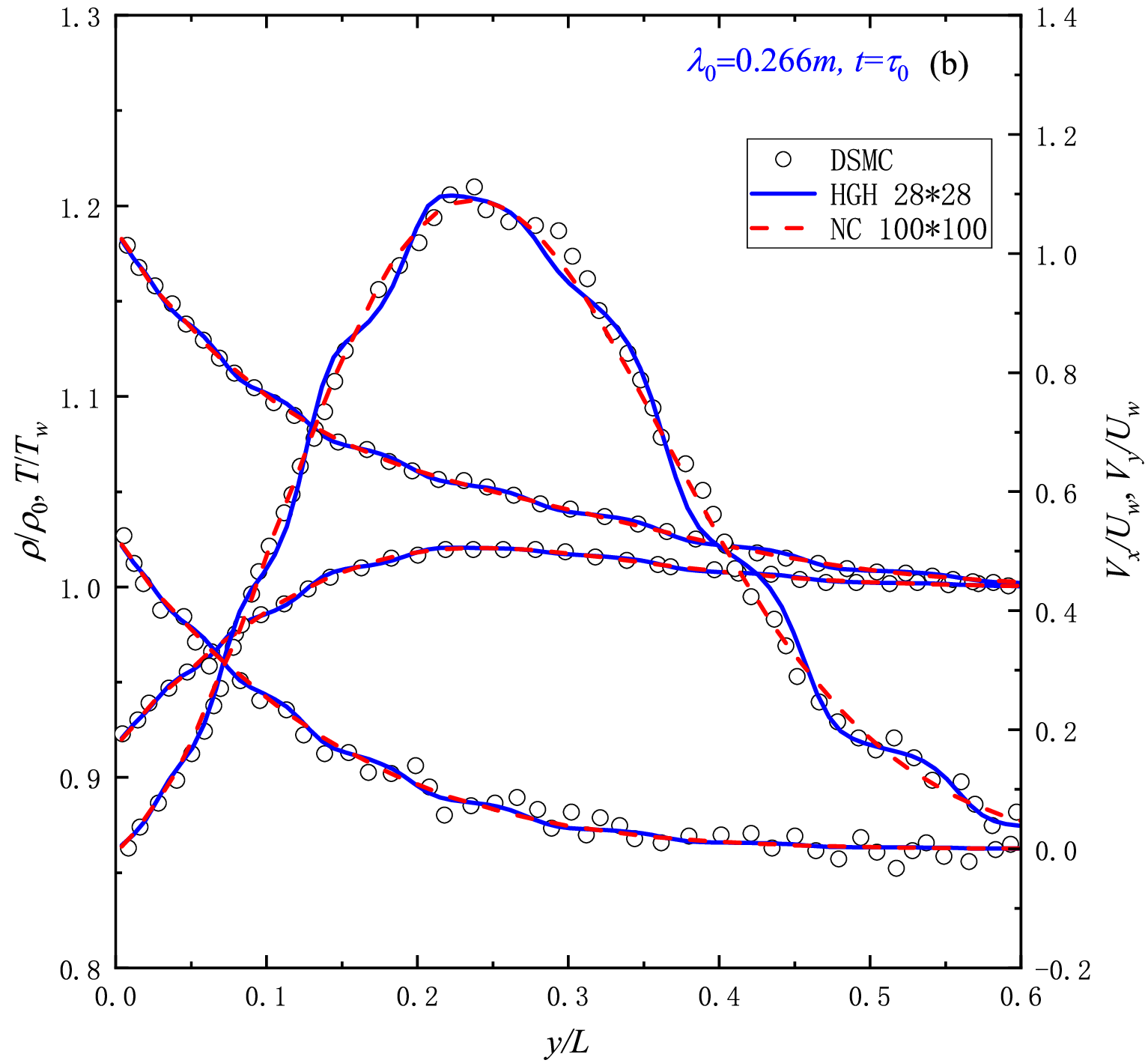}
	\includegraphics[width=5.4cm,height=4.5cm]{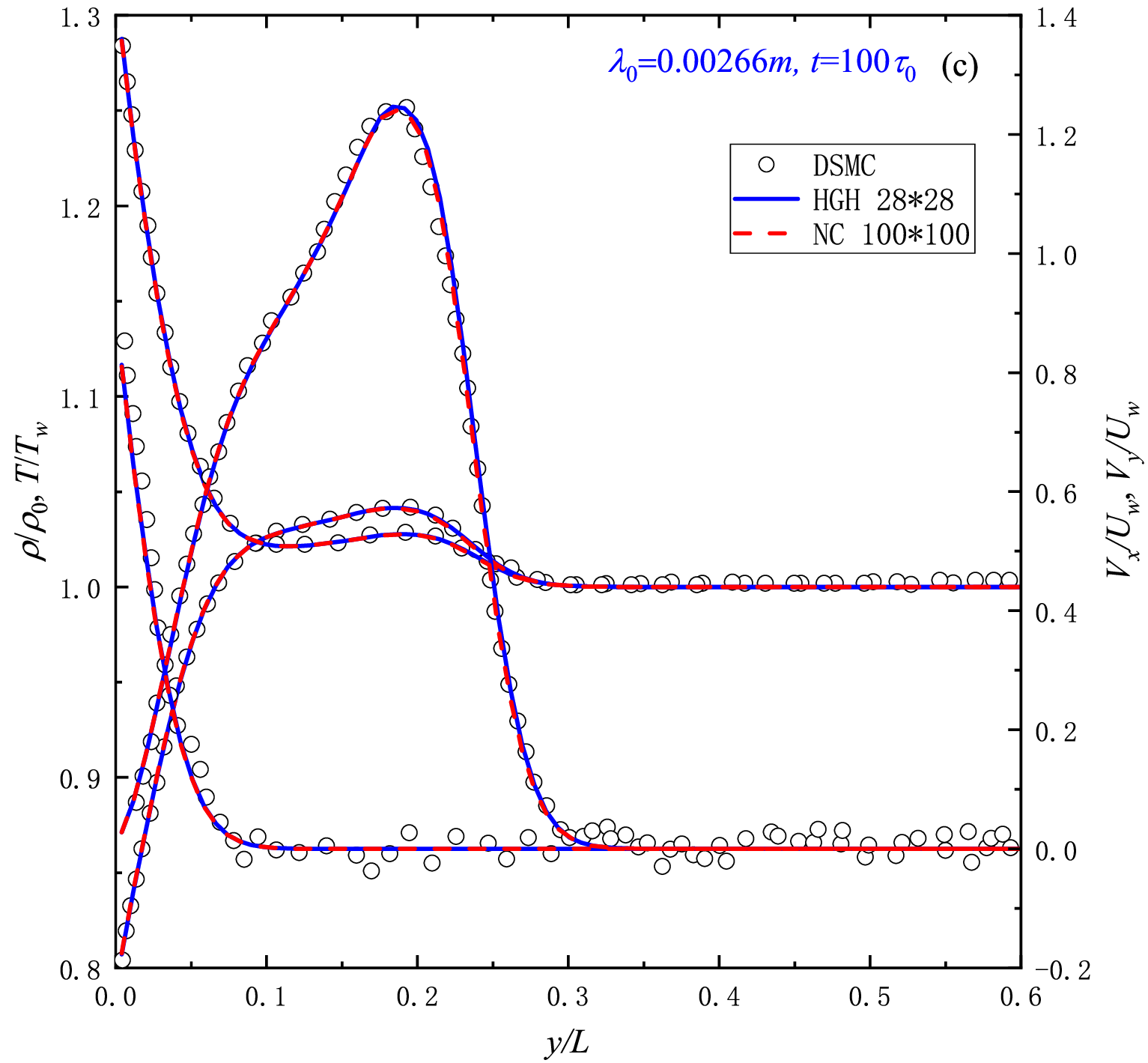}
	\caption{\label{fig:RayHGH} \centering Profiles for the Rayleigh flow by HGH.}
\end{figure*}

\begin{figure*}[!t]
	\centering
	\includegraphics[width=5.4cm,height=4.5cm]{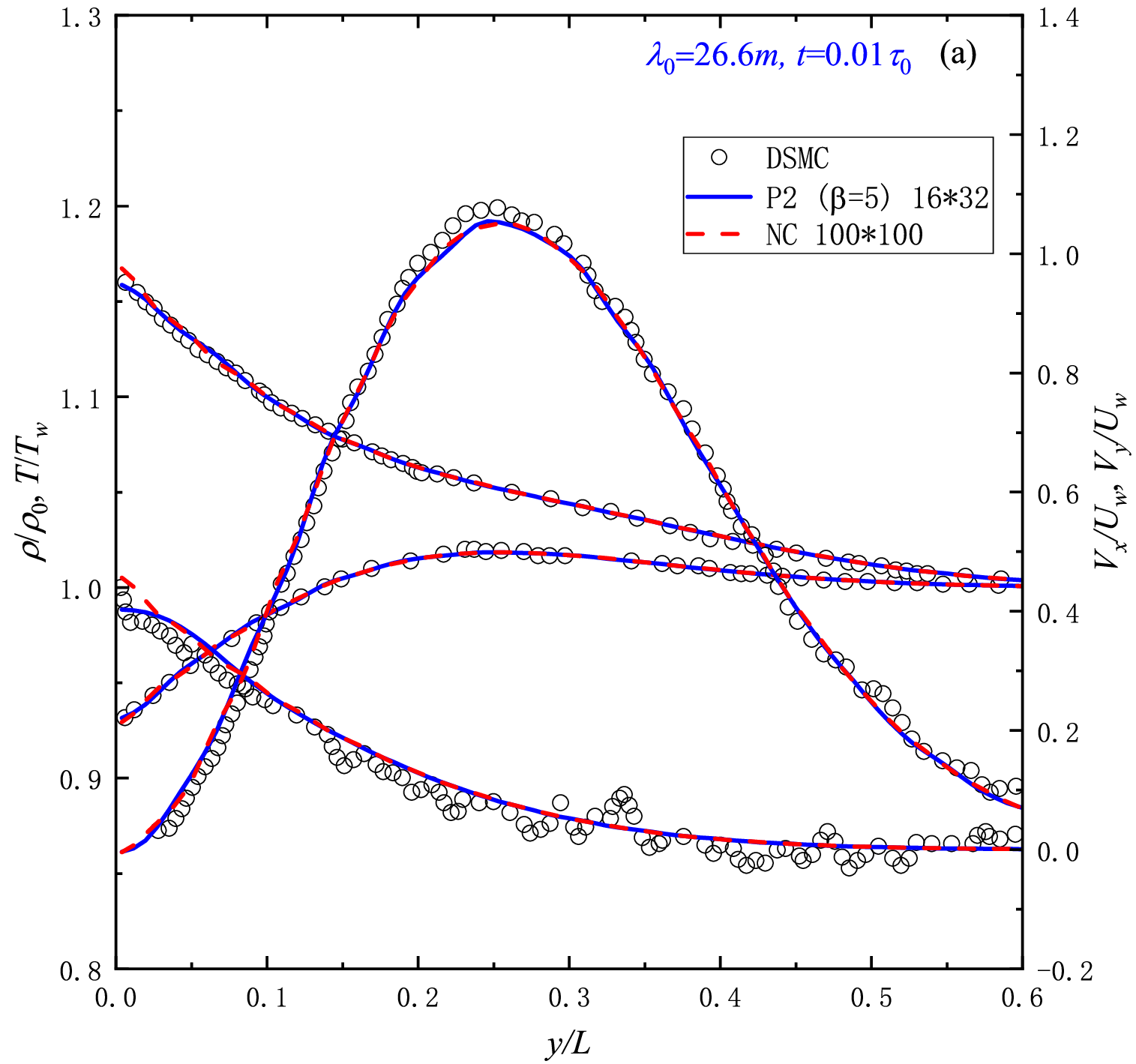}
	\includegraphics[width=5.4cm,height=4.5cm]{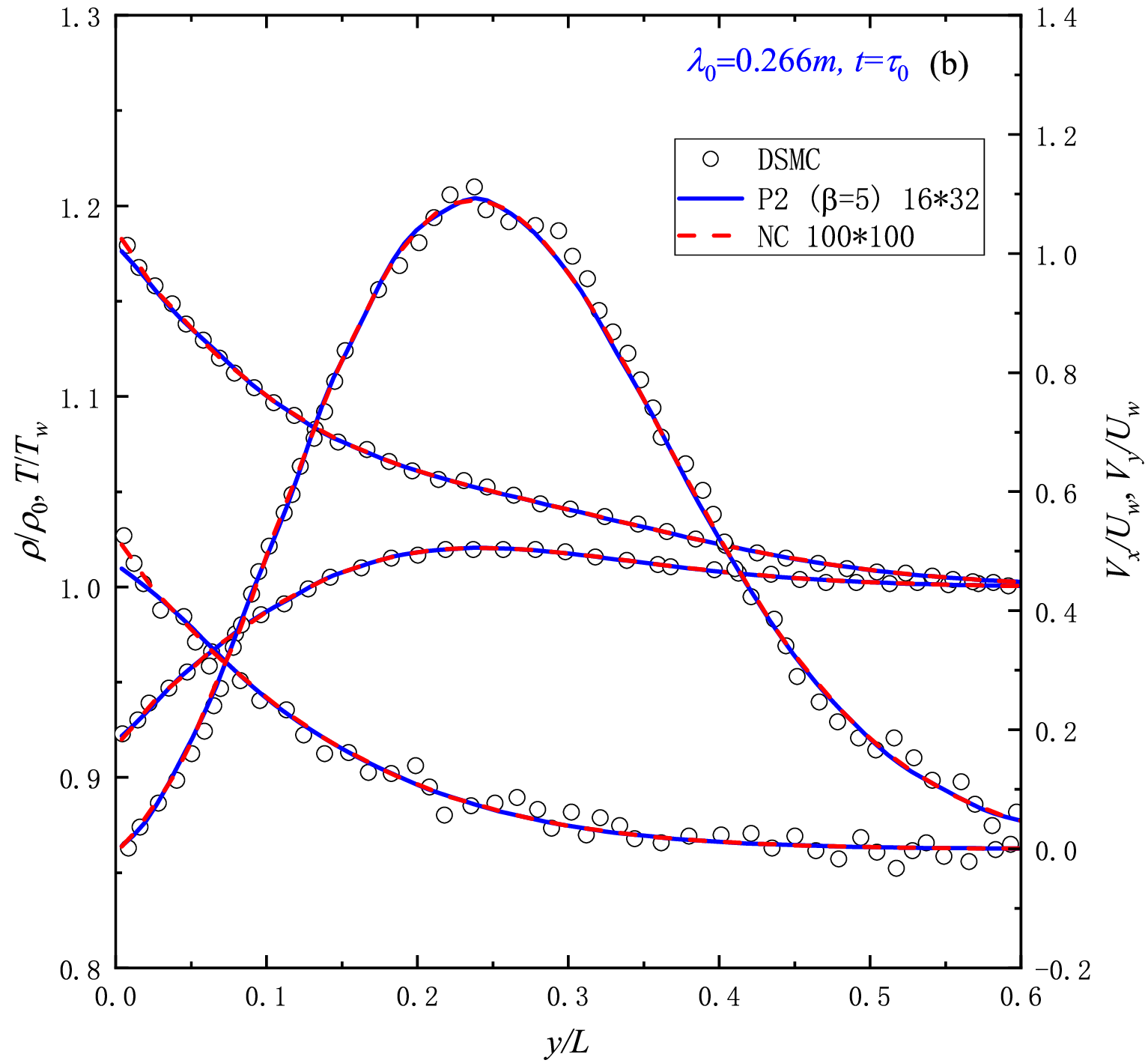}
	\includegraphics[width=5.4cm,height=4.5cm]{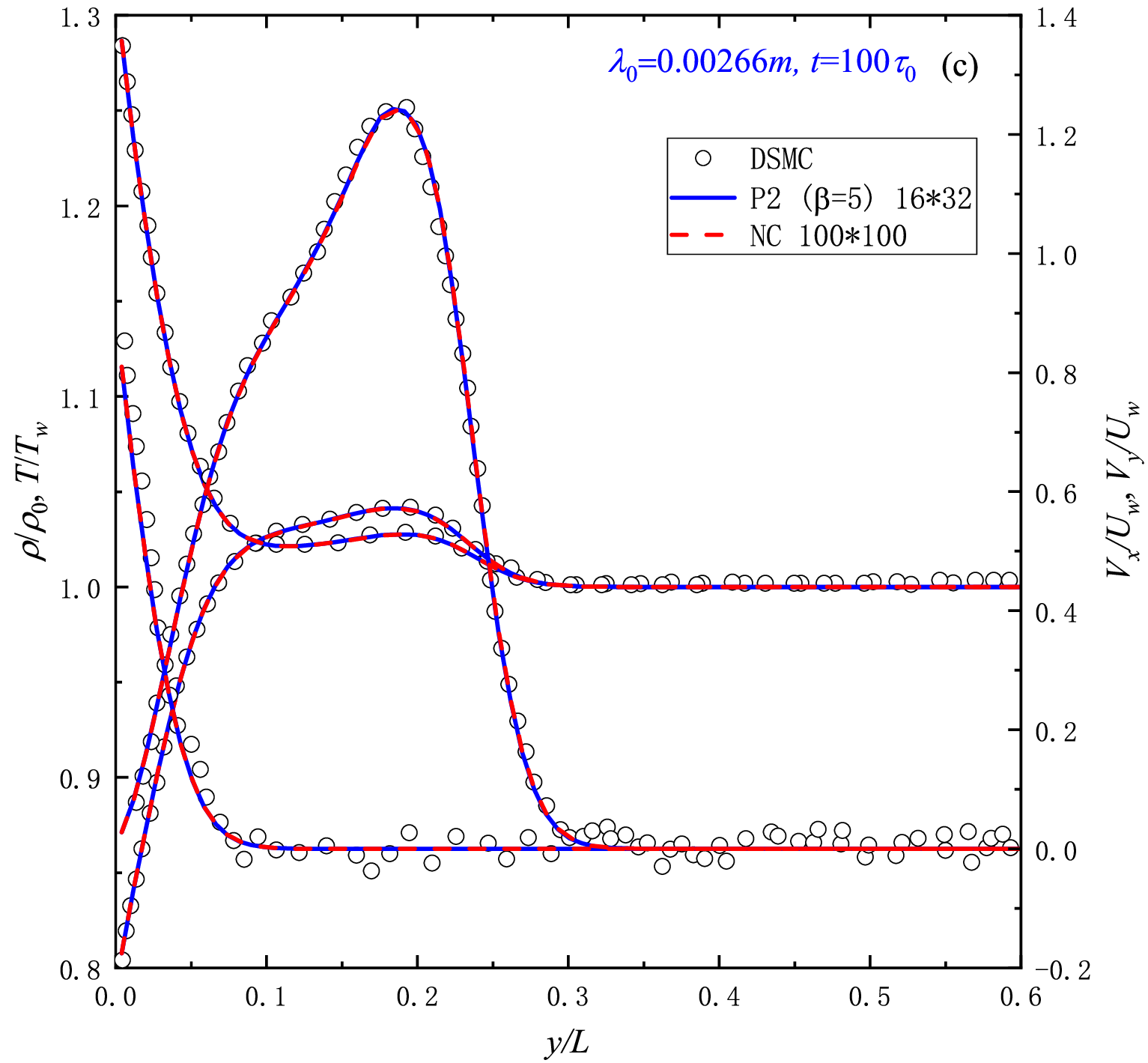}
	\caption{\label{fig:RayP25} \centering Profiles for the Rayleigh flow by P2 ($\beta=5$).}
\end{figure*}

This section applies the PGQ to simulate unsteady flows, specifically examining the Rayleigh flow introduced by Sun et al. \citep{SUN2002}. Previous studies by Huang et al. \citep{Huang2013} and Zhu et al. \citep{ZHU2019} have also utilized this case to validate their developments of the UGKS and implicit UGKS methods. As depicted in Fig.~\ref{fig:GeoRay}, the scenario involves an infinitely long plate, initially at rest in argon gas at $T_0=273K$, which abruptly adjusts to a constant temperature of $373K$ and a velocity of $U_w=10 m/s$. The computational domain, with a characteristic length of 1m, is divided into $10\times 128$ uniform grids. This study presents computational results at three specific time instances $t=0.01\tau_0, \tau_0, 100\tau_0$, corresponding to Knudsen numbers of Kn=26.6, 0.266, and 0.00266, respectively. Comparative results across four distinct quadrature rules are shown in Figs.~\ref{fig:RayHGH}$\sim$\ref{fig:RayP25}, including the half-range Gauss-Hermite (HGH) rule with $28 \times 28$ discrete velocities, the P1 and P2 rules with $16 \times 32$ discrete velocities, and the Newton-Cotes (NC) rule with $101 \times 101$ dimensionless velocities uniformly distributed in $[-4,~4]$. The findings suggest that the NC rule's limited computational accuracy necessitates finer velocity discretization for accurately capturing transitions from continuum to free molecular domains. Although the HGH rule generally provides superior accuracy over the NC rule, the $28 \times 28$ discrete velocities appear inadequate for comprehensive domain calculations, particularly evident at $t=0.01\tau_0$ (Kn=26.6), where non-physical oscillations emerge prominently. As Knudsen numbers increase, discrete velocities tend to cluster more significantly. As shown in Fig.~\ref{fig:RadDis}, configurations with larger $\alpha$ values and smaller $\beta$ values concentrate discrete velocities more effectively. Notably, the computational performance of P1($\alpha=2$) with $16 \times 32$ discrete velocities surpasses that of P1($\alpha=1$). Likewise, the P2($\beta=5$) displays greater accuracy. This observation underscores the importance of selecting appropriate discrete points or adjusting parameter values to better match the discrete velocity space with the particles' velocity distributions, particularly for high Kn Rayleigh flows.

To evaluate the computational efficacy of various quadrature rules, calculations were performed until $t=100\tau$. The time step was determined by $\Delta t= \mathcal{C} \Delta x_{\min}/\xi_{\max}$, the Courant-Friedrichs-Lewy (CFL) number $\mathcal{C}=1$. Among the three assessed PGQs, P2 ($\beta=5$) exhibited the most concentrated discrete velocity distribution with the smallest $\xi_{\max}$, allowing for the largest time step. Consequently, P2 ($\beta=5$) demonstrated superior computational efficiency. In contrast, the NC rule, which truncates the discrete velocity space to the interval $[-4,~4]$, theoretically permitted a larger time step but resulted in significantly lower computational efficiency; the CPU time required for the NC computations was nearly twenty times greater than that for P2($\beta=5$). The HGH rule also permitted relatively large time steps, yet its computational efficiency remained comparatively low. Overall, P2($\beta=5$) has the highest computational accuracy and efficiency in this simulation.

\begin{figure*}[!t]
	\centering
	\includegraphics[width=5.4cm,height=4.5cm]{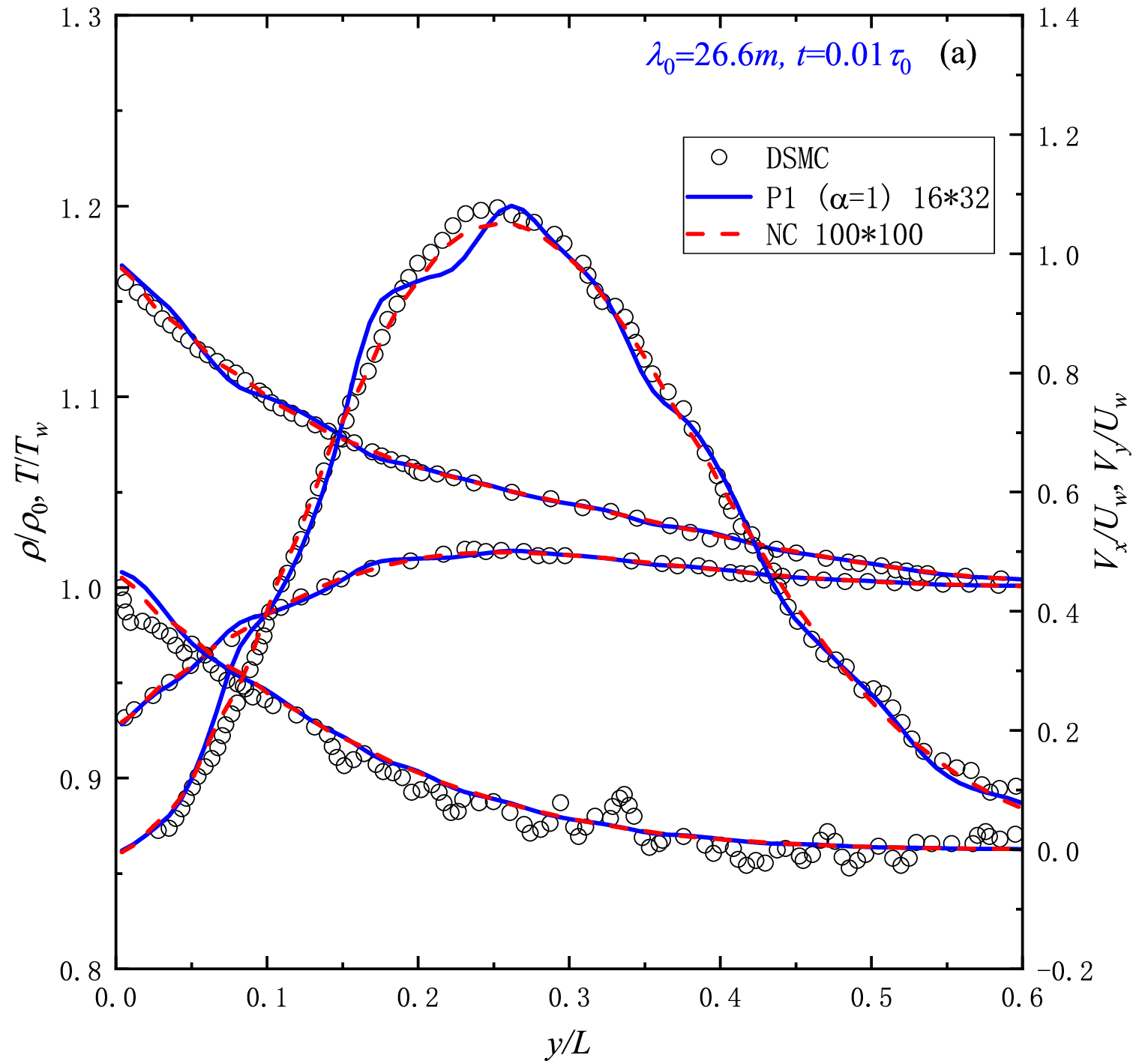}
	\includegraphics[width=5.4cm,height=4.5cm]{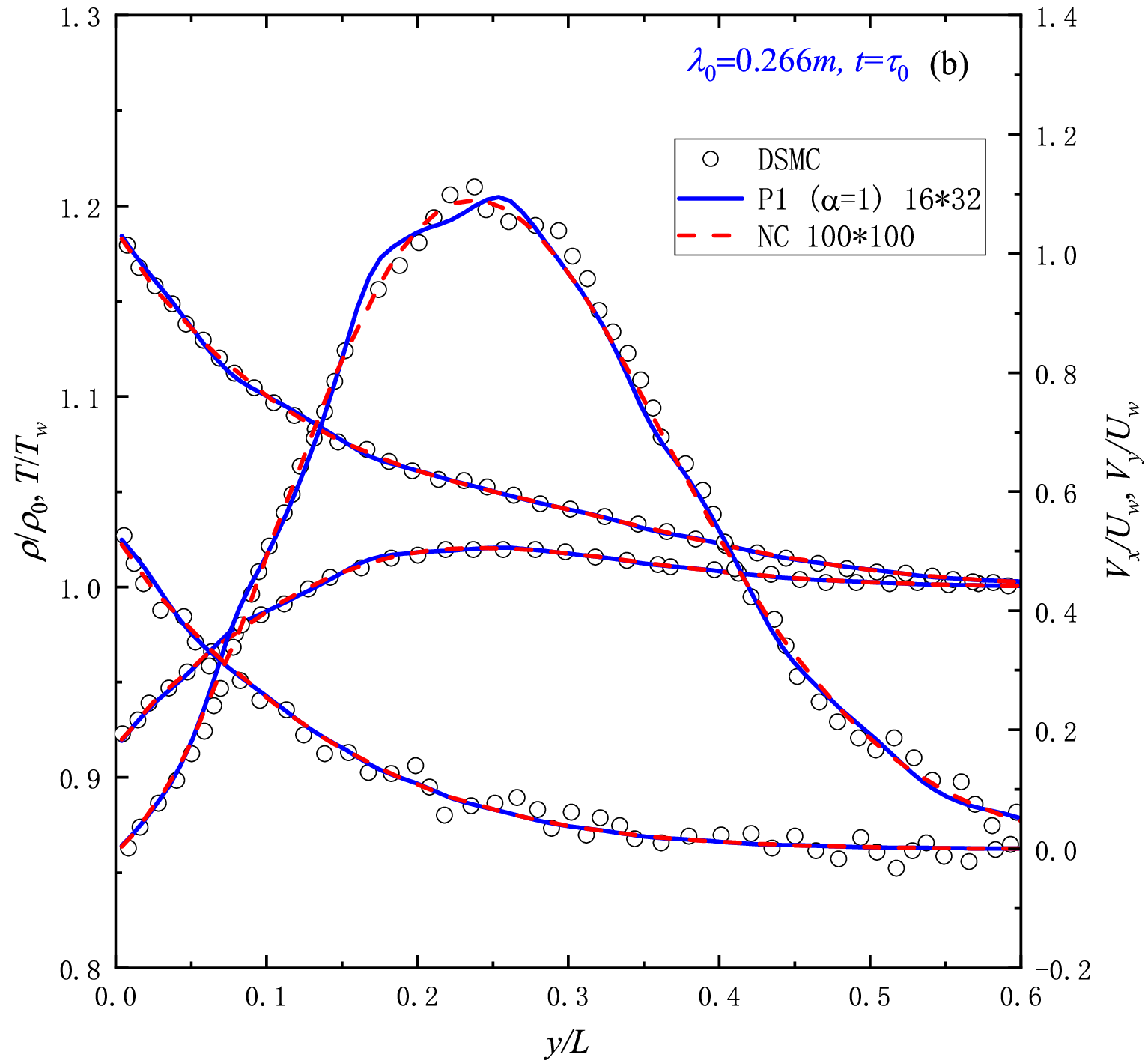}
	\includegraphics[width=5.4cm,height=4.5cm]{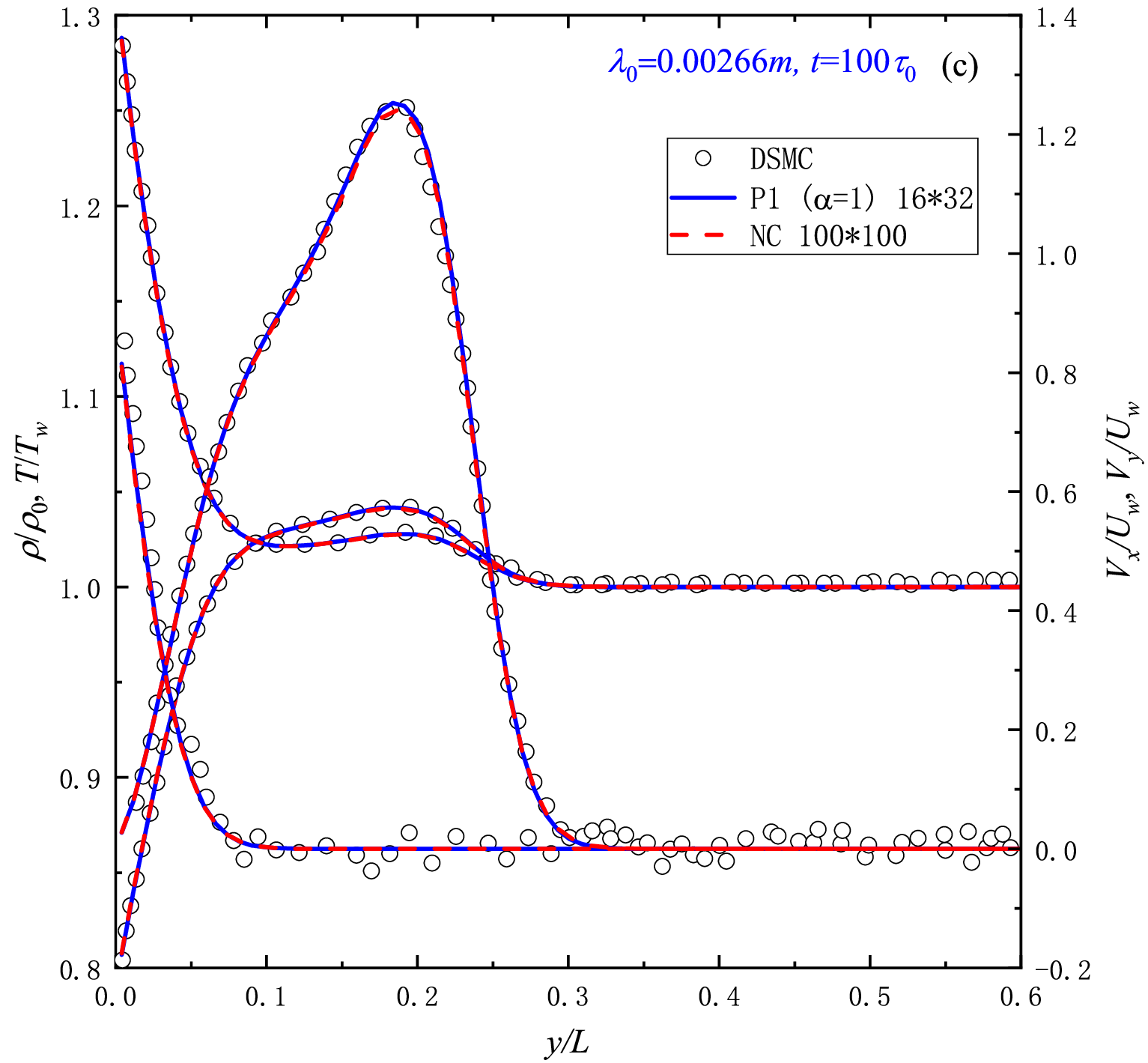}
	\caption{\label{fig:RayP11} \centering Profiles for the Rayleigh flow by P1 ($\alpha=1$).}
\end{figure*}
\begin{figure*}[!t]
	\centering
	\includegraphics[width=5.4cm,height=4.5cm]{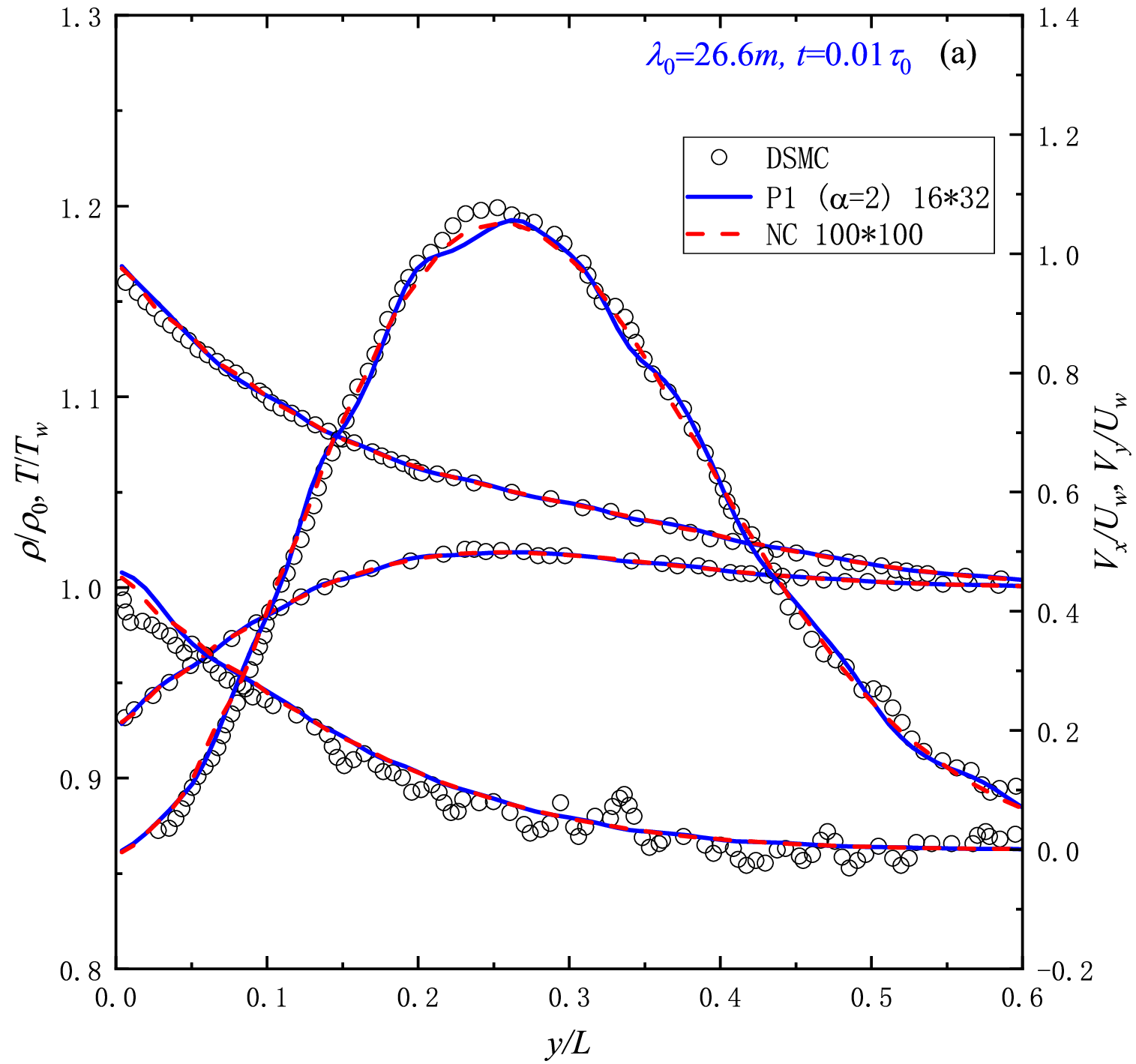}
	\includegraphics[width=5.4cm,height=4.5cm]{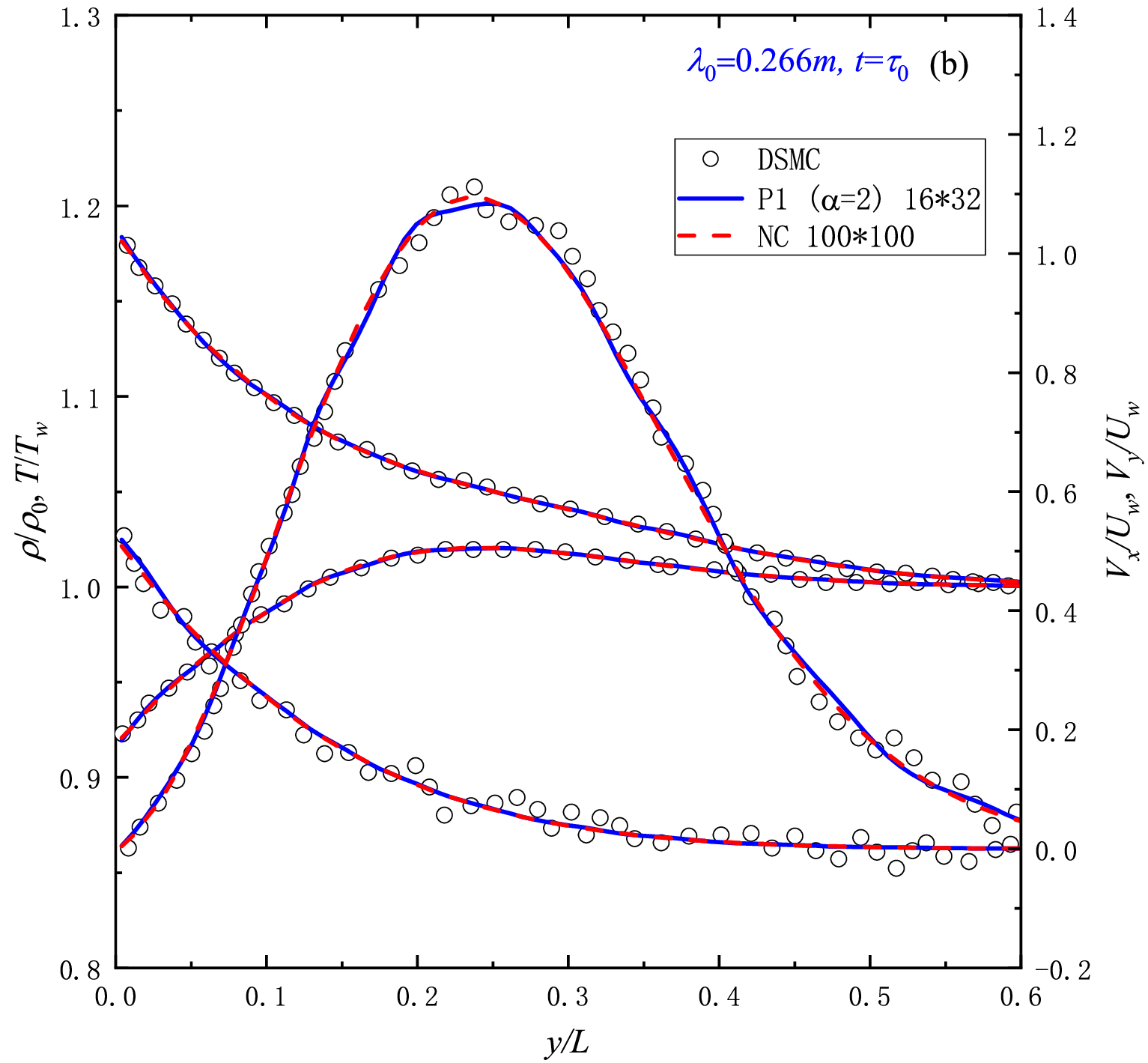}
	\includegraphics[width=5.4cm,height=4.5cm]{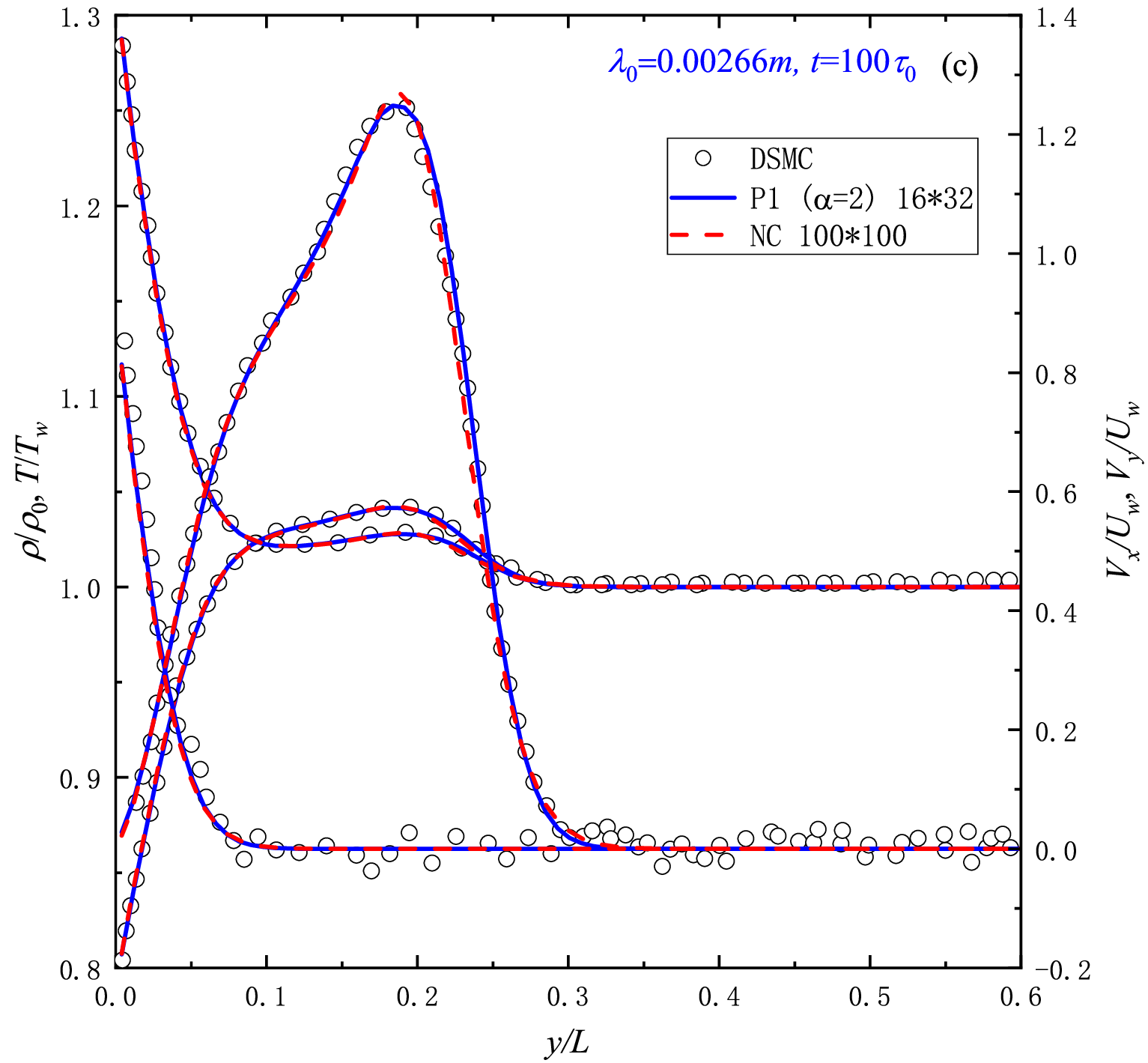}
	\caption{\label{fig:RayP12} \centering Profiles for the Rayleigh flow by P1 ($\alpha=2$).}
\end{figure*}

\subsection{Cavity flow}
\label{sec4.4}

\begin{figure*}[!t]
	\centering
	\includegraphics[width=6.5cm,height=6.5cm]{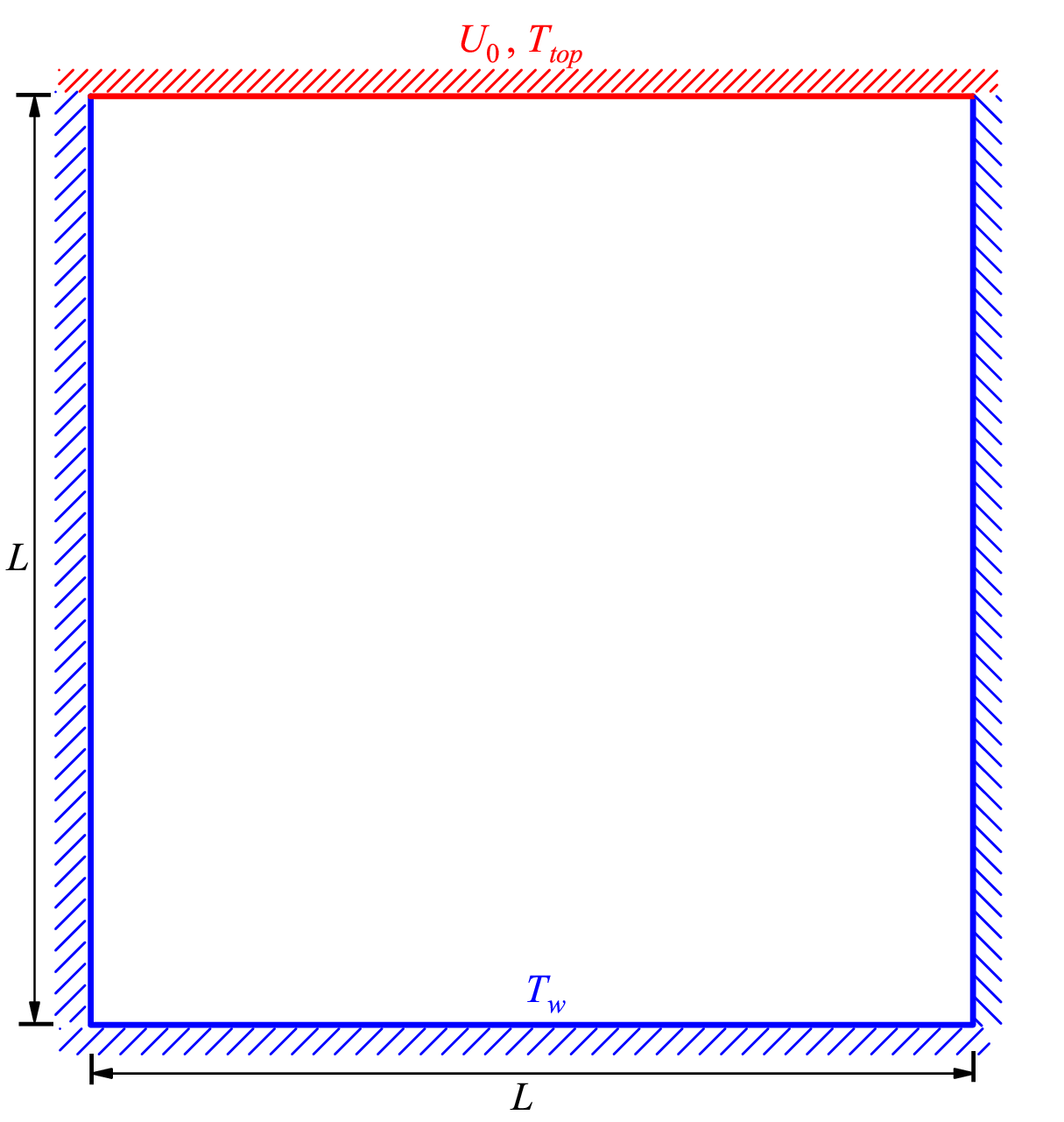}
	\caption{\label{fig:geoCavity} \centering Schematic diagram of the cavity flow.}
\end{figure*}

Cavity flow serves as a benchmark case to validate computational methods for fluid dynamics problems. This section presents numerical simulations of two types of cavity flows: lid-driven cavity flow and TDI cavity flow. These simulations aim to achieve three primary objectives: (1) further validation of PGQ’s high performance and efficiency in simulating multiscale flows, (2) confirmation of PGQ’s flexibility in velocity discretization using the B-type velocity distribution within PGQ, and (3) validation of the 3D PGQ rule outlined in Section \ref{sec3.2}.

\begin{figure*}[!t]
	\centering
	\includegraphics[width=5.4cm,height=4.5cm]{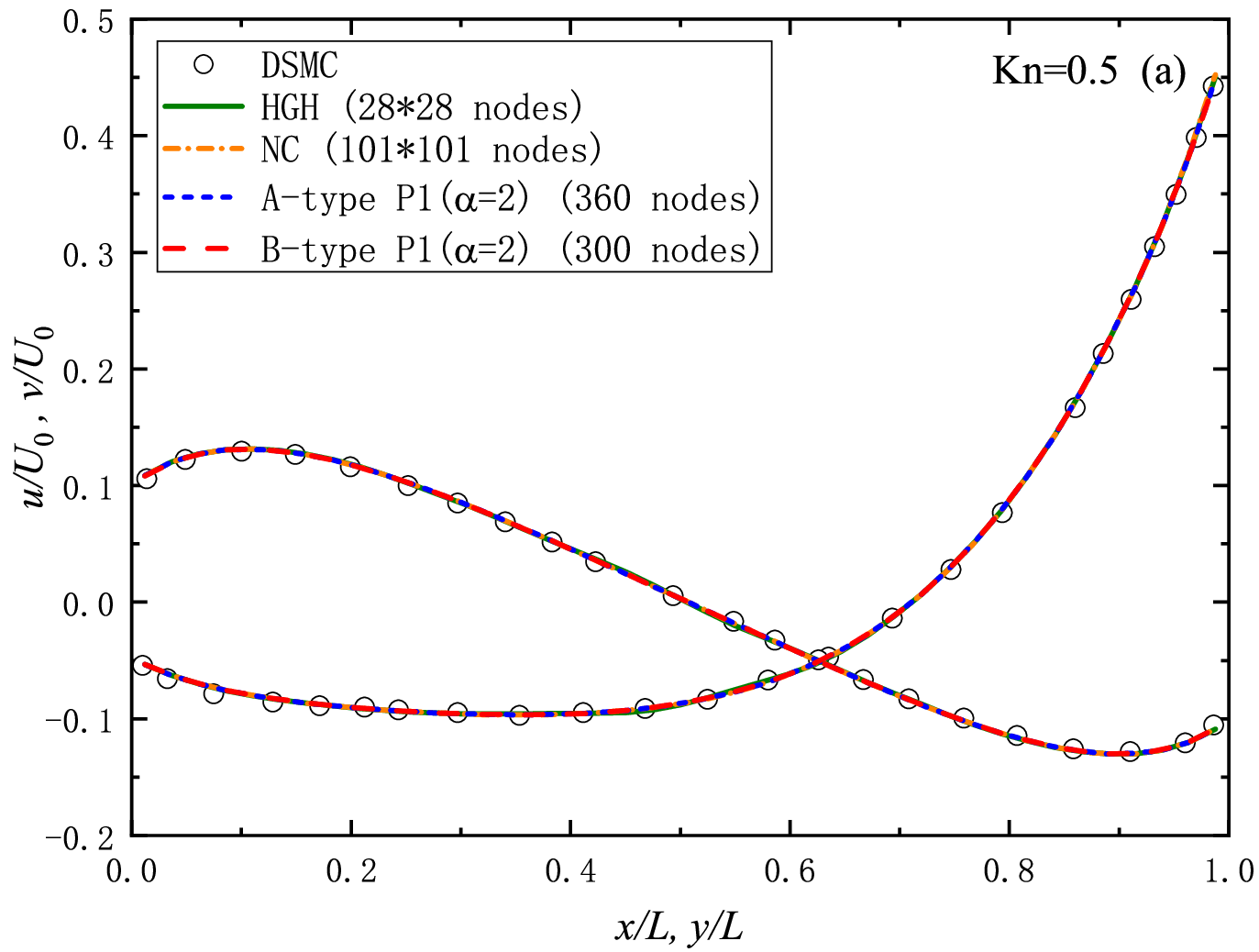}
	\includegraphics[width=5.4cm,height=4.5cm]{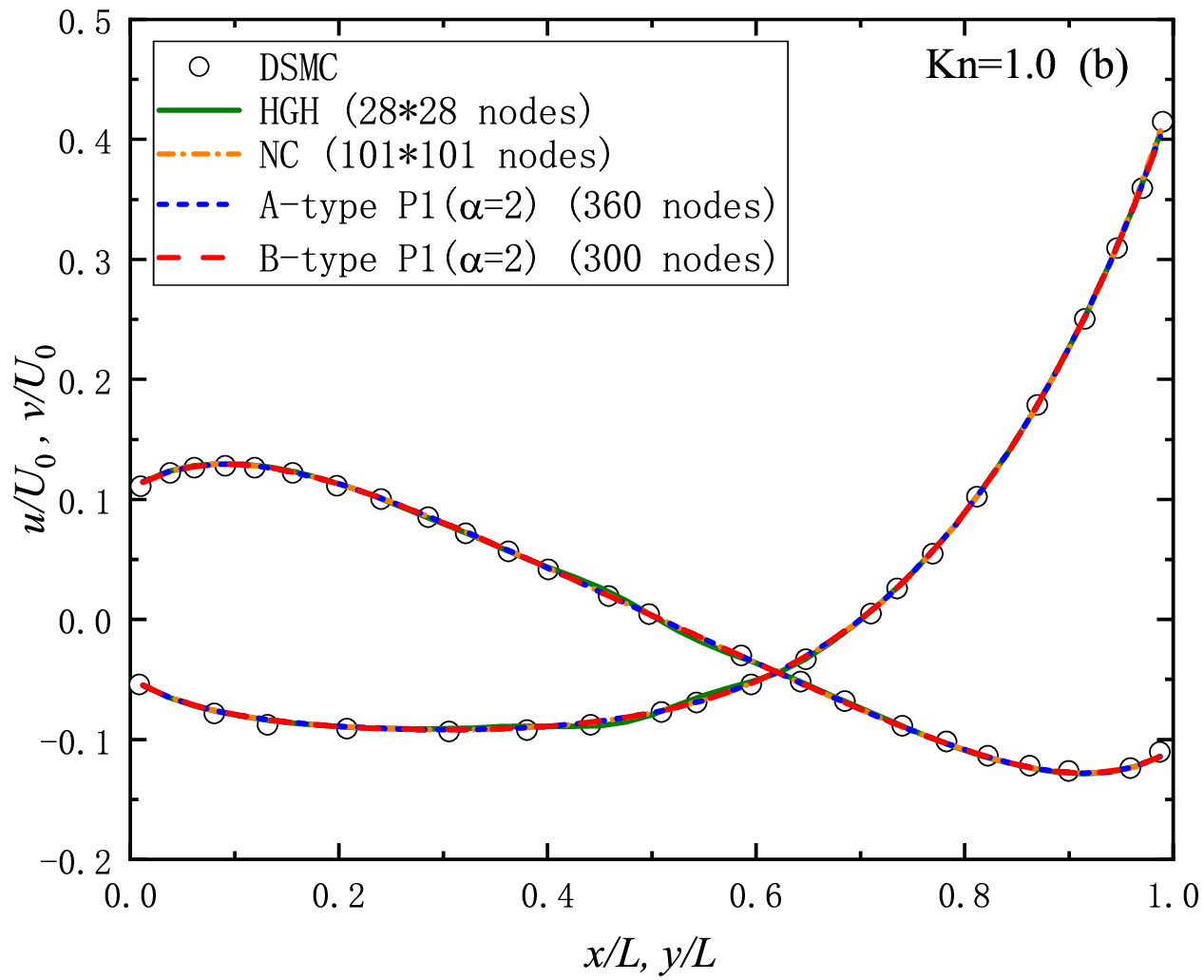}
	\includegraphics[width=5.4cm,height=4.5cm]{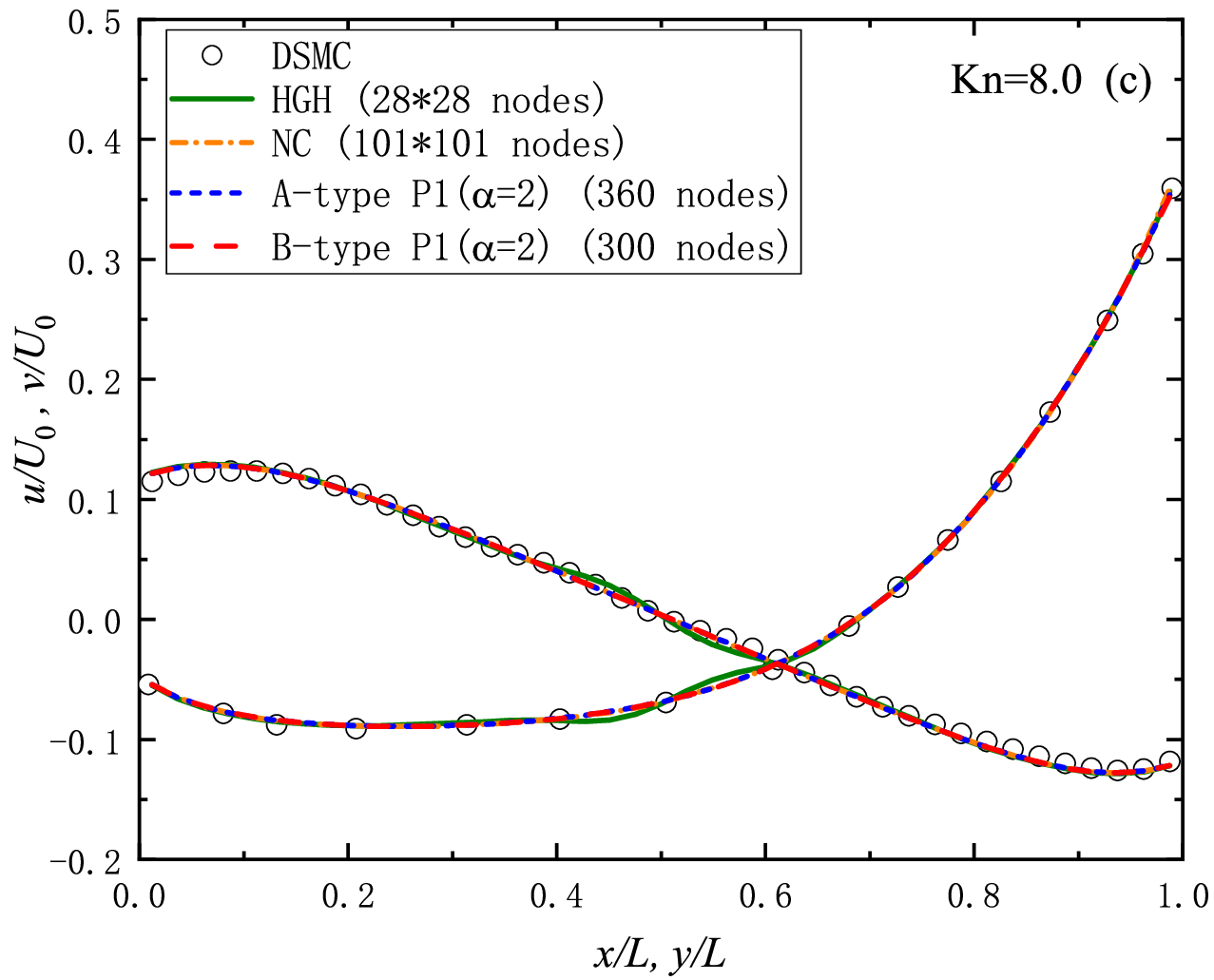}
	\caption{\label{fig:CavityU} \centering  Velocity profiles of the lid-driven cavity flow.}
\end{figure*}

\begin{figure*}[!t]
	\centering
	\includegraphics[width=5.4cm,height=5.4cm]{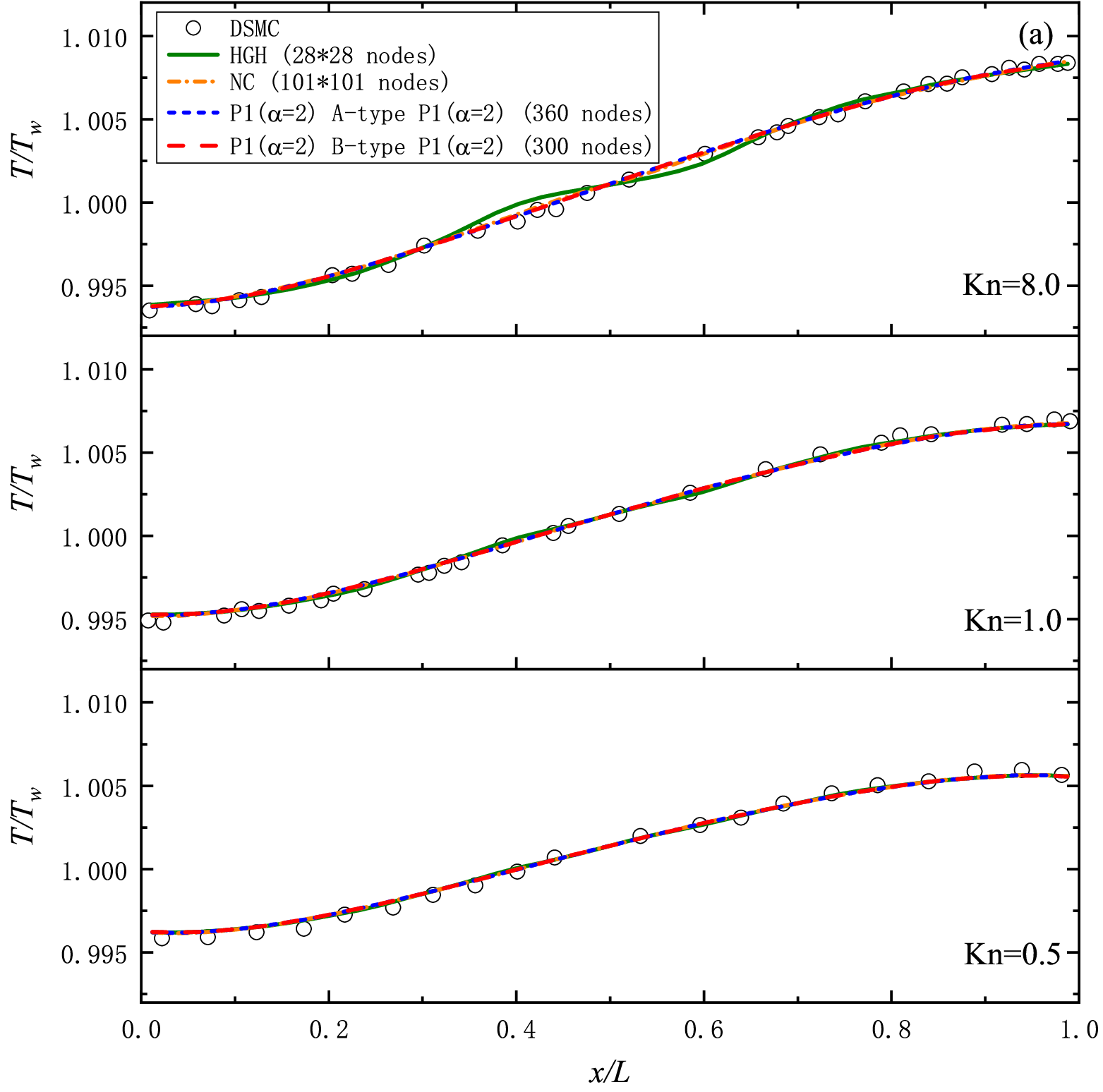}
	\includegraphics[width=5.4cm,height=5.4cm]{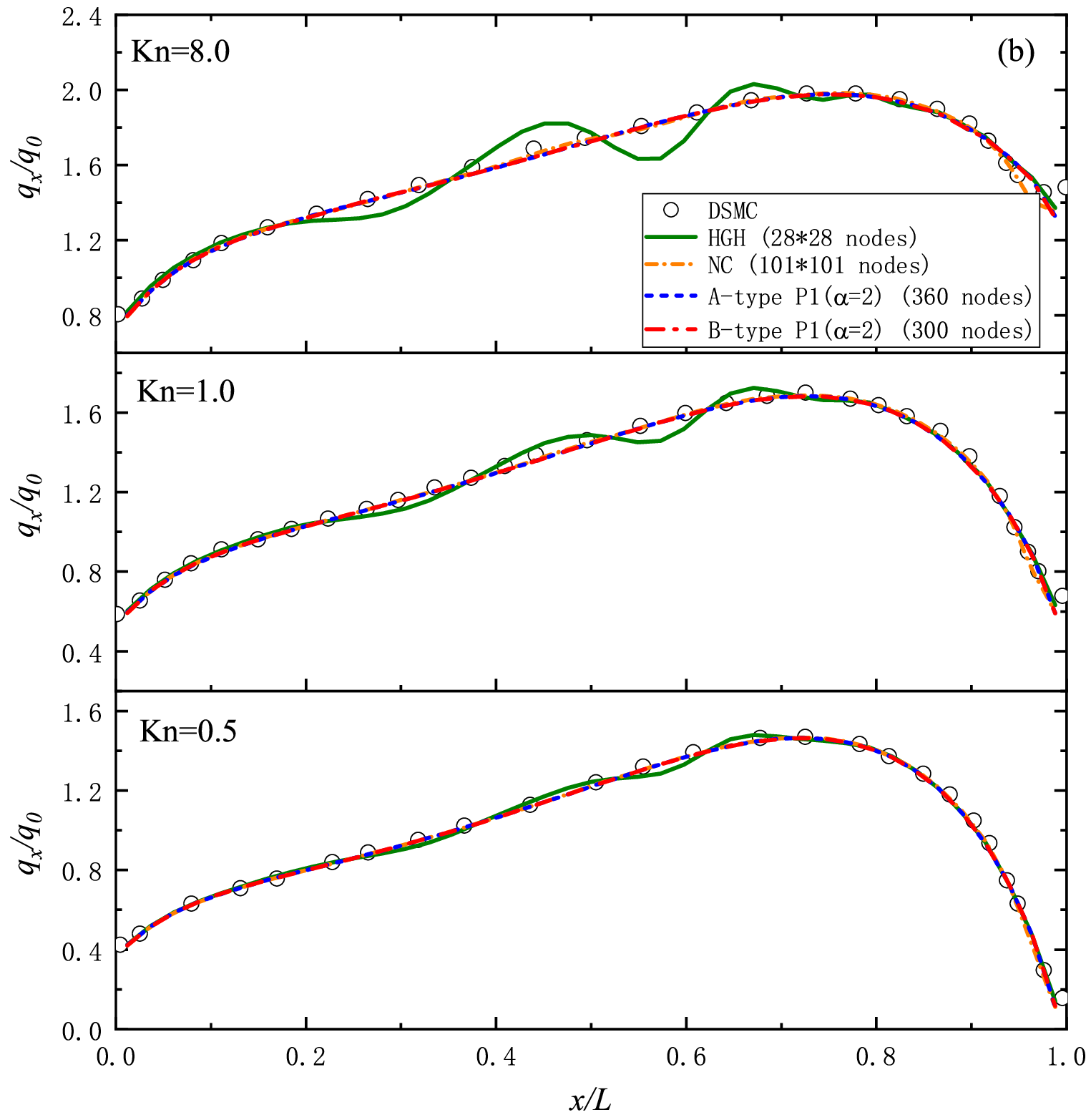}
	\includegraphics[width=5.4cm,height=5.4cm]{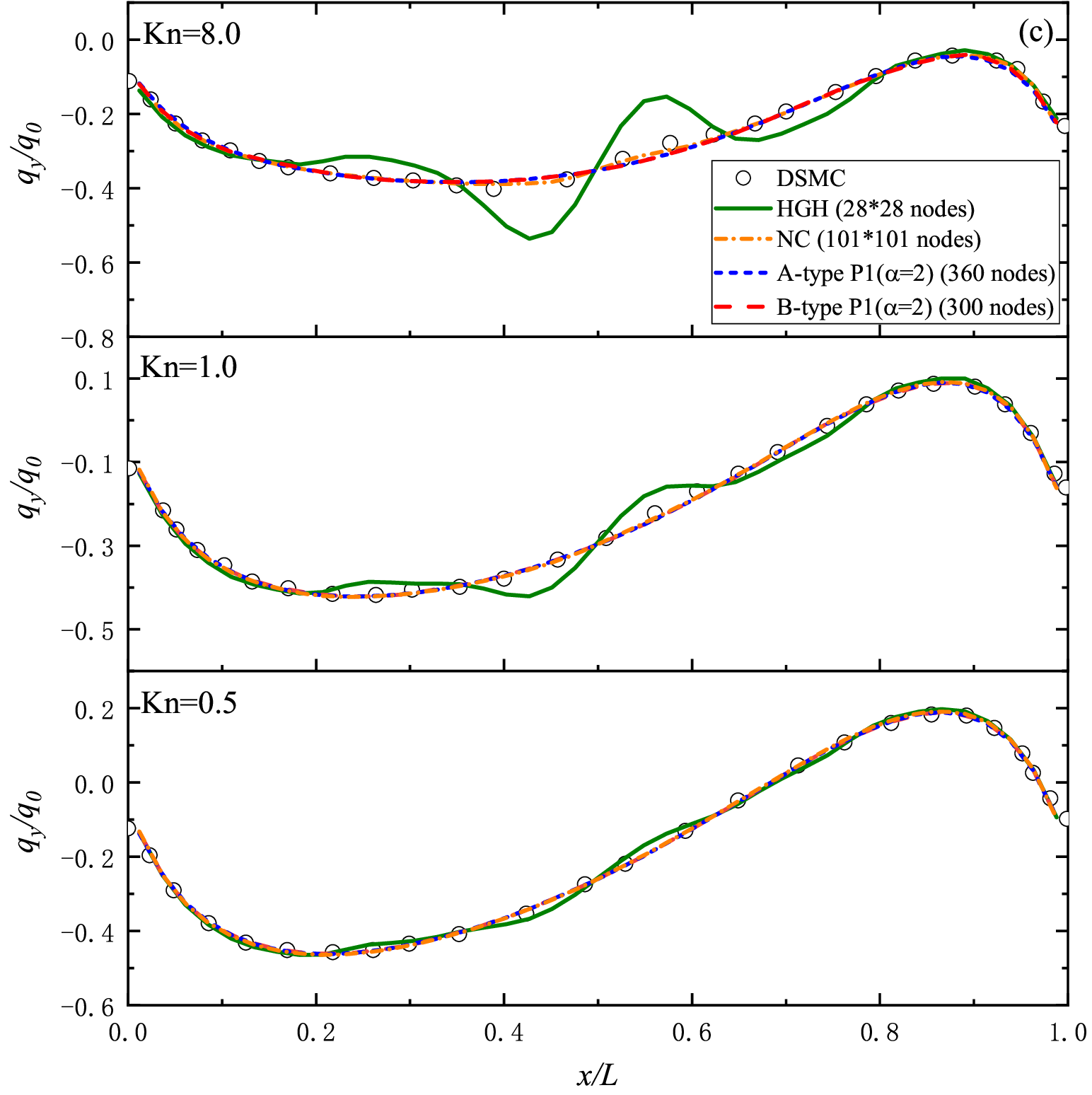}
	\caption{\label{fig:CavityT} \centering  Temperature and Heat flux profiles of the lid-driven cavity flow.}
\end{figure*}

Initially, PGQ is employed to simulate lid-driven cavity flow under isothermal wall conditions, as depicted in Fig.~\ref{fig:geoCavity}. The wall temperature is set at \(T_w=T_{top}=273 \text{K}\), with the top lid moving at a constant velocity of \(U_0=50 \text{m/s}\). The square cavity’s characteristic size is \(L=1 \text{m}\), and the Knudsen numbers range from \( \text{Kn}=0.5 \) to \( \text{Kn}=8.0 \). Calculations are performed on a \(41 \times 41\) grid, comparing three Gaussian quadratures: HGH with \( 28 \times 28 \) nodes, and two PGQ variations employing A-type and B-type velocity discretization, as illustrated in Fig.~\ref{fig:NorDis}. The A-type distribution allocates an equal number of discrete points on each orbit, resulting in \( 4 \times 90 \) velocities in this instance. In contrast, the B-type distribution allows for a variable number of discrete points on different orbits, reducing the number of discrete velocities based on the A-type distribution. Here, the B-type distribution comprises 60, 70, 80, and 90 points arranged from the inner to the outer orbits. Computational results, depicted in Fig.~\ref{fig:CavityU} and Fig.~\ref{fig:CavityT}, are compared with benchmark solutions derived from NC and DSMC in Ref. \citep{zhao2020}.

\begin{figure*}[!t]
	\centering
	\subfigure[Kn=0.5]{
		\label{CavityT0.5C}
		\includegraphics[width=5.3cm,height=5.35cm]{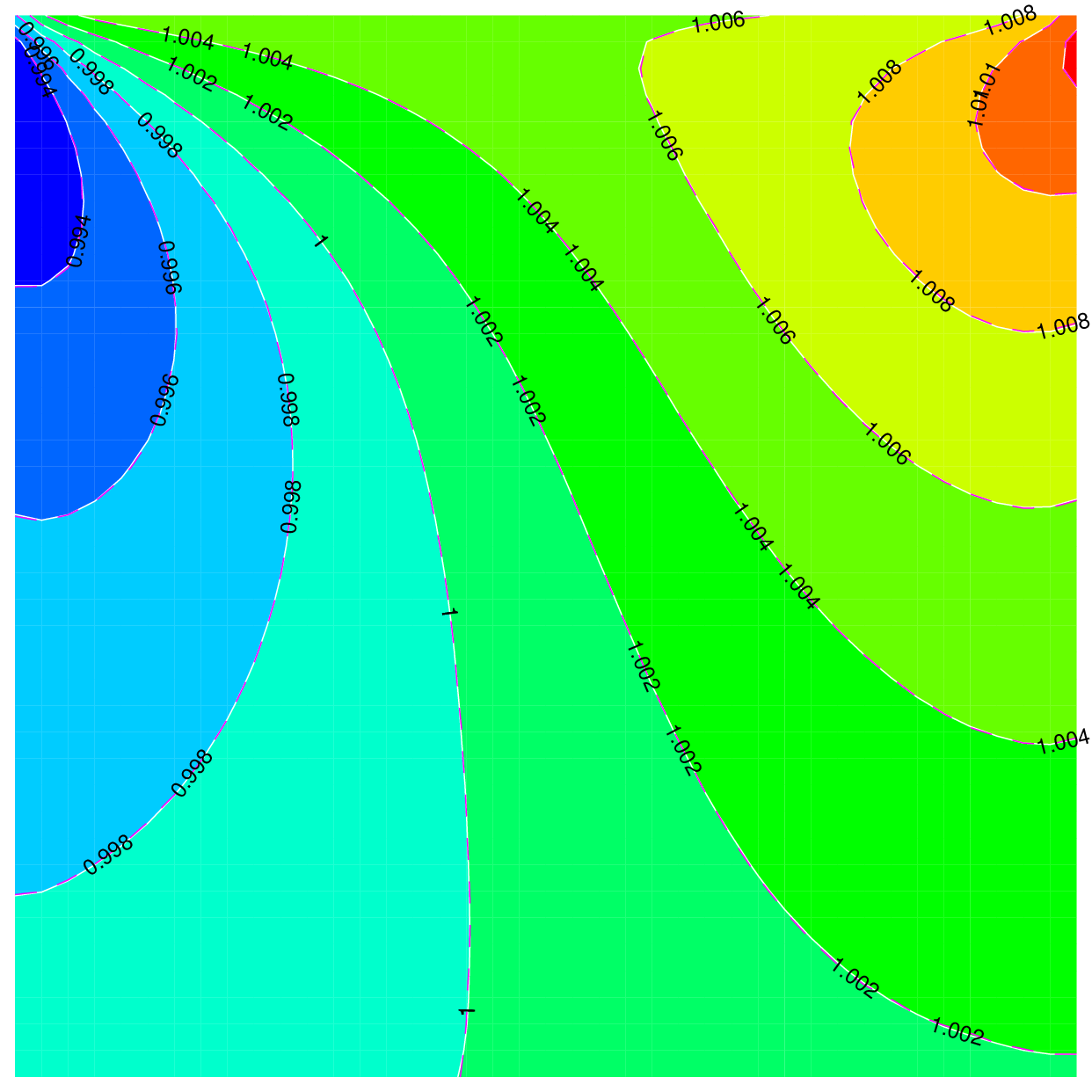}}
	\subfigure[Kn=1.0]{
		\label{CavityT1.0C}
		\includegraphics[width=5.3cm,height=5.35cm]{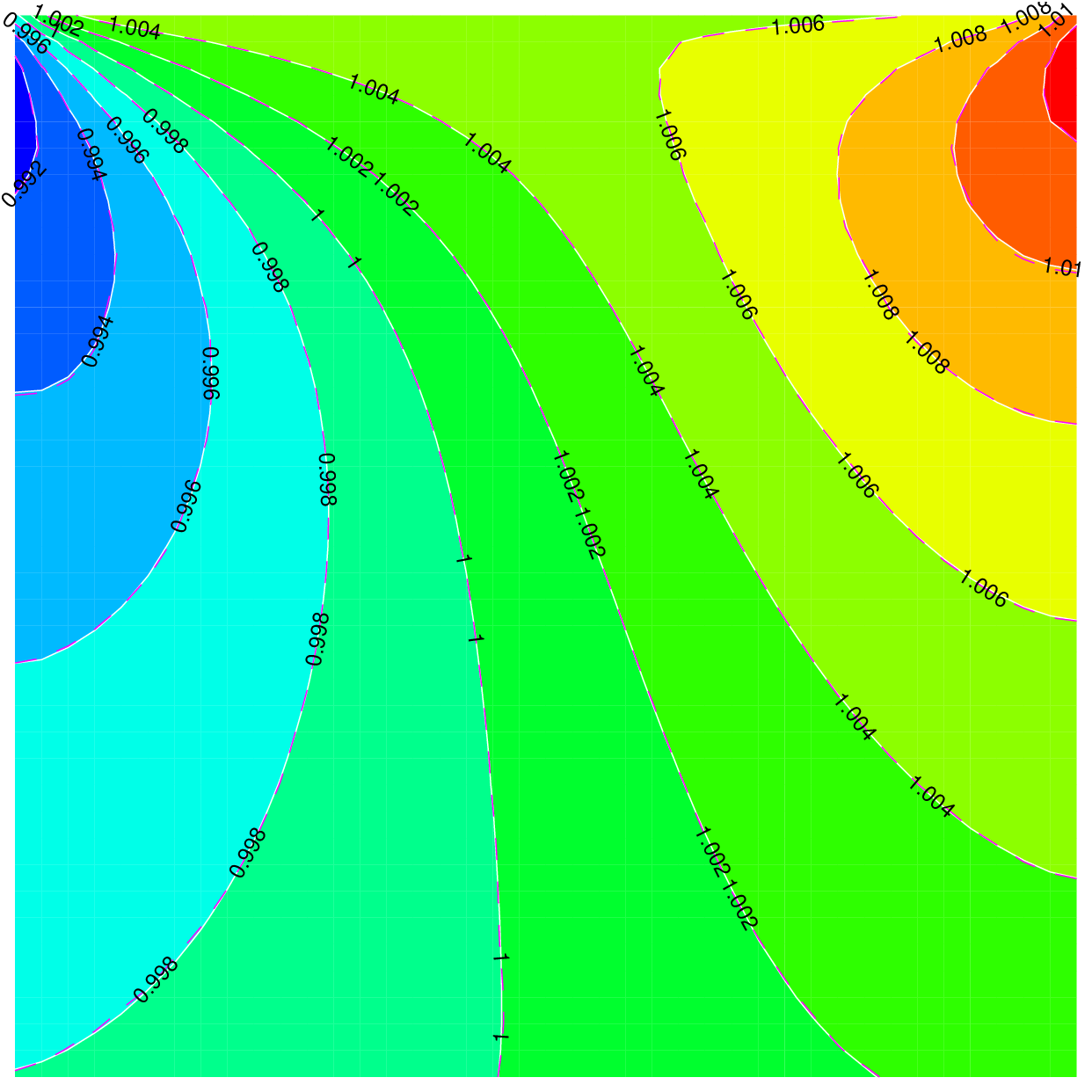}}
	\subfigure[Kn=8.0]{
		\label{CavityT8.0C}
		\includegraphics[width=5.3cm,height=5.35cm]{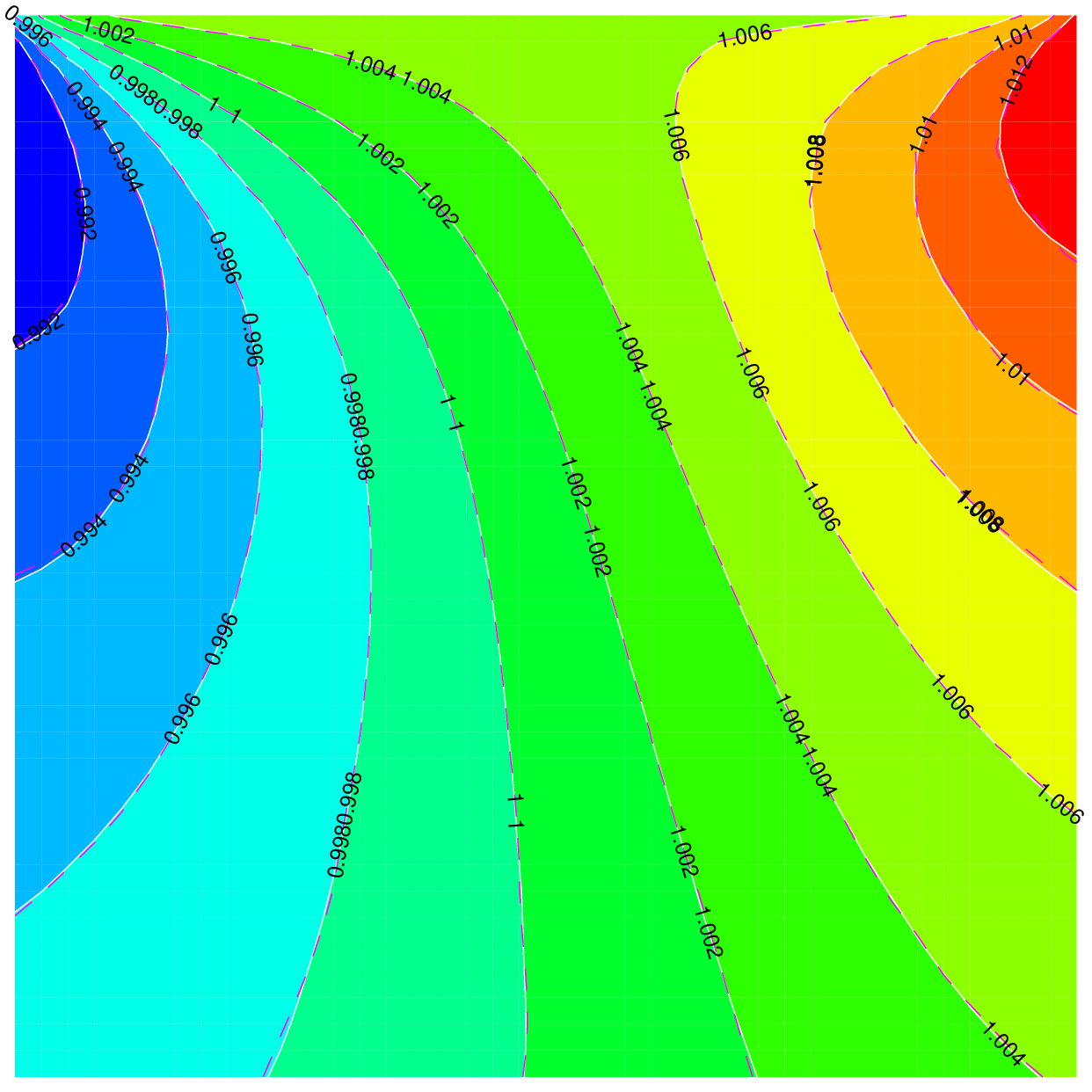}}
	\caption{\label{fig:CavityTC} \centering Temperature contours of the lid-driven cavity flow calculated by A-type and B-type PGQ.}
\end{figure*}

\begin{figure*}[!t]
	\centering
	\subfigure[Kn=0.5]{
		\label{CavityQx0.5C}
		\includegraphics[width=5.3cm,height=5.35cm]{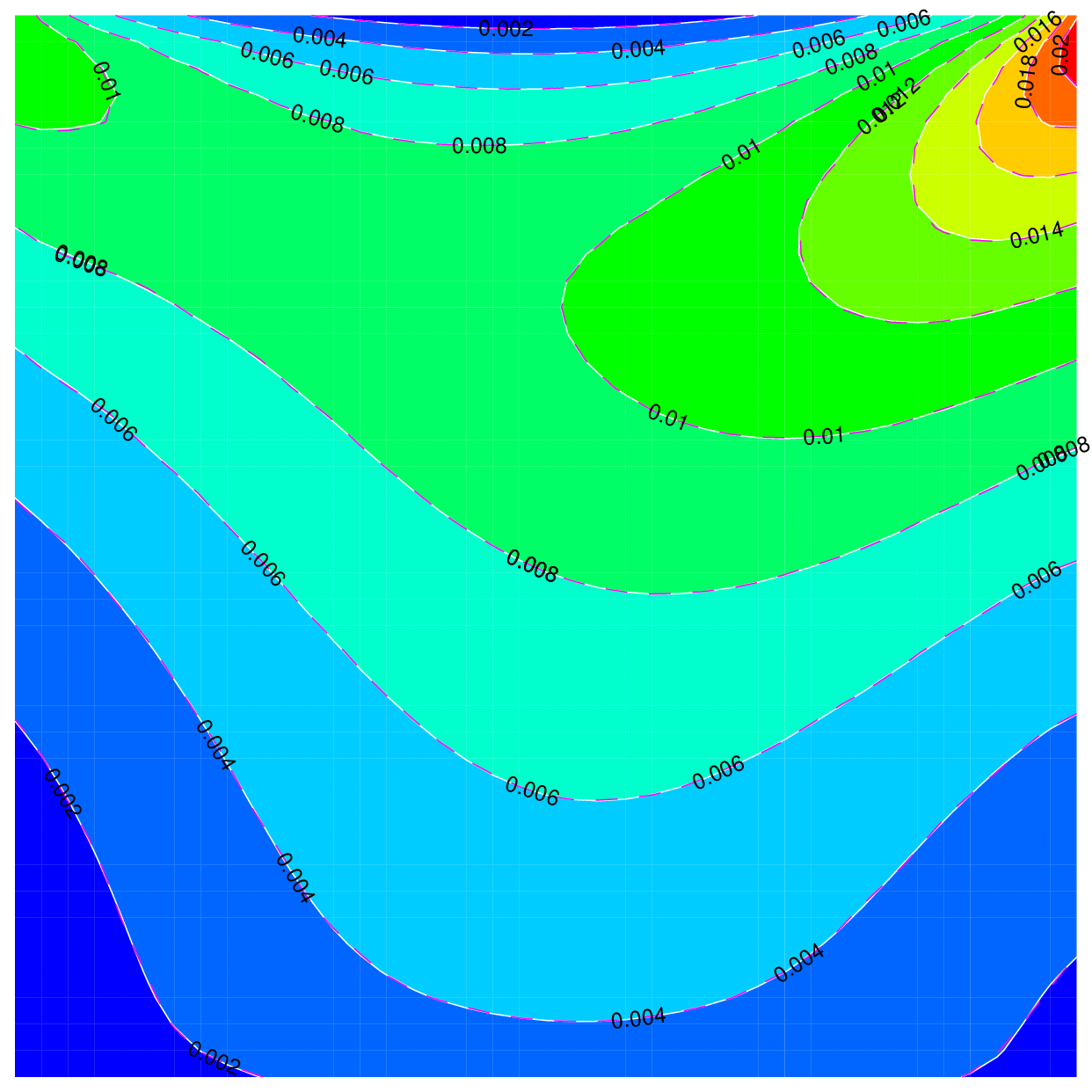}}
	\subfigure[Kn=1.0]{
		\label{CavityQx1.0C}
		\includegraphics[width=5.3cm,height=5.35cm]{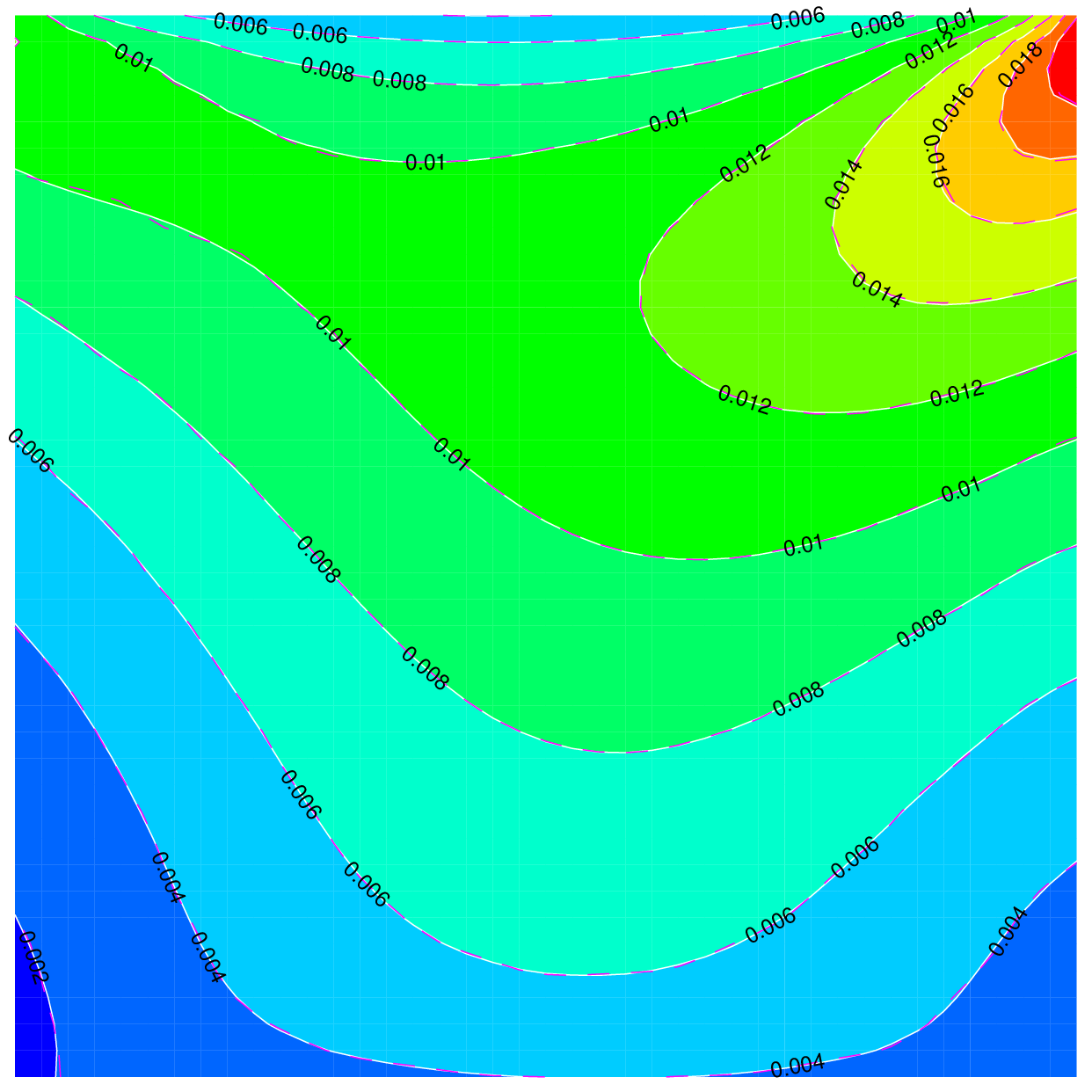}}
	\subfigure[Kn=8.0]{
		\label{CavityQx8.0C}
		\includegraphics[width=5.3cm,height=5.35cm]{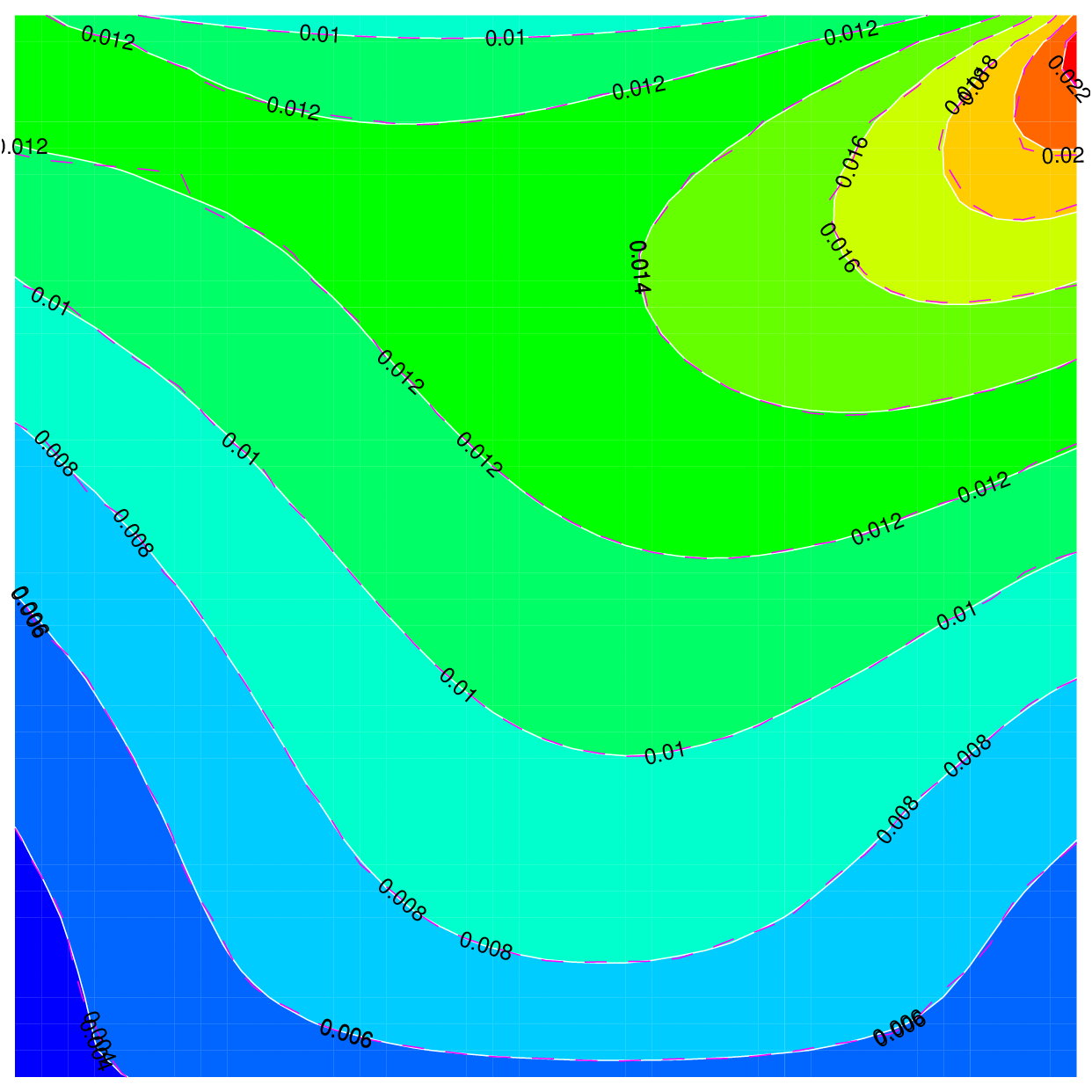}}
	\caption{\label{fig:CavityQxC} \centering  Heat flux (Qx) contours of the lid-driven cavity flow calculated by A-type and B-type PGQ.}
\end{figure*}

\begin{figure*}[!t]
	\centering
	\subfigure[Kn=0.5]{
		\label{CavityQy0.5C}
		\includegraphics[width=5.3cm,height=5.35cm]{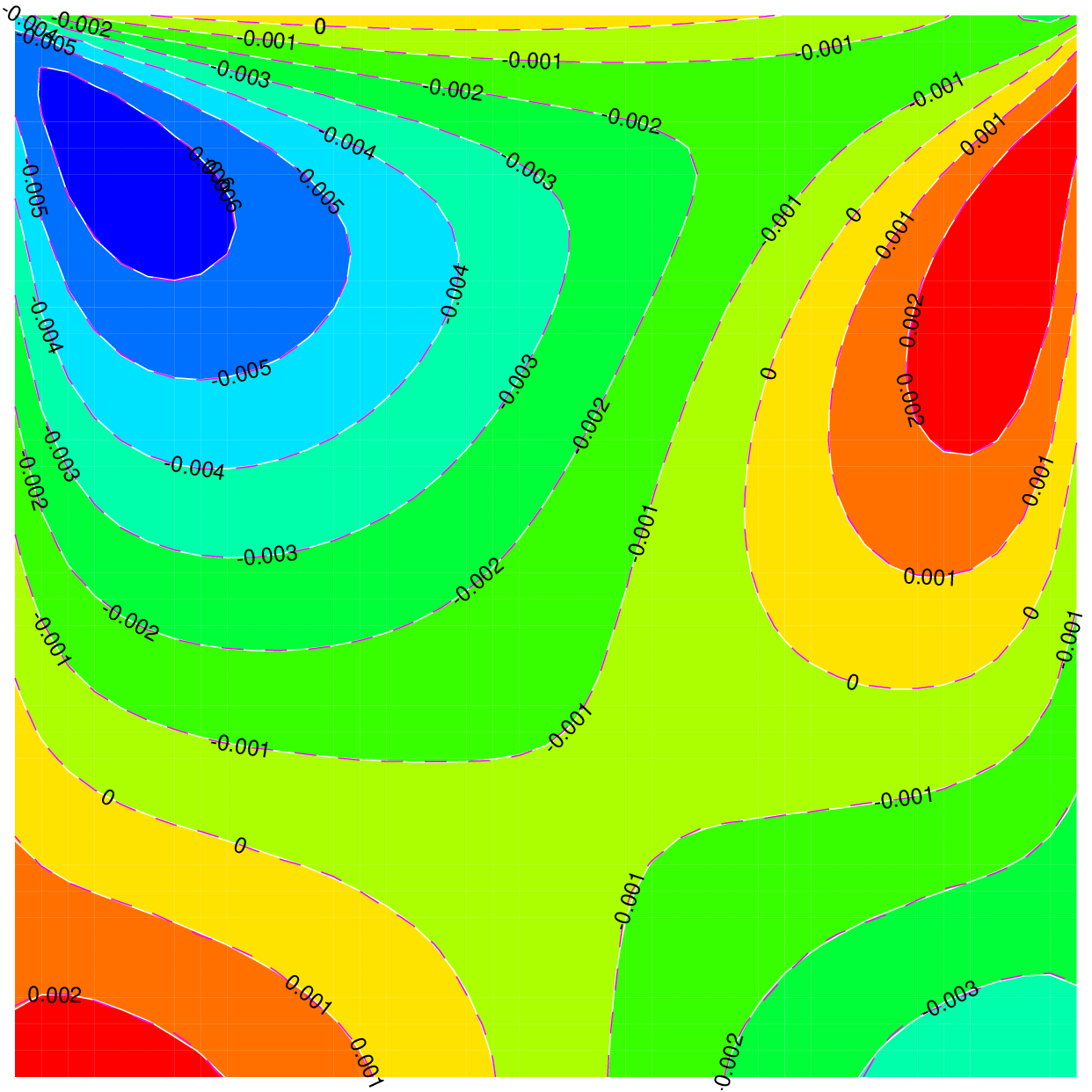}}
	\subfigure[Kn=1.0]{
		\label{CavityQy1.0C}
		\includegraphics[width=5.3cm,height=5.35cm]{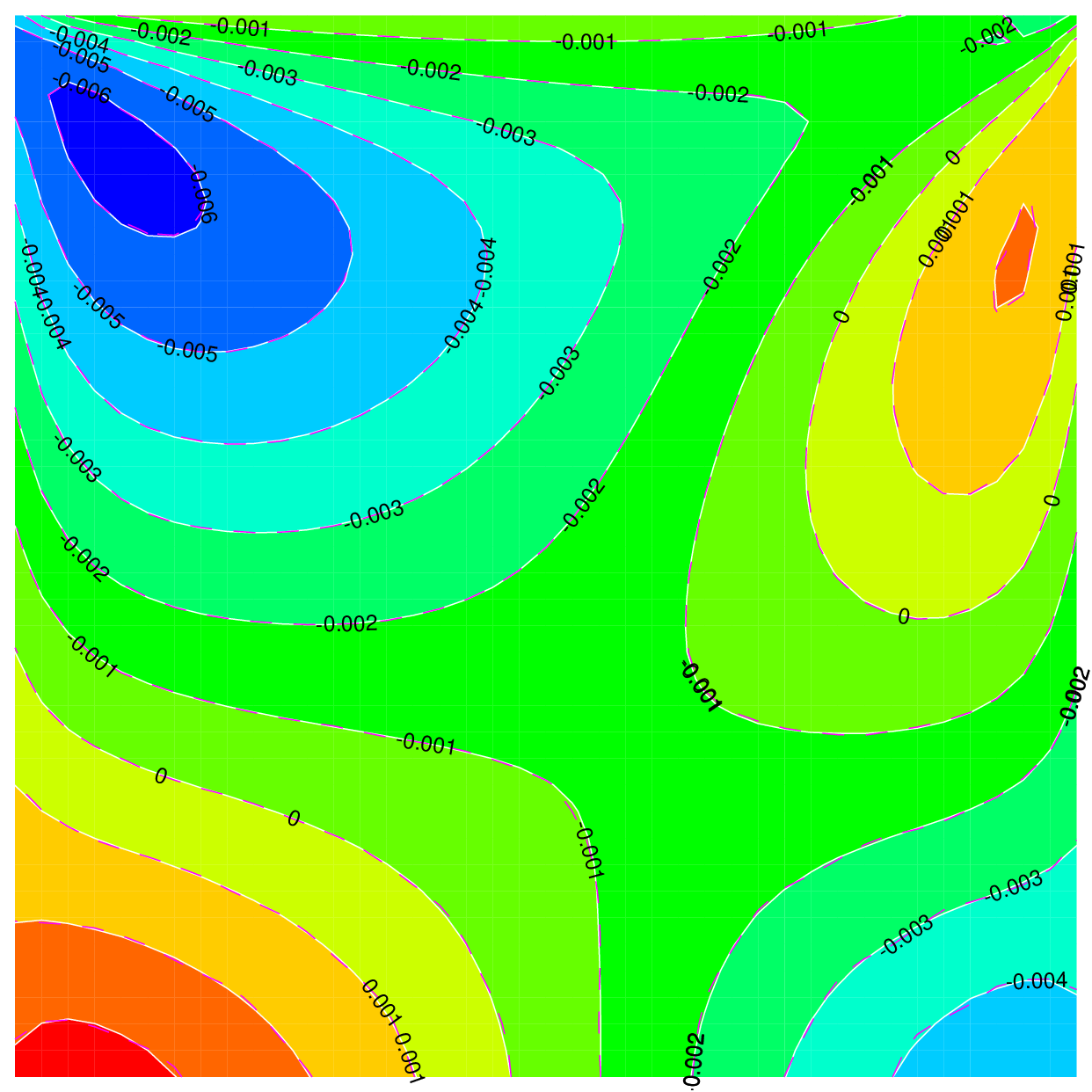}}
	\subfigure[Kn=8.0]{
		\label{CavityQy8.0C}
		\includegraphics[width=5.3cm,height=5.35cm]{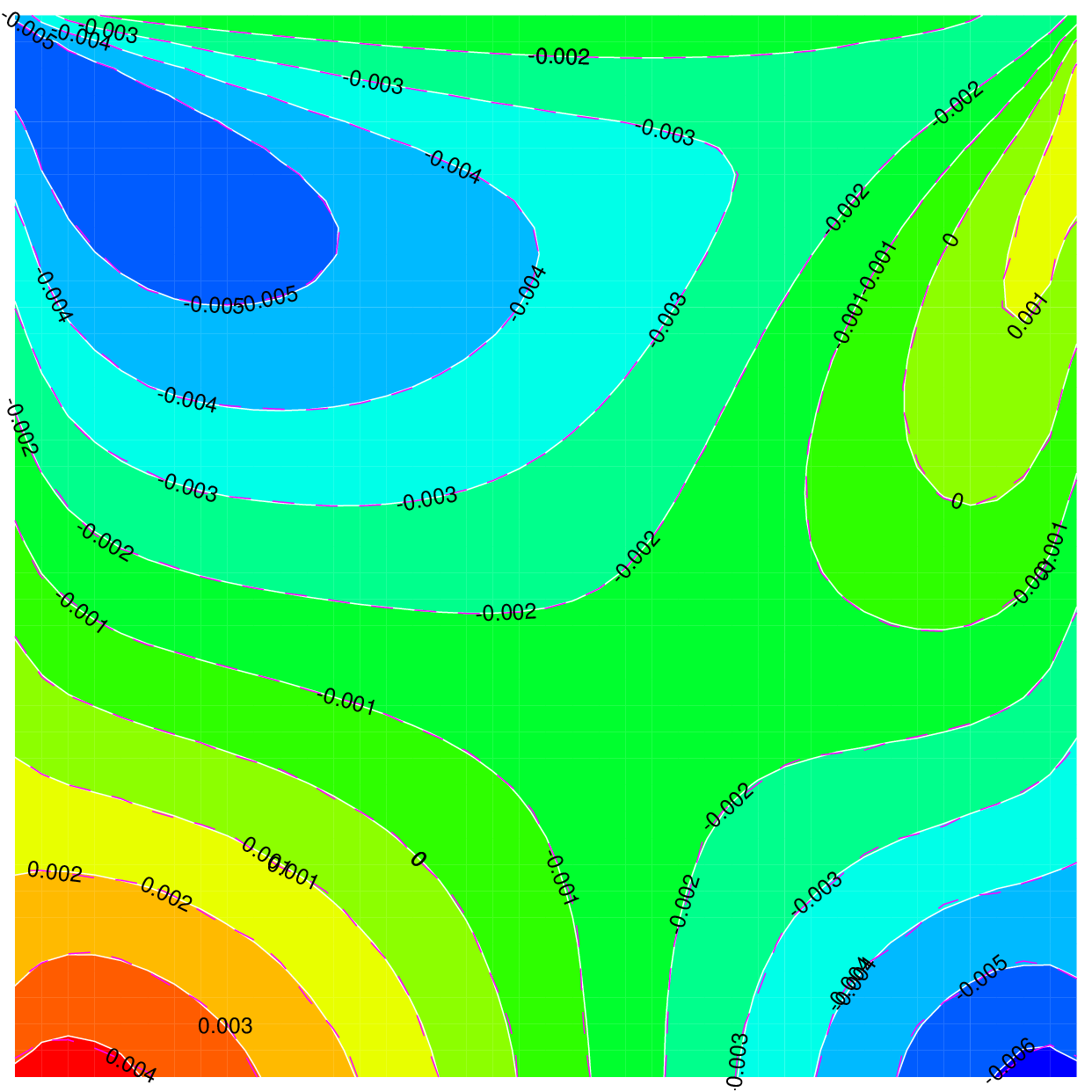}}
	\caption{\label{fig:CavityQyC} \centering  Heat flux (Qy) contours of the lid-driven cavity flow calculated by A-type and B-type PGQ.}	
\end{figure*}

Due to the limitations of conventional Gaussian quadratures in high Knudsen number flows, the HGH quadrature with \( 28 \times 28 \) discrete velocities fails to accurately predict lid-driven flow across these Knudsen numbers. As the Knudsen number increases, the accuracy of the calculations deteriorates. Discrepancies between HGH predictions and DSMC results are evident in the velocity (Fig.~\ref{fig:CavityU}) and temperature distributions (Fig.~\ref{fig:CavityT}(a)), with pronounced oscillations near the center of the square cavity. Additionally, HGH’s prediction of heat flux (Fig.~\ref{fig:CavityT}(b)(c)) is notably deficient. In contrast, both A-type and B-type PGQ distributions, despite using a substantially lower number of discrete velocities, accurately predict lid-driven flow across these Knudsen numbers, yielding nearly indistinguishable computational results. A subsequent comparison of temperature and heat flux distributions in Fig.~\ref{fig:CavityTC} to Fig.~\ref{fig:CavityQyC} further confirms the marginal differences between the results based on these two velocity distribution types. Thus, PGQ proves to be significantly more accurate and effective than HGH in simulating multiscale lid-driven square cavity flows, with PGQ’s discrete velocity points offering flexible adjustment according to specific requirements. Furthermore, as illustrated in Table 3, PGQ employs the minimum number of discrete velocities and utilizes the maximum time step in this simulation, achieving a computational speed exceeding that of NC by over 60 times.

\begin{table*}[!h]
	\caption{\label{tab:tab3} Comparison between computation time for lid-driven cavity flow}
	\centering
	\begin{tabular}{c|cccc}
		\hline
		\multirow{2}{*}{Method}& NC & HGH & A-type P1$(\alpha=2)$ & B-type P1$(\alpha=2)$ \\
		& $(101\times 101)$ & $(28\times 28)$ & (360 nodes) & (300 nodes) \\
		\hline
		$\xi_{max}^{\ast}$ & 4.0 & 5.2 & 2.46 & 2.46 \\
		$\Delta t~(10^{-3})$ & 3.05 & 2.33 & 4.95 & 4.95 \\
		Total steps & 9000 & 16000 & 5000 & 5000 \\
		CPU time (min) & 74.5 & 9.9 & 1.5 & 1.2 \\		
		\hline
	\end{tabular}
\end{table*}

In the TDI cavity flow configuration, all four walls remain stationary. The top wall temperature is set at \( T_{\text{top}} = T_h = 400 \, \text{K} \), and the other three walls are maintained at \( T_w = T_c = 200 \, \text{K} \). The reference and median temperatures are both established at \( T_{\text{ref}} = T_m = 300 \, \text{K} \). This analysis is executed within a 2D framework. Diverging from previous practices, the Shakhov model is employed over the reduced Shakhov model for computation, as detailed in Section \ref{sec2.1}. Even for flow problems in 1D and 2D physical spaces, the reduced Shakhov model necessitates discretization of the 3D velocity space, serving as a tool for testing 3D velocity discretization methods in lower dimensions. Simulations are carried out in three distinct cases with \( Kn = 0.1,~1,~10 \) representing varying degrees of rarefaction, with the computational domain divided into a uniform grid of \( 61 \times 61 \) cells. For velocity discretization, strategies \( S1(\alpha=2,\gamma=1) \) and \( S2(\beta=10,\gamma=1) \) are adopted as specified in Section \ref{sec3.2}; the number of discrete velocity points is given in Table~\ref{tab:tab4}. For \( \alpha=2, \gamma=1 \), the radial and azimuthal distribution functions are \( w^S(r) = 2r^2 e^{-r^2} \) and \( w^S(\phi) = 1 \), respectively, with corresponding Gauss points and weight coefficients listed in Table~\ref{tab:tabA2}. For \( \beta=10, \gamma=1 \), we utilize \( w^S(r) = 11\sqrt{\ln r^{-11}} r^{10} \) and \( w^S(\phi) = 1 \), with associated Gauss points and weight coefficients provided in Table~\ref{tab:tabB1}.

\begin{table*}[!h]
	\caption{\label{tab:tab4} Velocity discretization settings for TDI cavity flow}
	\centering
	\begin{tabular}{cccc}
		\hline
		-- & Kn=0.1 & Kn=1.0 & Kn=10.0 \\
		\hline
		Zhu et al. \citep{ZHU2019} & $28 \times 28 \times 2$ HGH & $161 \times 161 \times 2$ NC & $201 \times 201 \times 2$ NC \\
		Present & $8 \times 45 \times 3$ PGQ & $8 \times 60 \times 3$ PGQ & $8 \times 90  \times 3$ PGQ \\		
		\hline
	\end{tabular}
\end{table*}

\begin{figure*}[!t]
	\footnotesize
	\centering
	\includegraphics[width=7.5cm,height=6.5cm]{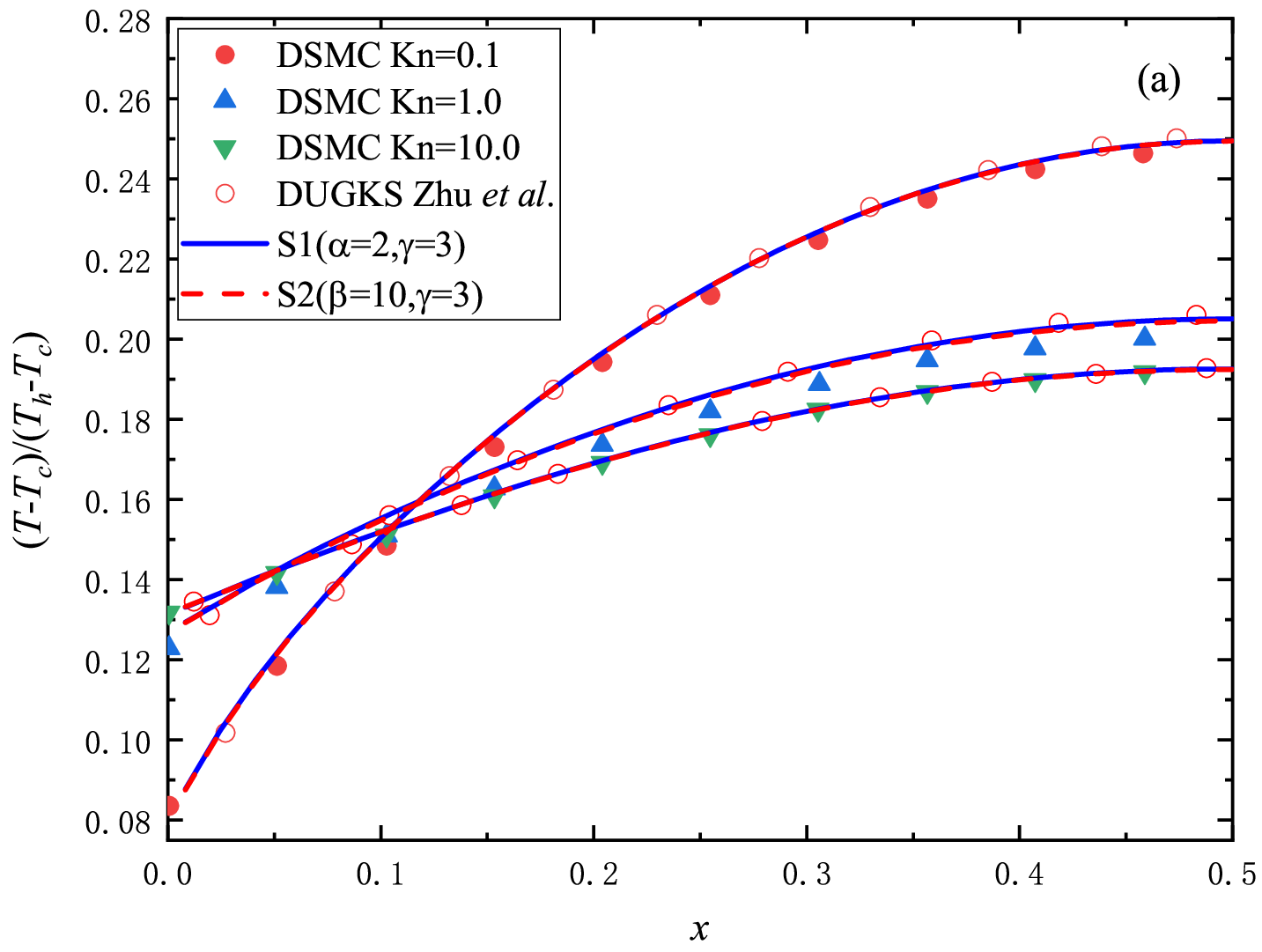}
	\includegraphics[width=7.5cm,height=6.5cm]{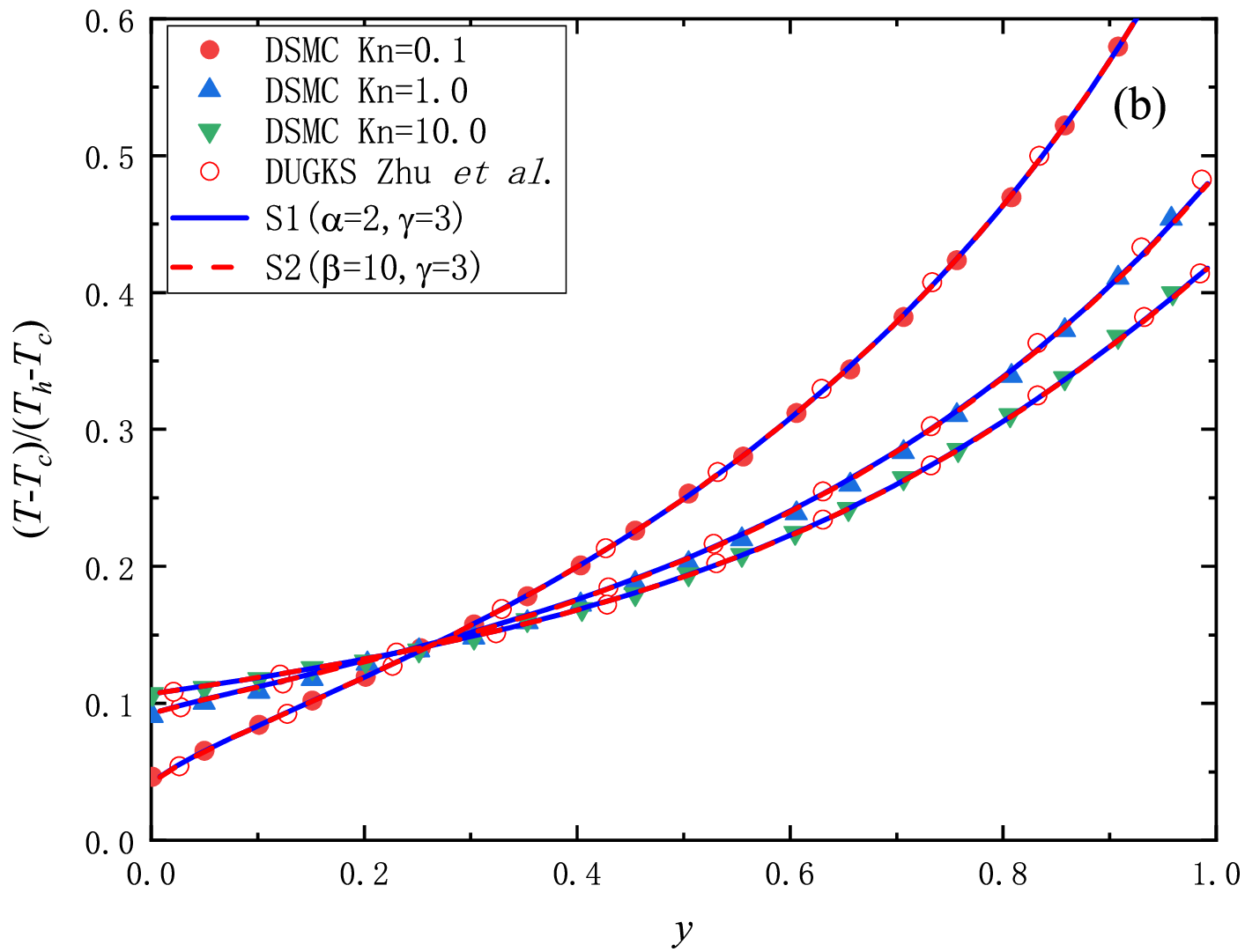}
	\includegraphics[width=7.5cm,height=6.5cm]{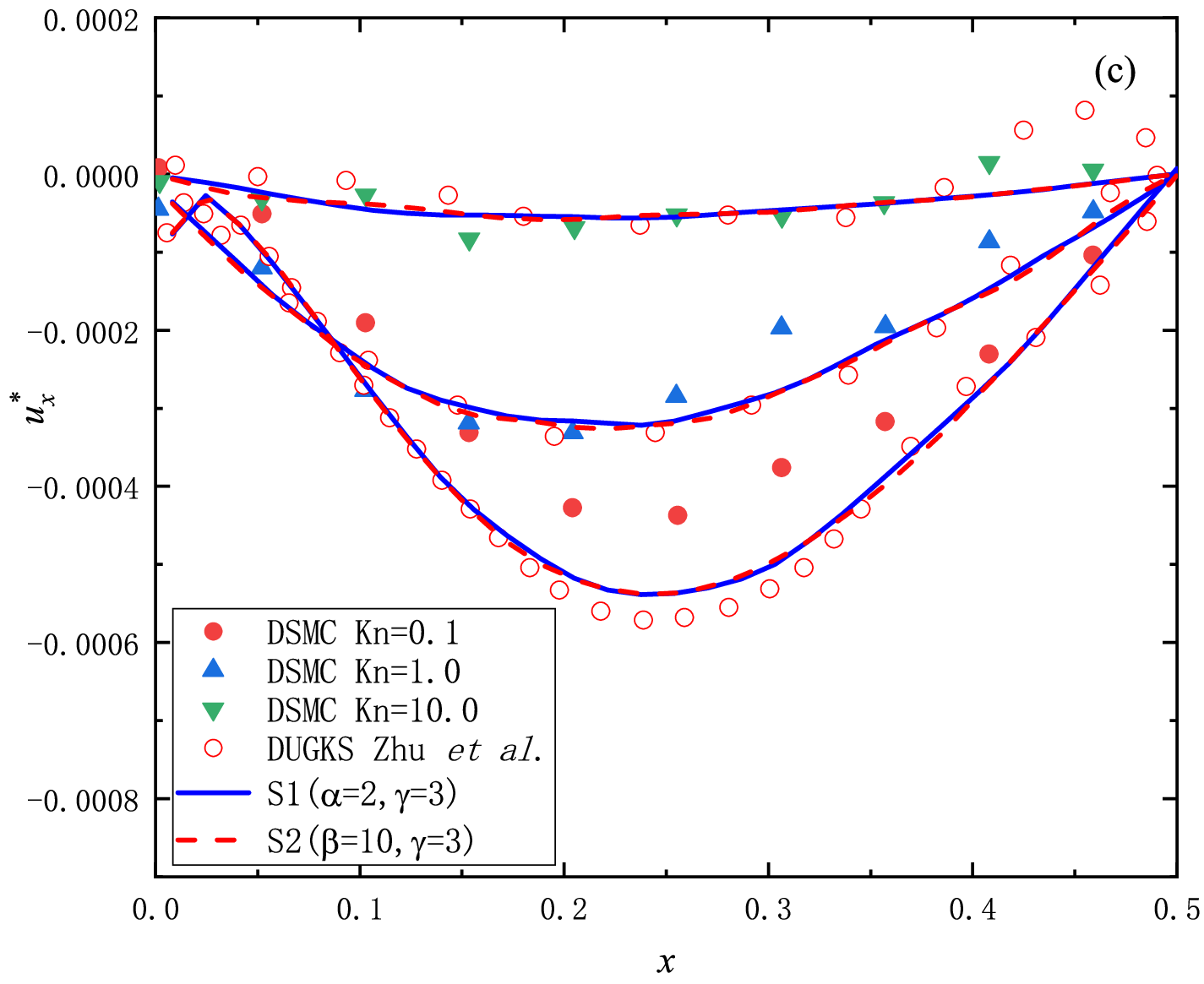}
	\includegraphics[width=7.5cm,height=6.5cm]{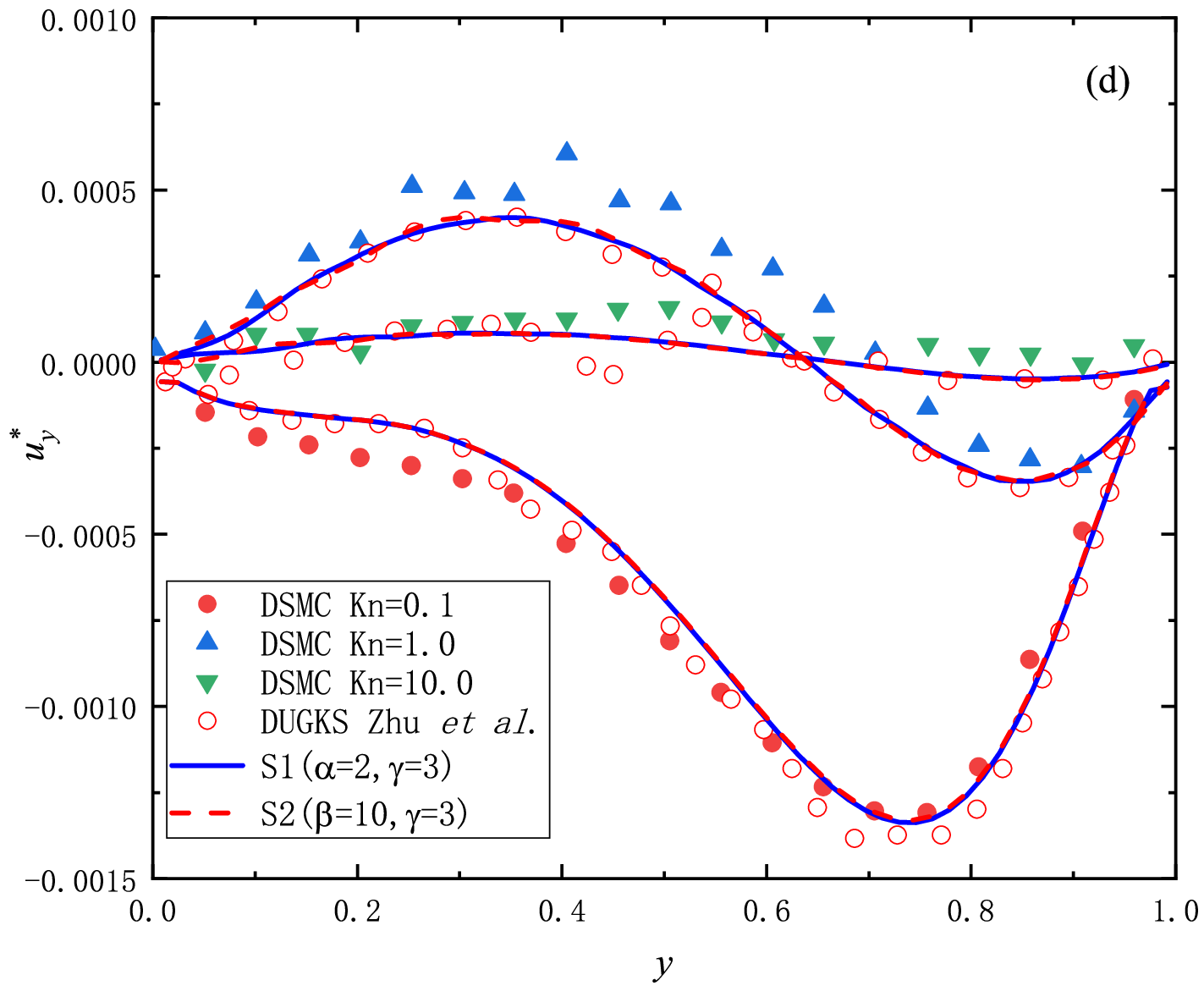}
	\caption{\label{fig:TCavityT} \centering  Temperature and velocity profiles of the TDI cavity flow.}
\end{figure*}

\begin{figure*}[!t]
	\centering
	\subfigure[Kn=0.1]{
		\label{Tcavity0.1}
		\includegraphics[width=5.3cm,height=5.3cm]{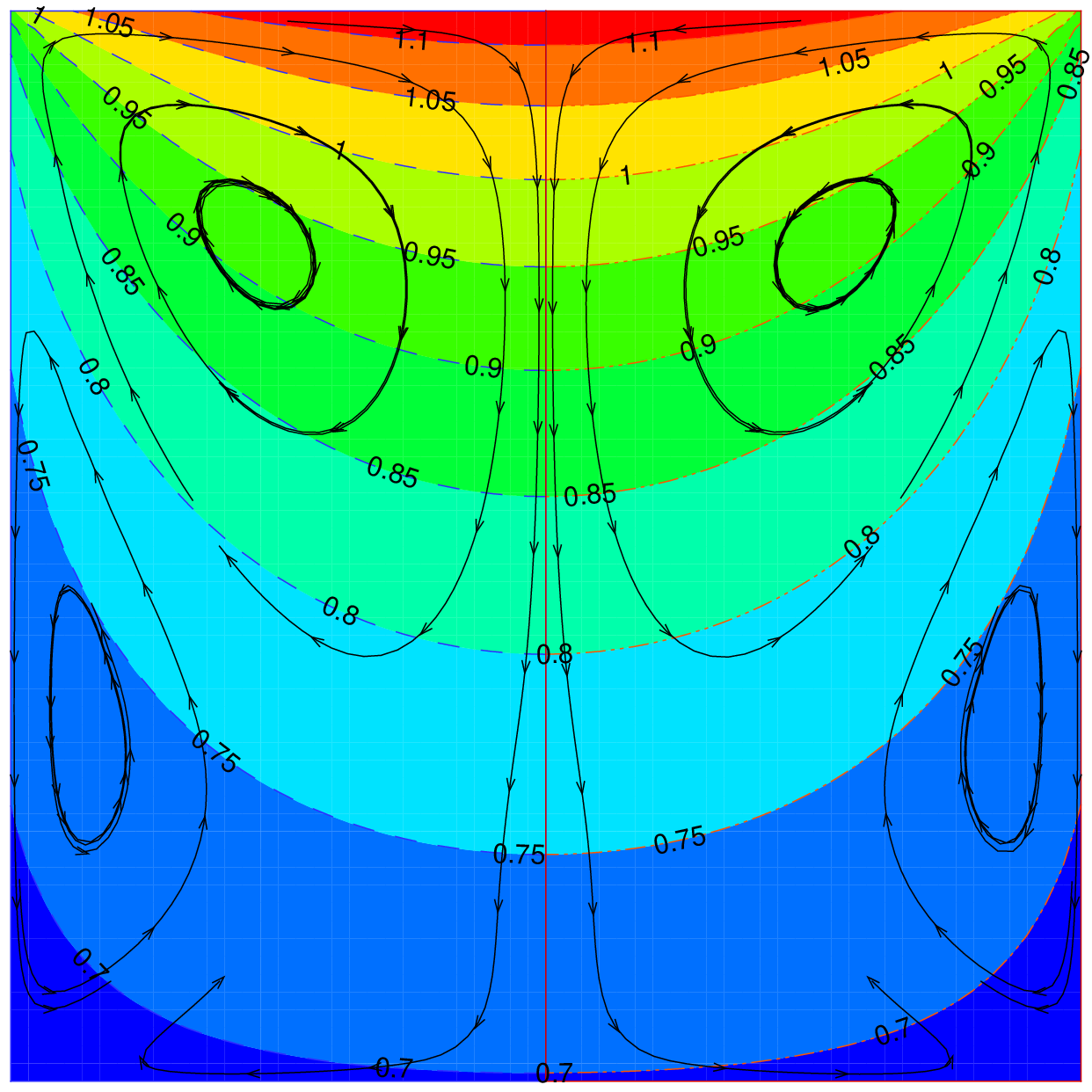}}
	\subfigure[Kn=1.0]{
		\label{Tcavity1.0}
		\includegraphics[width=5.3cm,height=5.3cm]{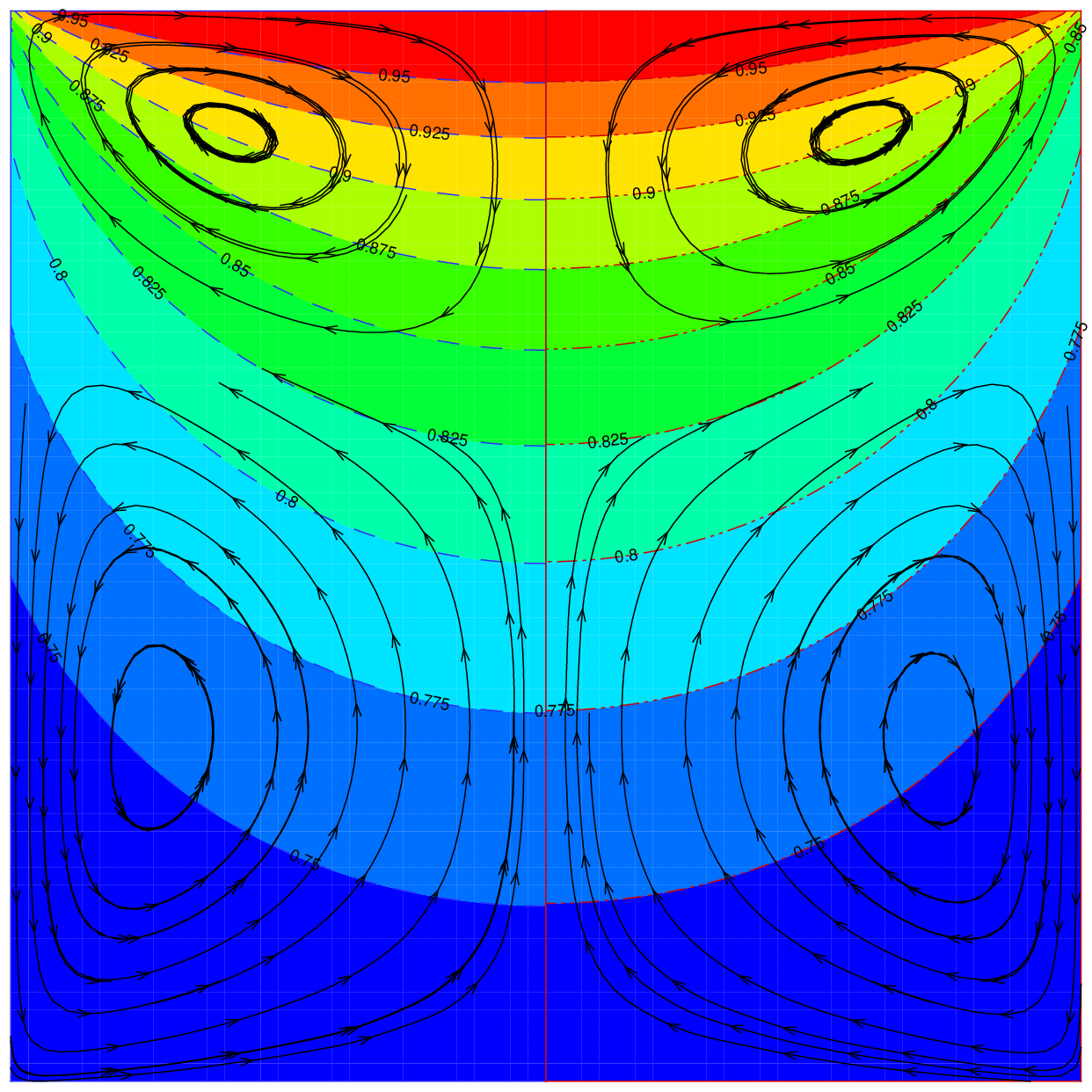}}
	\subfigure[Kn=10.0]{
		\label{Tcavity10.0}
		\includegraphics[width=5.3cm,height=5.3cm]{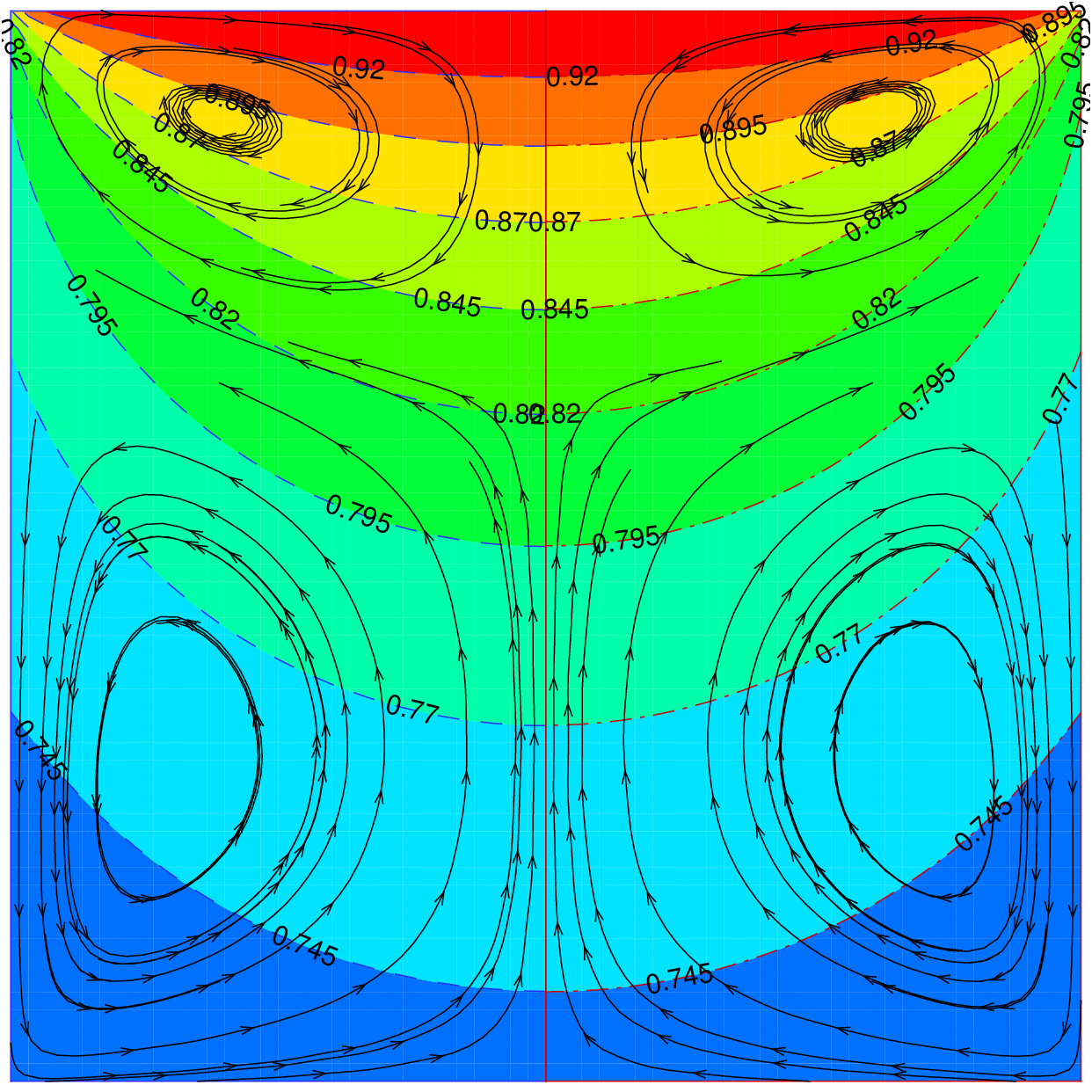}}
	\caption{\label{fig:TCavityTU} \centering  Temperature field and streamlines of the TDI cavity flow. In each sub-figure, left and right half are results using the S1 and S2, respectively.}	
\end{figure*}

In the TDI cavity flow, the flow intensity declines as the Knudsen number increases. The three Knudsen numbers examined in this study demonstrate significantly low flow velocities, complicating the measurement of precise velocity distributions. For instance, at Kn=10, Zhu et al. \citep{ZHU2019} employed an exceptionally fine grid for velocity discretization. Yet, the resulting velocity field lacked smoothness and exhibited non-physical vortices in several areas. Conversely, predicting the temperature field appears to be comparatively less complex. Fig.~\ref{fig:TCavityT} (a) and (b) display the temperature distributions along the vertical and horizontal midlines, respectively. The results achieved using two 3D PGQ rule align closely with those obtained via the DUGKS and DSMC methods, as documented by Zhu et al. \citep{ZHU2019}. Fig.~\ref{fig:TCavityTU} further contrasts the two 3D PGQ rule established in Section \ref{sec3.2}, demonstrating their capability to generate smooth temperature fields and streamline profiles. Notably, at a Knudsen number of 10, the proposed methodology effectively mitigates the occurrence of oscillating streamlines and non-physical vortices seen in the reference study by Zhu et al., substantiating the effectiveness of the suggested approach in this research context.

\subsection{Cylinder flow}

To further evaluate the performance of the PGQ method in hypersonic rarefied flows, we present calculations for the flow past a cylinder at a Mach number of 5 and a Knudsen number of 1. The computational domain is defined as an annular region between two concentric circles, with an inner radius of \( r = 0.01 \, \text{m} \) and an outer radius \( R = 10r \). The outer boundary represents the free-stream conditions, with a velocity \( U_\infty = 1538.18 \, \text{m/s} \) and a temperature \( T_\infty = 273 \, \text{K} \). The inner wall is maintained at a temperature \( T_\infty \) and treated as a diffusely reflecting surface. In the simulations, the dimensionless temperature and density at both the inner and outer boundaries are set to \( T_0 = 1 \) and \( \rho_0 = 1 \), respectively. The characteristic velocity is defined as \( u_\infty = \sqrt{2 R T_\infty} \), yielding a free-stream velocity of 4.56.

The computational grid consists of \( 64 \times 64 \) cells. Velocity discretization is carried out using both the PGQ method (as described in this paper) and the Newton-Cotes (NC) rule. For supersonic flow, discrete velocities are typically distributed over a broad range. Therefore, for the PGQ method, a \( P_2 \) approximation with \( 20 \times 45 \) grid points and a parameter \( \beta = 1000 \) is employed, resulting in a discrete velocity distribution confined to a circular region with a radius of 8. In contrast, the NC rule utilizes \( 89 \times 89 \) points, covering a velocity range of \( [-10, 10] \times [-10, 10] \), which corresponds to approximately 8.8 times the number of discrete velocities used by the PGQ method.

\begin{figure*}[!t]
	\centering
	\includegraphics[width=8cm,height=8cm]{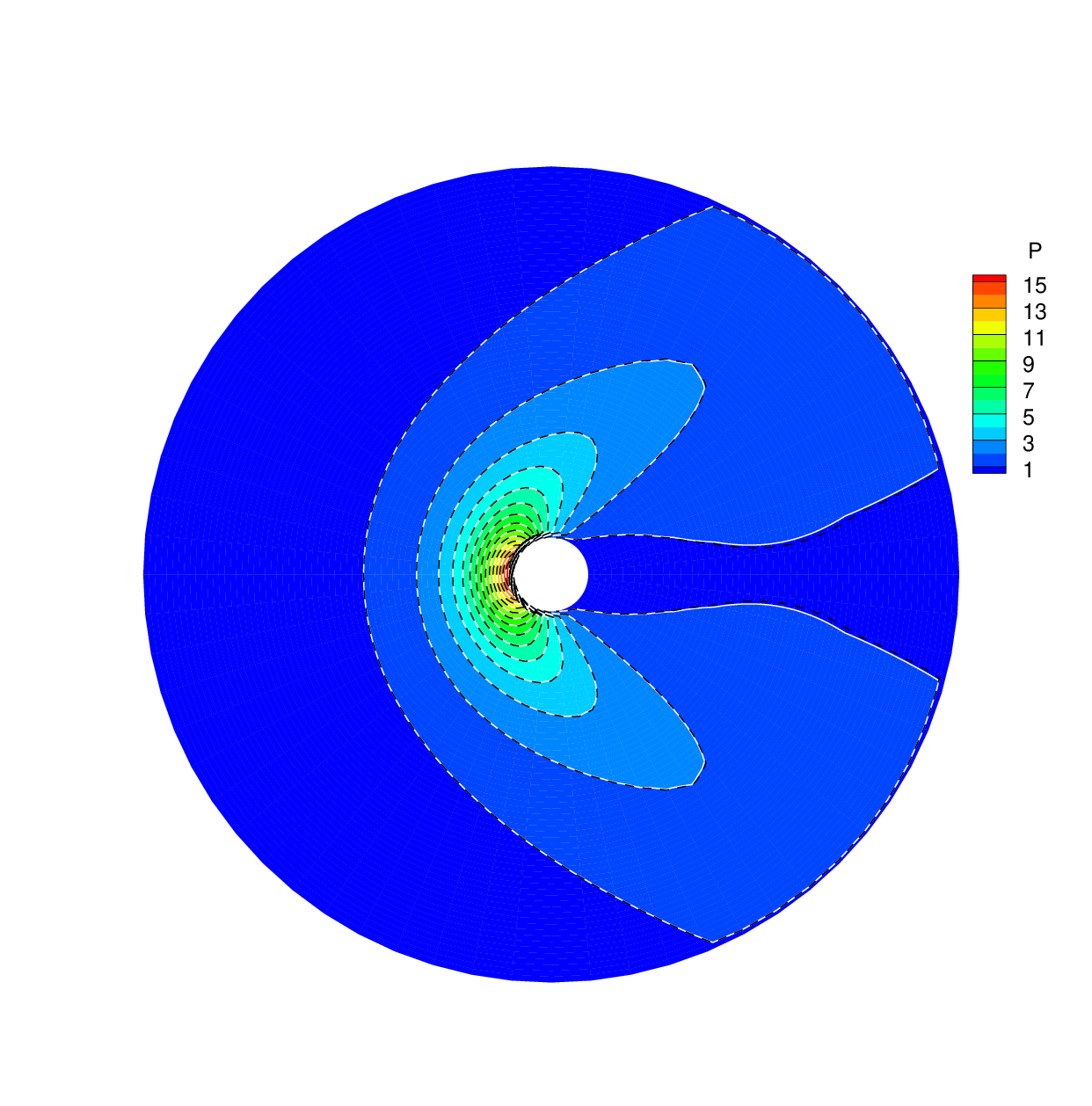}
	\includegraphics[width=8cm,height=8cm]{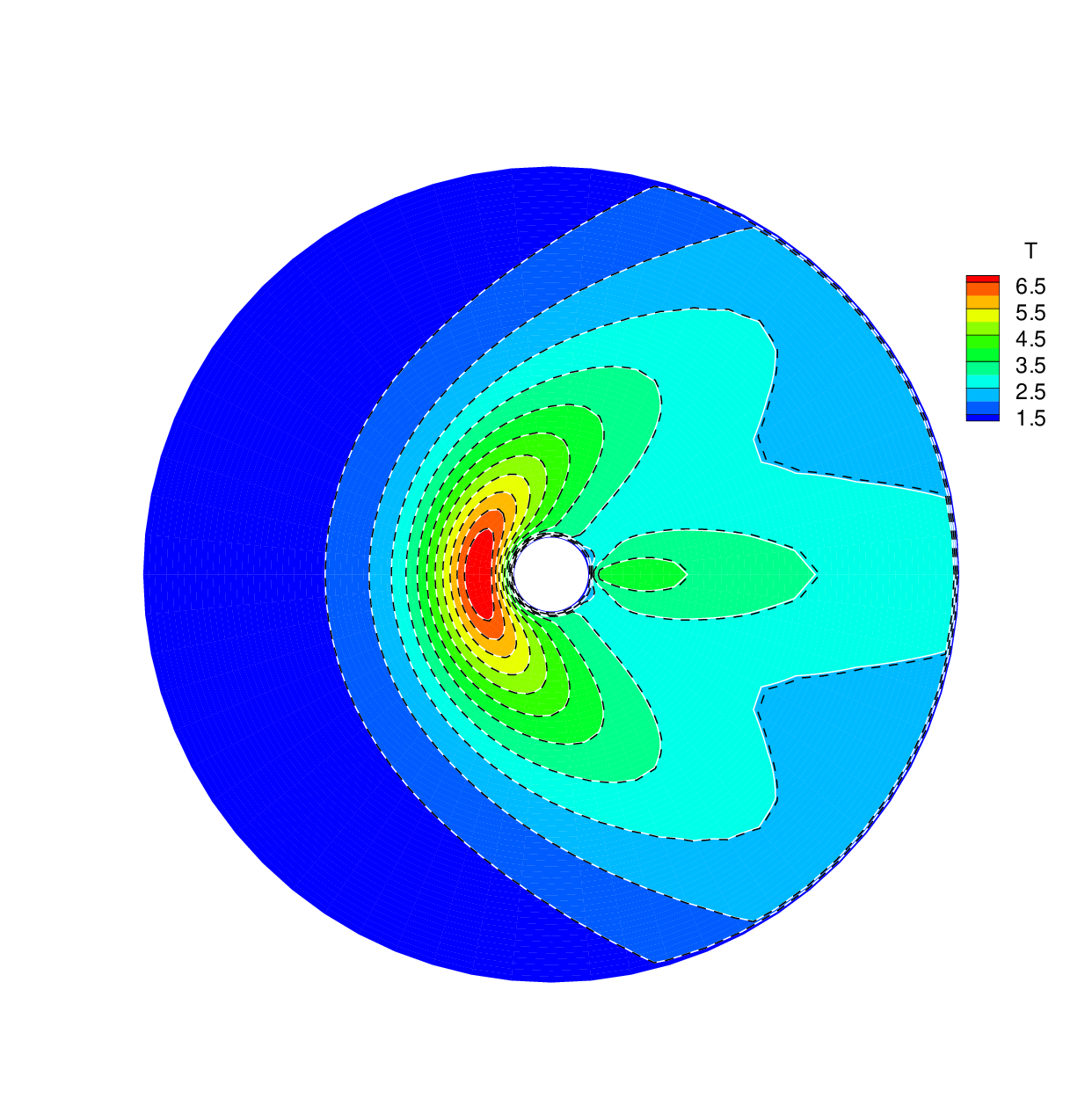}
	\includegraphics[width=8cm,height=8cm]{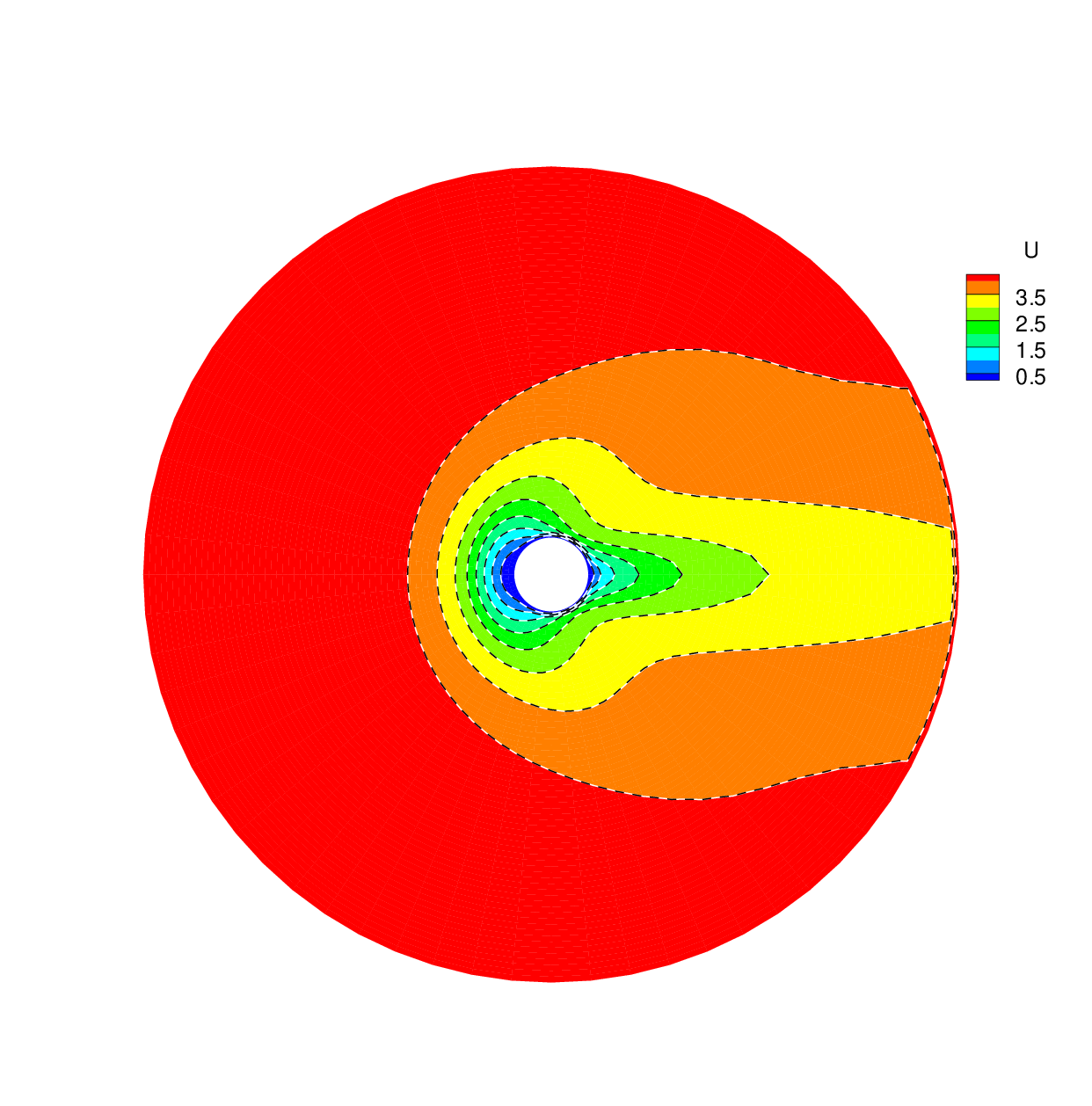}
	\includegraphics[width=8cm,height=8cm]{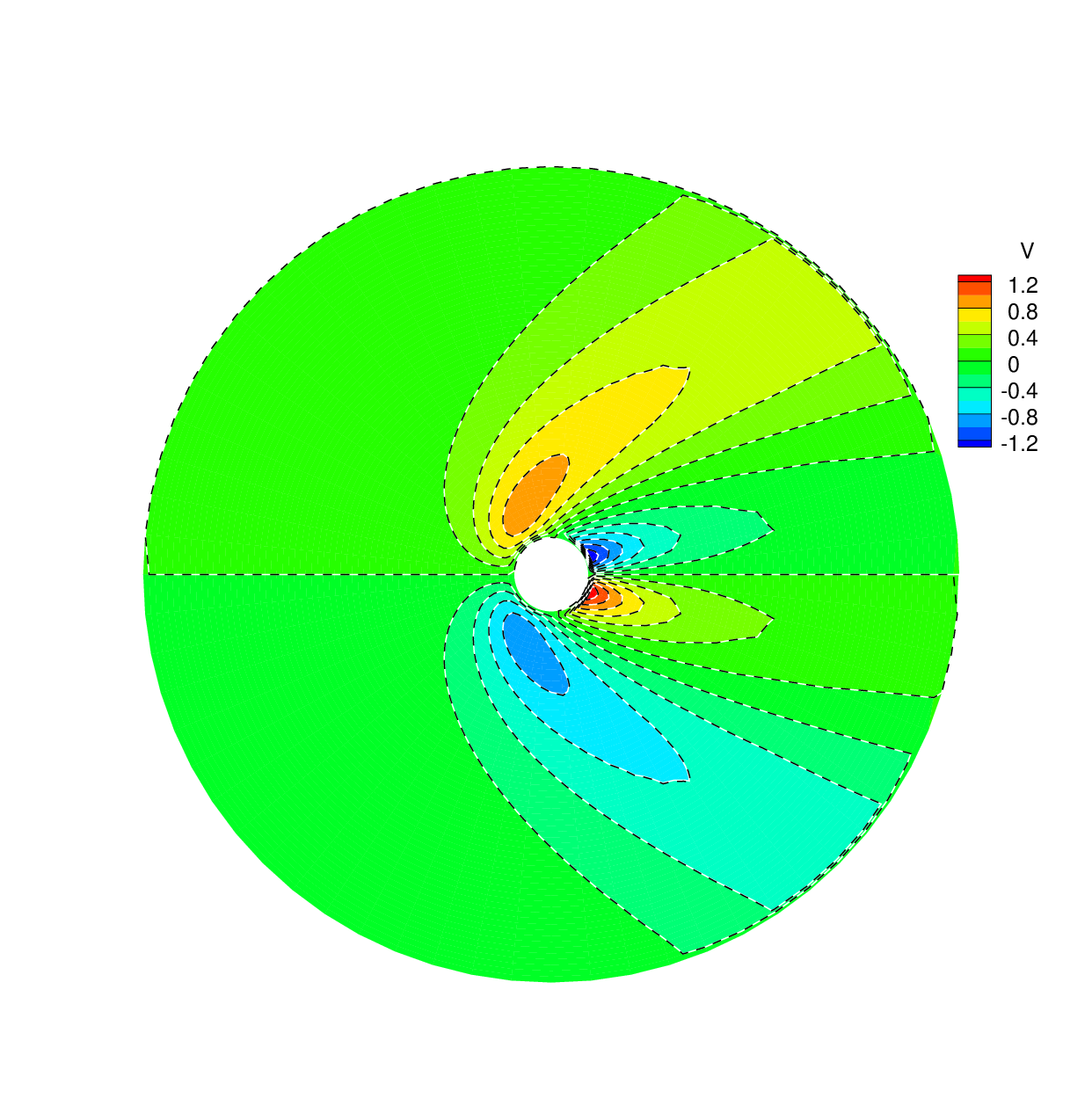}
	\caption{\label{fig:CylinderC} \centering  Comparison of pressure, temperature, u-velocity, and v-velocity contours for cylinder flow at Ma = 5 and Kn = 1 (NC: colored background with white solid line; PGQ: black dashed line).}
\end{figure*}

\begin{figure*}[!t]
	\centering
	\includegraphics[width=5.3cm,height=4.5cm]{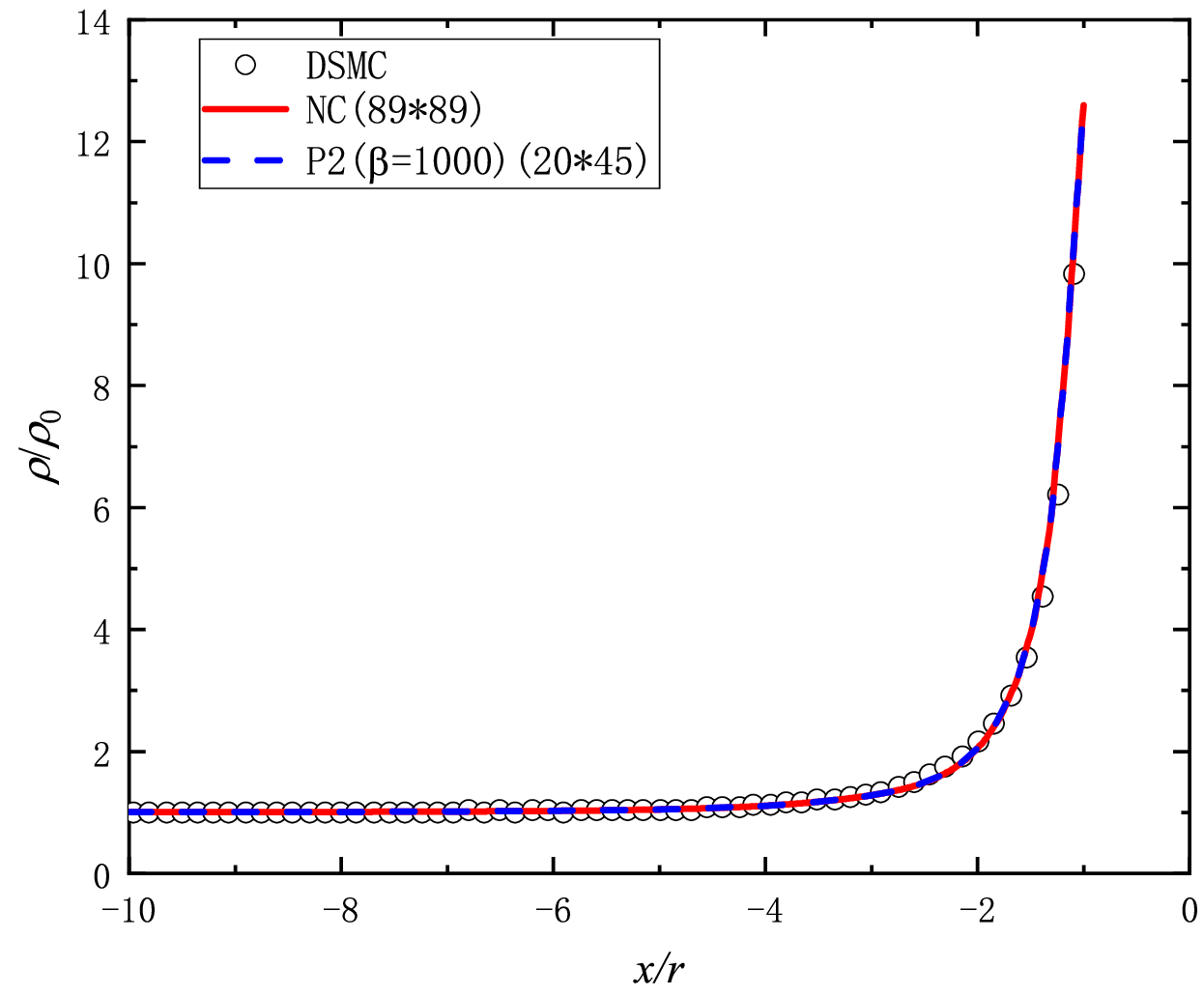}
	\includegraphics[width=5.3cm,height=4.5cm]{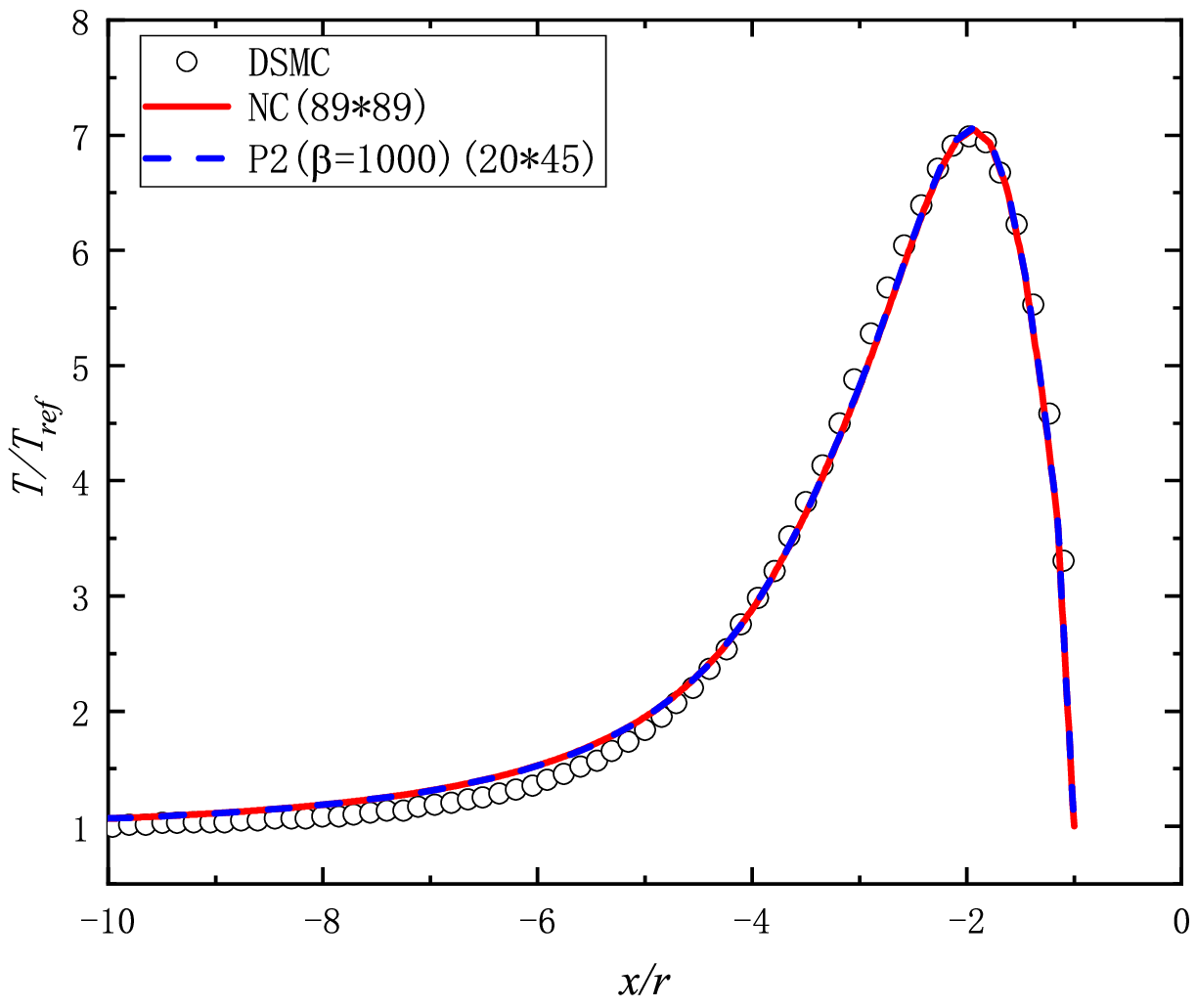}
	\includegraphics[width=5.3cm,height=4.5cm]{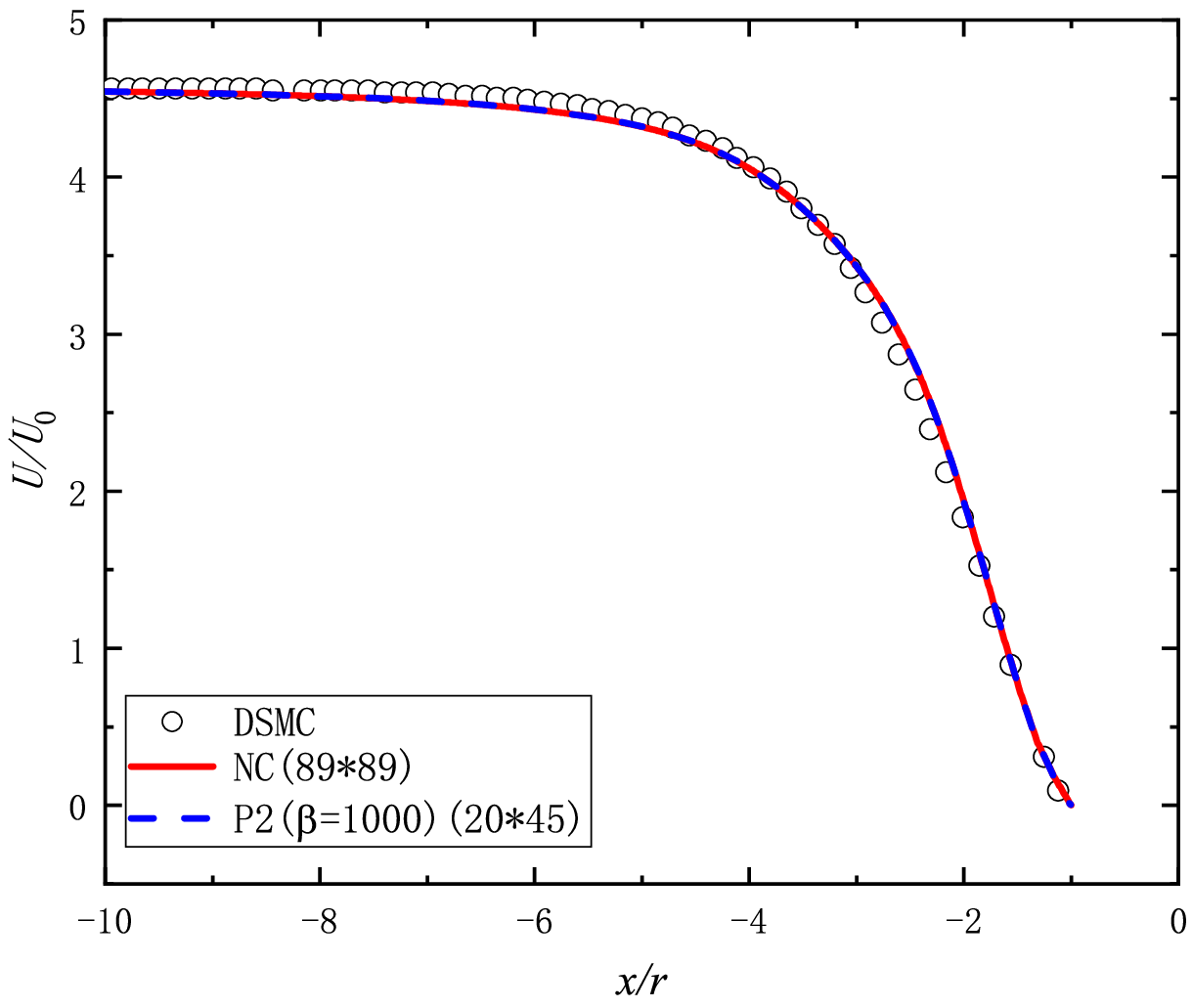}
	\caption{\label{fig:CylinderL} \centering  Density, u-velocity, and temperature profiles along the stagnation line of the Cylinder flow.}
\end{figure*}

Fig.~\ref{fig:CylinderC} compares the pressure, temperature, and velocity profiles obtained using both methods. Despite the more sparse velocity distribution employed by the PGQ method, its computational accuracy is found to be in excellent agreement with that of the NC method. Additionally, Fig.~\ref{fig:CylinderL} presents the density, temperature, and velocity distributions along the stagnation streamline, showing perfect agreement between the results from both methods. These findings highlight the efficiency and accuracy of the PGQ method in simulating hypersonic flows.

\label{sec4.5}

\section{Conclusion}
\label{sec5}
In this study, we introduce a parametric Gaussian quadrature (PGQ) for velocity discretization within the framework of DUGKS, an emerging computational fluid dynamics method capable of solving Boltzmann equations across various Knudsen numbers within a unified framework. Despite its versatility, DUGKS encounters a common computational challenge wherein conventional velocity discretization methods require an extensive number of nodes, particularly for problems with high Knudsen and Mach numbers. Currently, velocity space discretization relies mainly on Newton-Cotes quadrature and Gaussian quadrature rules, each presenting inherent limitations within the DUGKS framework. Newton-Cotes quadrature suffers from low accuracy, while the conventional Gaussian quadrature struggles to align with particle velocity distribution characteristics. Thus, there is a pressing need for the development of high-precision and flexible velocity discretization methods.
Motivated by this need, we propose an approach utilizing integral variation and parameterized weight functions to achieve the high accuracy of Gaussian methods in velocity discretization, addressing the inflexibility of fixed Gaussian points. We present two types of PGQs tailored for 2D and 3D velocity spaces, enabling adjustments to discrete velocity points through parameters. Numerical experiments validate the efficacy of our method across various Knudsen numbers, showcasing its superiority in computational efficiency and accuracy compared to traditional methods such as Newton-Cotes and half-range Gauss-Hermite quadratures.
Our primary focus is on validating the 2D PGQ; however, future work will pay more attention to the 3D PGQ and investigate its effectiveness in multi-scale flows with high Mach numbers.

\appendix

\section{Gaussian quadratures for semi-infinite interval integrals}
\label{APPA}
\setcounter{table}{0}
\renewcommand\theequation{A.\arabic{equation}}

\begin{table*}[!t]
	\caption{\label{tab:tabA1} Quadrature Abscissas and Weights;~$w(x)=x^{\gamma}e^{-x}$}
	\footnotesize
	\centering
	\begin{tabular}{l|ll|ll}
		\hline
		~ & $x_i~(\gamma=0)$ & $w_i~(\gamma=0)$  & $x_i~(\gamma=\frac{1}{2})$  & $w_i~(\gamma=\frac{1}{2})$\\ \hline
		N=2   & 0.5857864376269050 & 8.535533905932737(-1) & 0.91886116991581 & 7.233630235462755(-1) \\
		~ & 3.4142135623730954 & 1.464466094067262(-1) & 4.08113883008419 & 1.628639019064826(-1) \\ \hline
		N=4   & 0.3225476896193923 & 6.031541043416333(-1) & 0.5235260767382691 & 4.530087465586076(-1) \\
		~ & 1.7457611011583467 & 3.5741869243779995(-1) & 2.1566487632690943 & 3.816169601717997(-1) \\
		~ & 4.5366202969211280 & 3.888790851500541(-2) & 5.1373875461767110 & 5.07946275722408(-2) \\
		~ & 9.3950709123011330 & 5.392947055613295(-4) & 10.182437613815926 & 8.06591150110031(-4) \\ \hline
		N=8   & 0.1702796323051010 & 3.691885893416378(-1) & 0.2826336481165991 & 2.2713936195247167(-1)  \\
		~ & 0.9037017767993799 & 4.187867808143427(-1) & 1.1398738015816137 & 3.9359454280361506(-1) \\
		~ & 2.2510866298661307 & 1.757949866371718(-1) & 2.6015248434060294 & 2.1290897086722818(-1) \\
		~ & 4.2667001702876590 & 3.334349226121566(-2) & 4.724114537527790 & 4.7877483203138180(-2) \\
		~ & 7.0459054023934655 & 2.794536235225675(-3) & 7.605256299231614 & 4.5425174747626330(-3) \\
		~ & 10.758516010180996 & 9.076508773358110(-5) & 11.41718207654583 & 1.6240460018532575(-4) \\
		~ & 15.740678641278006 & 8.485746716272541(-7) & 16.49941079765582 & 1.6423774138061169(-6) \\
		~ & 22.863131736889260 & 1.048001174871512(-9) & 23.73000399593471 & 2.1739431266309080(-9) \\ \hline
		N=16  & 0.0876494104789278 & 2.0615171495780069(-1) & 0.1473991846163111  & 9.774098913713067(-2)  \\
		~ & 0.4626963289150808 & 3.310578549508840(-1) & 0.5909018112431884 & 2.523079012122726(-1) \\
		~ & 1.1410577748312270 & 2.6579577764421436(-1) & 1.3344875116145762 & 2.724198251520787(-1) \\
		~ & 2.1292836450983810 & 1.3629693429637774(-1) & 2.3850115520046535 & 1.7166350712629086(-1) \\
		~ & 3.4370866338932067 & 4.7328928694125220(-2) & 3.752567873874768 & 6.9540261026554210(-2) \\
		~ & 5.0780186145497680 & 1.1299900080339450(-2) & 5.451062939568397 & 1.8734808778456017(-2) \\
		~ & 7.0703385350482350 & 1.8490709435263094(-3) & 7.499085532907372 & 3.3812292389549530(-3) \\
		~ & 9.4383143363919380 & 2.0427191530827824(-4) & 9.921219136072429 & 4.0525569008016644(-4) \\
		~ & 12.214223368866160 & 1.4844586873981333(-5) & 12.750055460117064 & 3.1561318148868834(-5) \\
		~ & 15.441527368781617 & 6.8283193308712460(-7) & 16.029386360375128 & 1.5413042593641668(-6) \\
		~ & 19.180156856753136 & 1.8810248410796997(-8) & 19.819512877102024 & 4.4749378027620335(-8) \\
		~ & 23.515905693991908 & 2.862350242973859(-10) & 24.206680643468307 & 7.136037163751593(-10) \\
		~ & 28.578729742882140 & 2.127079033224097(-12) & 29.321456103352332 & 5.532849784136933(-12) \\
		~ & 34.583398702286620 & 6.297967002517801(-15) & 35.37955078717556 & 1.703727514691122(-14) \\
		~ & 41.940452647688330 & 5.050473700035547(-18) & 42.79325597075464 & 1.418249588844844(-17) \\
		~ & 51.701160339543320 & 4.161462370372804(-22) & 52.618366255753244 & 1.213712303922957(-21) \\ \hline
	\end{tabular}
\end{table*}

\begin{table*}[!t]
	\caption{\label{tab:tabA2} Quadrature Abscissas and Weights;~$w(x)=x^{\alpha}e^{-x^2}$}
	\footnotesize
	\centering
	\begin{tabular}{l|ll|ll}
		\hline
		~ & $x_i~(\alpha=1)$ & $w_i~(\alpha=1)$  & $x_i~(\alpha=2)$  & $w_i~(\alpha=2)$\\ \hline
		N=2 & 0.5466400565221693 & 0.3252320794479061(0) & 0.7539869079898871 & 0.2738413467266824(0) \\
		~ & 1.518176674506265 & 0.1747679205520937(0) & 1.734055298879163 & 0.1692721159996965(0) \\ \hline
		N=4 & 0.2800995401403832 & 0.1139990543365298(0) & 0.4238628193900528 & 0.7649092266787873(-l) \\
		~ & 0.8320770658174104 & 0.2692323797134971(0) & 1.014332104566760 & 0.2435439494642453(0) \\
		~ & 1.556389870300421 & 0.1107889566826584(0) & 1.742437375162050 & 0.1162953035510695(0) \\
		~ & 2.463284959722103 & 0.5979609267314704(-2) & 2.639813333635586 & 0.6783287043185401(-2) \\ \hline
		N=8 & 0.1218127678061463 & 0.2397877317765308(-1) & 0.1990000637984294 & 0.9599144336400067(-2)  \\
		~ & 0.3882449491473571 & 0.1092506819189940(0) & 0.5059526450205794 & 0.7072944976303661(-l) \\
		~ & 0.7651497067658092 & 0.1797622678433810(0) & 0.9041682182040568 & 0.157366887003943l(0) \\
		~ & 1.224690624761160 & 0.1351751653621029(0) & 1.372615723971598 & 0.1429322724003870(0) \\
		~ & 1.751398297664409 & 0.4552181928573556(-1) & 1.900969572329702 & 0.5431444004253597(-l) \\
		~ & 2.343383197810315 & 0.6064921853788935(-2) & 2.490479841967435 & 0.7835224153141577(-2) \\
		~ & 3.016608849956826 & 0.2448536436477049(-3) & 3.158780677105240 & 0.3338952597020048(-3) \\
		~ & 3.831371300820741 & 0.1516914696753451(-5) & 3.966720403265353 & 0.2149767232664775(-5) \\ \hline
		N=16 & 0.04775799543737674 & 0.3795307814831678(-2) & 0.0817491338998452  & 0.7050727473210895(-3)  \\
		~ & 0.1575643611266753 & 0.2136808301992049(-1) & 0.2154761962759740 & 0.7107111654073120(-2) \\
		~ & 0.3236556568455920 & 0.5595857089379011(-1) & 0.4003530517087630 & 0.2844188515941899(-l) \\
		~ & 0.5391473546675038 & 0.9587168277747507(-1) & 0.6298538771405607 & 0.6660235171398239(-l) \\
		~ & 0.7970053979972014 & 0.1169082070371872(0) & 0.8976124329697087 & 0.1025785712747278(0) \\
		~ & 1.090958307363892 & 0.1029363012162623(0) & 1.198149260240153 & 0.1077502032531791(0) \\
		~ & 1.415975970714936 & 0.6468246716393942(-1) & 1.527188184719962 & 0.7747156370638879(-l) \\
		~ & 1.768437030466615 & 0.2831911613754905(-1) & 1.881745606015598 & 0.3763106373385135(-l) \\
		~ & 2.146149962010079 & 0.8362647991652432(-2) & 2.260132964654088 & 0.1204873635560290(-l) \\
		~ & 2.548365652625752 & 0.1597736202726321(-2) & 2.661980315279350 & 0.2453208613776865(-2) \\
		~ & 2.975896592510777 & 0.1870134647150351(-3) & 3.088376381635592 & 0.3020309847850189(-3) \\
		~ & 3.431483868308089 & 0.1243935496206526(-4) & 3.542256017930265 & 0.2092121075871870(-4) \\
		~ & 3.920694119664905 & 0.4208466925294357(-6) & 4.029312272760483 & 0.7314637349679360(-6) \\
		~ & 4.454120573510955 & 0.6051847030054333(-8) & 4.560203031431090 & 0.1080646863902574(-7) \\
		~ & 5.053674269642785 & 0.2643406562982473(-10) & 5.156826768007481 & 0.4828081616137754(-10) \\
		~ & 5.778478847939104 & 0.1524594098604790(-13) & 5.8781144889155572 & 0.2840126937112534(-13) \\ \hline
	\end{tabular}
\end{table*}

\lstset{
	language=Python,
	basicstyle=\ttfamily,
	keywordstyle=\color{cyan}\ttfamily,
	numbers=left,
	numberstyle=\tiny\color{gray},
	numbersep=5pt,
	showspaces=false,
	showstringspaces=false,
	frame=single
}

The \(n\)-th order Gaussian quadrature formula for semi-infinite interval integral presented in this paper takes the form:
\begin{equation}
	\int\limits_{0}^{+\infty}{w\left( x \right) f\left( x \right) dx}\approx\sum_{i=0}^{n-1}{A_if\left( x_i \right)}=\boldsymbol{A}^Tf(\boldsymbol{x}),
	\label{eq:GGL}
\end{equation}
Here, the weight function \(w(x)=\alpha x^{\beta-1}e^{-x^{\alpha}}\), \(\boldsymbol{A}^T=[A_0, A_1, \cdots, A_{n-1}]\) and \(\boldsymbol{x}^T=[x_0, x_1, \cdots, x_{n-1}]\) represent the abscissas and weights of GGL, respectively. Particularly, for \(f(x)=x^k\), we have:
\begin{eqnarray}
	I_{k}=\int\limits_0^{+\infty}{w\left( x \right)x^kdx}=\varGamma \left(\frac{\beta+k}{\alpha} \right),
	\label{eq:IK}
\end{eqnarray}
where \(\varGamma\) denotes the Gamma function. According to the formula Eq.~(\ref{eq:IK}), we can establish a system of equations for determining Gaussian points and weight coefficients. Assuming $\boldsymbol{x}^T$ is the root of the $n$-th orthogonal polynomial $R_n(x)=x^n+\sum_{m=0}^{n-1}{a_{n,m}x^m}$ with weight $w(x)$ on $(0,+\infty)$, and $R_n$ is orthogonal to $x^m$ for $0\leq m\textless n$, that is
\begin{equation}
	\int\limits_0^{+\infty}{w\left( x \right) R_n\left( x \right) x^mdx}=0,~0\leq m\textless n.
	\label{eq:orthogonal}
\end{equation}
According to Eq.~(\ref{eq:orthogonal}), we can establish a system of equations for solving the coefficients of polynomial \(R_n\) \citep{doman2015}:
\begin{equation}
	\left( \begin{array}{c}
		I_0\\
		I_1\\
		\cdots\\
		I_{n-1}\\
	\end{array}\begin{array}{c}
		I_1\\
		I_2\\
		\cdots\\
		I_n\\
	\end{array}\begin{array}{c}
		I_2\\
		I_3\\
		\cdots\\
		I_{n+1}\\
	\end{array}\begin{array}{c}
		\cdots\\
		\cdots\\
		\cdots\\
		\cdots\\
	\end{array}\begin{array}{c}
		I_{n-1}\\
		I_n\\
		\cdots\\
		I_{2n-2}\\
	\end{array} \right) \left( \begin{array}{c}
		a_{n,0}\\
		a_{n,1}\\
		\cdots\\
		a_{n,n-1}\\
	\end{array} \right) =-\left( \begin{array}{c}
		I_n\\
		I_{n+1}\\
		\cdots\\
		I_{2n-1}\\
	\end{array} \right) 
	\label{eq:orSE}
\end{equation}
Furthermore, by combining Eq.~(\ref{eq:GGL}) and Eq.~(\ref{eq:IK}), we can easily derive the following relationship:
\begin{equation}
	\left( \begin{array}{c}
		1\\
		x_0\\
		x_{0}^{2}\\
		\cdots\\
		x_{0}^{n-1}\\
	\end{array}\begin{array}{c}
		1\\
		x_1\\
		x_{1}^{2}\\
		\cdots\\
		x_{1}^{n-1}\\
	\end{array}\begin{array}{c}
		\cdots\\
		\cdots\\
		\cdots\\
		\cdots\\
		\cdots\\
	\end{array}\begin{array}{c}
		1\\
		x_{n-1}\\
		x_{n-1}^{2}\\
		\cdots\\
		x_{n-1}^{n-1}\\
	\end{array} \right) \left( \begin{array}{c}
		A_0\\
		A_1\\
		A_2\\
		\cdots\\
		A_{n-1}\\
	\end{array} \right) =\left( \begin{array}{c}
		I_0\\
		I_1\\
		I_2\\
		\cdots\\
		I_{n-1}\\
	\end{array} \right) 
	\label{eq:sovA}
\end{equation}

In general, orthogonal polynomials and the coefficients of Gaussian quadrature can be obtained through Eq.~(\ref{eq:orSE}) and Eq.~(\ref{eq:sovA}). However, when the order \(n\) or parameter \(\beta\) is large, the condition number of the coefficient matrix of these two equations becomes large, making the solution of the equations exceedingly difficult. Fortunately, researchers in computational mathematics have discovered some general forms of orthogonal polynomials and coefficients. For the generalized Gauss-Laguerre quadrature with the weight function \(w(x)=x^{\alpha}e^{-x}\), the \(n\)-th order Laguerre polynomial and the corresponding weight coefficient are given by:
\begin{equation}
	L_{n}^{(\alpha)}\left( x \right) =\sum_{k=0}^n{\left( \begin{array}{c}
			n+\alpha\\
			n-k\\
		\end{array} \right) \frac{\left( -x \right) ^k}{k!}},
	\label{subeq:PLaguerre}
\end{equation}
\begin{equation}
	\omega _{\mathcal{L} ,i}=\frac{\varGamma \left( n+\alpha+1 \right)}{n!x_i\left[ \frac{d}{dx}L_{n}^{\alpha}\left( x_i \right) \right] ^2},
	\label{subeq:wLaguerre}
\end{equation}

As a classic Gaussian quadratures, the roots of Laguerre polynomials its corresponding weights can be directly computed using the \texttt{\color{teal}roots\_genlaguerre} functions in Python.

\begin{lstlisting}
	from scipy.special import roots_genlaguerre
	roots, weights = roots_genlaguerre(n, alpha)
\end{lstlisting}

In this study, we employed Gaussian quadrature with weight functions \(w(x) = x^\alpha e^{-x}\), where \(\alpha = 1, 2\), for simulation purposes. The corresponding abscissas and weights are detailed in Table~\ref{tab:tabA1}. Additionally, Shizgal developed another Gaussian quadrature scheme utilizing \(w(x) = x^\alpha e^{-x^2}\), with its respective abscissas and weights provided in Table~\ref{tab:tabA2}.

\section{Gaussian quadratures for finite interval integrals}
\label{APPB}
\setcounter{table}{0}
\renewcommand\theequation{B.\arabic{equation}}
\begin{table*}[!t]
	\caption{\label{tab:tabB1} Quadrature Abscissas and Weights;~$w(x)=\sqrt{-\left( \beta +1 \right) ^3\ln x}x^{\beta}$}
	\footnotesize
	\centering
	\begin{tabular}{c|cc|cc|cc}
		\hline
		~ & $x_i~(\beta=5)$ & $w_i~(\beta=5)$  & $x_i~(\beta=10)$  & $w_i~(\beta=10)$& $x_i~(\beta=20)$  & $w_i~(\beta=20)$\\ \hline
		N=2 & 0.57629132 & 0.29530203 & 0.72038002 & 0.24657219 & 0.83390397 & 0.21718414\\
		~ & 0.88460605 & 0.70469797 & 0.92910804 & 0.75342781 & 0.95998085 & 0.78281586\\ \hline
		N=4 & 0.32687150 & 0.01644316 & 0.48936401 & 0.0064105 & 0.65828009 & 0.00298003\\
		~ & 0.57616736 & 0.18867383 & 0.6996249 & 0.12952583 & 0.81035271 & 0.09427616\\
		~ & 0.79573697 & 0.48320350 & 0.86144333 & 0.4805269 & 0.91565869 & 0.46548481\\
		~ & 0.94640386 & 0.31167951 & 0.96454445 & 0.38353676 & 0.97885944 & 0.4372590\\ \hline
		N=8 & 0.14210207 & 1.58763404(-4) & 0.25947998 & 9.30424950(-6)  & 0.44683993 & 1.41643758(-6)\\
		~ & 0.27007867 & 3.72985550(-3) & 0.39759869 & 5.32996239(-4) & 0.57745599 & 1.64201979(-4)\\
		~ & 0.41277725 & 2.74523303(-2) & 0.53208644 & 7.83210768(-3) & 0.68925335 & 4.02765858(-3)\\
		~ & 0.56007311 & 1.01471606(-1) & 0.65915508 & 4.99945124(-2) & 0.78513061 & 3.71366747(-2)\\
		~ & 0.70063670 & 2.22864169(-1) & 0.77313028 & 1.68125955(-l) & 0.86439213 & 1.58542160(-1)\\
		~ & 0.82346092 & 3.05996280(-1) & 0.86843194 & 3.17908269(-1) & 0.92578666 & 3.36031562(-1)\\
		~ & 0.91886488 & 2.49524321(-1) & 0.94023843 & 3.24032578(-1) & 0.96841491 & 3.39693725(-1)\\
		~ & 0.97930636 & 8.88026753(-2) & 0.98486342 & 1.31564278(-1) & 0.99250076 & 1.24402601(-1)\\  \hline
	\end{tabular}
\end{table*}

\begin{table*}[!t]
	\caption{\label{tab:tabB2} Quadrature Abscissas and Weights;~$w(x)=(\beta+1)x^{\beta}$}
	\footnotesize
	\centering
	\begin{tabular}{l|ll|ll}
		\hline
		~ & $x_i~(\beta=5)$ & $w_i~(\beta=5)$  & $x_i~(\beta=1000)$  & $w_i~(\beta=1000)$\\ \hline
		N=2 & 0.6307915938297450 & 2.3002537642198062(-1) & 0.9966000912402799 & 1.469762770011098(-1) \\
		~ & 0.9247639617258105 & 7.6997462357801940(-1) & 0.9994158450147002 & 8.530237229989791(-1) \\ \hline
		N=4 & 0.3568937290501589 & 9.20687849031862(-3) & 0.9906905886331325 & 5.552852008870015(-4) \\
		~ & 0.6146693898553784 & 1.285704277869184(-1) & 0.9954938787780911 & 3.936967684862895(-2) \\
		~ & 0.8310790038601141 & 4.323381849914543(-1) & 0.9982635679203677 & 3.583465332972617(-1) \\
		~ & 0.9665886464651178 & 4.2988450873130857(-1) & 0.9996789487953928 & 6.017285046533112(-1) \\ \hline
		N=8 & 0.1531506616095463 & 6.318986772426487(-5) & 0.9775842582719582 & 1.232592606927976(-9)  \\
		~ & 0.2872644038839036 & 1.6701516899719814(-3) & 0.9845129953718983 & 9.467213767103192(-7) \\
		~ & 0.4346274066992901 & 1.4012049001526895(-2) & 0.9893887815715767 & 9.758797732137589(-5) \\
		~ & 0.5845185665631613 & 6.026768632946612(-2) & 0.9930377952494749 & 2.922899870570007(-3) \\
		~ & 0.7251264097133869 & 1.5891180670646887(-1) & 0.9957781857321654 & 3.416227870648118(-2) \\
		~ & 0.8451894879311451 & 2.7531391920671877(-1) & 0.9977703722491161 & 1.774332399114046(-1) \\
		~ & 0.9350435074560832 & 3.0920534298966784(-1) & 0.9991043157836819 & 4.184756331090476(-1) \\
		~ & 0.9874605085244358 & 1.805558542084552(-1) & 0.9998311697858764 & 3.669074124712951(-1) \\ \hline
		N=16 & 0.0526300249207461 & 1.0808090660056195(-7) & 0.9504065523794794  & 8.9844662163020(-22)  \\
		~ & 0.1023279959501052 & 3.692254626039491(-6) & 0.9595777169622102 & 9.4069520818247(-18) \\
		~ & 0.1620647958527960 & 4.320054634061596(-5) & 0.9665486941860748 & 1.04932114694996(-l4) \\
		~ & 0.2305981139007020 & 2.831227033723278(-4) & 0.9722756704743601 & 3.23545825106545(-l2) \\
		~ & 0.3061601894894426 & 1.265639399936556(-3) & 0.9771306457000029 & 4.03173150155723(-10) \\
		~ & 0.3867283154576461 & 4.274300791549567(-3) & 0.9813076086387901 & 2.480567336366805(-8) \\
		~ & 0.4701224831599079 & 1.158174388520331(-2) & 0.9849236146134018 & 8.507095012029777(-7) \\
		~ & 0.5540771537028182 & 2.6154213640331478(-2) & 0.9880557508355463 & 1.760801598481376(-5) \\
		~ & 0.6363075425986467 & 5.04546421578546(-2) & 0.9907577483321799 & 2.322731598070508(-4) \\
		~ & 0.7145738468567876 & 8.446695946838328(-2) & 0.9930684976850878 & 2.028065483915047(-3) \\
		~ & 0.7867432443428619 & 1.237887203079341(-1) & 0.9950168324677634 & 1.202321759750449(-2) \\
		~ & 0.8508485426409111 & 1.591145057191218(-1) & 0.9966243996015842 & 4.911476461106695(-2) \\
		~ & 0.9051421181110701 & 1.7814611524839424(-1) & 0.9979074658050603 & 1.386479427103219(-1) \\
		~ & 0.9481438292687690 & 1.698212978875246(-1) & 0.9988780922184497 & 2.663371665482716(-1) \\
		~ & 0.9786819958664272 & 1.2929220912981512(-1) & 0.9995449175557176 & 3.283183125914964(-1) \\
		~ & 0.9959308889614428 & 6.130952877870559(-2) & 0.9999137770404158 & 2.032797733601271(-1) \\ \hline
	\end{tabular}
\end{table*}

The Gaussian quadrature rule employed for finite interval integration in this study is expressed as:
\begin{equation}
	\int\limits_0^{1}{w\left( x \right) f\left( x \right) dx} \approx \sum_{i=0}^{n-1}{A_i f\left( x_i \right)},
	\label{eq:GQFII}
\end{equation}
where the weight function \(w(x)=(\beta+1)x^{\beta}\) or \(w(x)=\sqrt{-\left( \beta +1 \right) ^3\ln x}x^{\beta}\), and \(\beta>-1\). The corresponding orthogonal polynomials and weight coefficients can be determined using Eq.~(\ref{eq:orSE}) and Eq.~(\ref{eq:sovA}). For instance, when \(w(x)=\sqrt{-\left( \beta +1 \right) ^3\ln x}x^{\beta}\), the weighted integration of \(x_k\) is expressed as:
\begin{equation}
	I_{k} = \int\limits_0^1{\underset{w\left( x \right)}{\underbrace{\sqrt{-\left( \beta +1 \right) ^3\ln x}x^{\beta}}}x^kdx} = \frac{\sqrt{\pi}}{2} \left( \frac{\beta +1}{\beta +k+1} \right) ^{\frac{3}{2}}.
	\label{eq:Ik3}
\end{equation}

However, these equations become ill-conditioned as the order increases, necessitating specialized methods for their solution. This section provides the abscissa and weight coefficients of Gaussian quadrature when \(\beta=5,~10,~20\), and \(n=2,~4,~8\), as shown in Table~\ref{tab:tabB1}.

When \(w(x)=(\beta+1)x^{\beta}\), the Gauss-Jacobi quadrature rule is applicable. Two commonly utilized forms of weight functions for Gauss-Jacobi quadrature are \(w(x) = (1-x)^\alpha (1+x)^\beta\) for \(x \in [-1,1]\), and \(w(x) = (1-x)^\alpha x^\beta\) for \(x \in [0,1]\). These two forms are mathematically equivalent and can be transformed into each other through simple variable substitution. For the first form, the Jacobi polynomials and the weight coefficients of the Gauss-Jacobi quadrature, are:
\begin{equation}
	J_{n}^{\left( \alpha ,\beta \right)}\left( x \right) = \frac{\varGamma \left( \alpha +n+1 \right)}{n!\varGamma \left( \alpha +\beta +n+1 \right)}\sum_{k=0}^n{\left( \begin{array}{c}
			n\\
			k\\
		\end{array} \right) \frac{\varGamma \left( \alpha +\beta +n+k+1 \right)}{\varGamma \left( \alpha +k+1 \right)}\left( \frac{x-1}{2} \right) ^k},
	\label{eq:PJacobi}
\end{equation}

\begin{equation}
	\omega _{\mathcal{J} ,i} = \frac{2^{\alpha +\beta +1}\varGamma \left( n+\alpha +1 \right) \varGamma \left( n+\beta +1 \right)}{n!\varGamma \left( n+\alpha +\beta +1 \right) \left( 1-x_{i}^{2} \right) \left[ \frac{d}{dx}J_{n}^{\left( \alpha ,\beta \right)}\left( x_i \right) \right] ^2}.
	\label{eq:wJacobi}
\end{equation}

Table~\ref{tab:tabB2} presents partial abscissa and weight coefficients of Gauss-Jacobi quadrature for \(\beta = 5\) and \(\beta = 1000\). Similar to Gauss-Laguerre quadrature, Gauss-Jacobi quadrature represents a classic formula in numerical analysis, with its Gaussian points and weight coefficients readily obtainable through Python's function library. However, in P2 of this paper, the weight function utilized is \(w(x) = (\beta+1) x^\beta\) for \(x \in [0,1]\). This necessitates the variable substitution \(x^{\prime}=0.5(1+x)\), and \(\alpha=0\). The Python code for obtaining the roots of Jacobi polynomials and its corresponding weights is presented below:

\begin{lstlisting}
	from scipy.special import roots_jacobi
	alpha = 0
	roots, weights = roots_jacobi(n, alpha, beta)
	roots = 0.5 * (1 + roots)
	weights = 0.5**(beta+1) * (beta + 1) * weights
\end{lstlisting}

\bibliography{refs}

\begin{thebibliography}{43}
\expandafter\ifx\csname natexlab\endcsname\relax\def\natexlab#1{#1}\fi
\providecommand{\url}[1]{\texttt{#1}}
\providecommand{\href}[2]{#2}
\providecommand{\path}[1]{#1}
\providecommand{\DOIprefix}{doi:}
\providecommand{\ArXivprefix}{arXiv:}
\providecommand{\URLprefix}{URL: }
\providecommand{\Pubmedprefix}{pmid:}
\providecommand{\doi}[1]{\href{http://dx.doi.org/#1}{\path{#1}}}
\providecommand{\Pubmed}[1]{\href{pmid:#1}{\path{#1}}}
\providecommand{\bibinfo}[2]{#2}
\ifx\xfnm\relax \def\xfnm[#1]{\unskip,\space#1}\fi
\bibitem[{Fan and Shen(2001)}]{FAN2001}
\bibinfo{author}{J.~Fan}, \bibinfo{author}{C.~Shen},
\newblock \bibinfo{title}{{Statistical simulation of low-speed rarefied gas
  Flows}},
\newblock \bibinfo{journal}{J. Comput. Phys.} \bibinfo{volume}{167}
  (\bibinfo{year}{2001}) \bibinfo{pages}{393--412}.
\bibitem[{Li and Zhang(2009)}]{LI2009}
\bibinfo{author}{Z.~Li}, \bibinfo{author}{H.~Zhang},
\newblock \bibinfo{title}{{Gas-kinetic numerical studies of three-dimensional
  complex flows on spacecraft re-entry}},
\newblock \bibinfo{journal}{J. Comput. Phys.} \bibinfo{volume}{228}
  (\bibinfo{year}{2009}) \bibinfo{pages}{1116--1138}.
\bibitem[{Varoutis et~al.(2008)Varoutis, Valougeorgis, Sazhin, and
  Sharipov}]{var2008}
\bibinfo{author}{S.~Varoutis}, \bibinfo{author}{D.~Valougeorgis},
  \bibinfo{author}{O.~Sazhin}, \bibinfo{author}{F.~Sharipov},
\newblock \bibinfo{title}{{Rarefied gas flow through short tubes into vacuum}},
\newblock \bibinfo{journal}{J. Vac. Sci. Technol. A} \bibinfo{volume}{26}
  (\bibinfo{year}{2008}) \bibinfo{pages}{228--238}.
\bibitem[{Bird(1995)}]{Bird1995}
\bibinfo{author}{G.~A. Bird}, \bibinfo{title}{{Molecular gas dynamics and the
  direct simulation of gas flow}}, \bibinfo{publisher}{Clarendon Press},
  \bibinfo{year}{1995}.
\bibitem[{Myong et~al.(2019)Myong, Karchani, and Ejtehadi}]{Myong2019}
\bibinfo{author}{R.~S. Myong}, \bibinfo{author}{A.~Karchani},
  \bibinfo{author}{O.~Ejtehadi},
\newblock \bibinfo{title}{{A review and perspective on a convergence analysis
  of the direct simulation Monte Carlo and solution verification}},
\newblock \bibinfo{journal}{Phys. Fluids} \bibinfo{volume}{31}
  (\bibinfo{year}{2019}) \bibinfo{pages}{066101}.
\bibitem[{Mieussens(2000)}]{MIEUSSENS2000429}
\bibinfo{author}{L.~Mieussens},
\newblock \bibinfo{title}{{Discrete-velocity models and numerical schemes for
  the Boltzmann-BGK equation in plane and axisymmetric geometries}},
\newblock \bibinfo{journal}{J. Comput. Phys.} \bibinfo{volume}{162}
  (\bibinfo{year}{2000}) \bibinfo{pages}{429--466}.
\bibitem[{Yang et~al.(2016)Yang, Shu, Wu, and Wang}]{YANG2016291}
\bibinfo{author}{L.~Yang}, \bibinfo{author}{C.~Shu}, \bibinfo{author}{J.~Wu},
  \bibinfo{author}{Y.~Wang},
\newblock \bibinfo{title}{{Numerical simulation of flows from free molecular
  regime to continuum regime by a DVM with streaming and collision processes}},
\newblock \bibinfo{journal}{J. Comput. Phys.} \bibinfo{volume}{306}
  (\bibinfo{year}{2016}) \bibinfo{pages}{291--310}.
\bibitem[{Xu and Huang(2010)}]{Xu2010}
\bibinfo{author}{K.~Xu}, \bibinfo{author}{J.~C. Huang},
\newblock \bibinfo{title}{{A unified gas-kinetic scheme for continuum and
  rarefied flows}},
\newblock \bibinfo{journal}{J. Comput. Phys.} \bibinfo{volume}{229}
  (\bibinfo{year}{2010}) \bibinfo{pages}{7747--7764}.
\bibitem[{Xu and Huang(2011)}]{Xu2011}
\bibinfo{author}{K.~Xu}, \bibinfo{author}{J.~C. Huang},
\newblock \bibinfo{title}{{An improved unified gas-kinetic scheme and the study
  of shock structures}},
\newblock \bibinfo{journal}{IMA J. Appl. Math.} \bibinfo{volume}{76}
  (\bibinfo{year}{2011}) \bibinfo{pages}{698--711}.
\bibitem[{Huang et~al.(2012)Huang, Xu, and Yu}]{Huang2012}
\bibinfo{author}{J.~C. Huang}, \bibinfo{author}{K.~Xu},
  \bibinfo{author}{P.~Yu},
\newblock \bibinfo{title}{{A unified gas-kinetic scheme for continuum and
  rarefied flows II: multi-dimensional cases}},
\newblock \bibinfo{journal}{Commun. Computat. Phys.} \bibinfo{volume}{12}
  (\bibinfo{year}{2012}) \bibinfo{pages}{662–690}.
\bibitem[{Xu(2015)}]{Xu2015}
\bibinfo{author}{K.~Xu}, \bibinfo{title}{{Direct modeling for computational
  fluid dynamics: construction and application of unified gas-kinetic
  schemes}}, \bibinfo{publisher}{WORLD SCIENTIFIC}, \bibinfo{year}{2015}.
\bibitem[{Guo et~al.(2013)Guo, Xu, and Wang}]{dugks2013}
\bibinfo{author}{Z.~Guo}, \bibinfo{author}{K.~Xu}, \bibinfo{author}{R.~Wang},
\newblock \bibinfo{title}{{Discrete unified gas kinetic scheme for all Knudsen
  number flows: Low-speed isothermal case}},
\newblock \bibinfo{journal}{Phys. Rev. E} \bibinfo{volume}{88}
  (\bibinfo{year}{2013}) \bibinfo{pages}{033305}.
\bibitem[{Guo et~al.(2015)Guo, Wang, and Xu}]{dugks2015}
\bibinfo{author}{Z.~Guo}, \bibinfo{author}{R.~Wang}, \bibinfo{author}{K.~Xu},
\newblock \bibinfo{title}{{Discrete unified gas kinetic scheme for all Knudsen
  number flows. II. Thermal compressible case}},
\newblock \bibinfo{journal}{Phys. Rev. E} \bibinfo{volume}{91}
  (\bibinfo{year}{2015}) \bibinfo{pages}{033313}.
\bibitem[{Guo and Xu(2021)}]{dugks2021}
\bibinfo{author}{Z.~Guo}, \bibinfo{author}{K.~Xu},
\newblock \bibinfo{title}{{Progress of discrete unified gas-kinetic scheme for
  multiscale flows}},
\newblock \bibinfo{journal}{Adv. Aerodyn.} \bibinfo{volume}{3}
  (\bibinfo{year}{2021}) \bibinfo{pages}{111--152}.
\bibitem[{Kr{\"u}ger et~al.(2016)Kr{\"u}ger, Kusumaatmaja, Kuzmin, Shardt, and
  Viggen}]{TK2016}
\bibinfo{author}{T.~Kr{\"u}ger}, \bibinfo{author}{H.~Kusumaatmaja},
  \bibinfo{author}{A.~Kuzmin}, \bibinfo{author}{O.~Shardt},
  \bibinfo{author}{E.~M. Viggen}, \bibinfo{title}{{The Lattice Boltzmann
  method: principles and practice}}, \bibinfo{publisher}{Springer},
  \bibinfo{year}{2016}.
\bibitem[{Kim et~al.(2008)Kim, Pitsch, and Boyd}]{KIM2008}
\bibinfo{author}{S.~H. Kim}, \bibinfo{author}{H.~Pitsch},
  \bibinfo{author}{I.~D. Boyd},
\newblock \bibinfo{title}{{Accuracy of higher-order lattice Boltzmann methods
  for microscale flows with finite Knudsen numbers}},
\newblock \bibinfo{journal}{J. Comput. Phys.} \bibinfo{volume}{227}
  (\bibinfo{year}{2008}) \bibinfo{pages}{8655--8671}.
\bibitem[{Song et~al.(2023)Song, Zhang, Zhou, Zhang, and Guo}]{SONG2023}
\bibinfo{author}{X.~Song}, \bibinfo{author}{Y.~Zhang},
  \bibinfo{author}{X.~Zhou}, \bibinfo{author}{C.~Zhang},
  \bibinfo{author}{Z.~Guo},
\newblock \bibinfo{title}{{Modified steady discrete unified gas kinetic scheme
  for multiscale radiative heat transfer}},
\newblock \bibinfo{journal}{Int. J. Heat Mass Tran.} \bibinfo{volume}{203}
  (\bibinfo{year}{2023}) \bibinfo{pages}{123799}.
\bibitem[{Liu and Guo(2023)}]{LIU2023}
\bibinfo{author}{P.~Liu}, \bibinfo{author}{Z.~Guo},
\newblock \bibinfo{title}{{A discrete unified gas kinetic scheme for simulating
  transient hydrodynamics in porous media with fractures}},
\newblock \bibinfo{journal}{Gas Sci. Eng.} \bibinfo{volume}{115}
  (\bibinfo{year}{2023}) \bibinfo{pages}{204997}.
\bibitem[{Yang et~al.(2017)Yang, Shu, Wu, and Wang}]{YANG2017}
\bibinfo{author}{L.~Yang}, \bibinfo{author}{C.~Shu}, \bibinfo{author}{J.~Wu},
  \bibinfo{author}{Y.~Wang},
\newblock \bibinfo{title}{{Comparative study of discrete velocity method and
  high-order lattice Boltzmann method for simulation of rarefied flows}},
\newblock \bibinfo{journal}{Comput. Fluids} \bibinfo{volume}{146}
  (\bibinfo{year}{2017}) \bibinfo{pages}{125--142}.
\bibitem[{Ball(2002)}]{Ball2002}
\bibinfo{author}{J.~S. Ball},
\newblock \bibinfo{title}{{Half-range generalized Hermite polynomials and the
  related Gaussian quadratures}},
\newblock \bibinfo{journal}{SIAM J. Numer. Anal.} \bibinfo{volume}{40}
  (\bibinfo{year}{2002}) \bibinfo{pages}{2311--2317}.
\bibitem[{Ambruş and Sofonea(2016)}]{Ambrus2016}
\bibinfo{author}{V.~E. Ambruş}, \bibinfo{author}{V.~Sofonea},
\newblock \bibinfo{title}{{Lattice Boltzmann models based on half-range
  Gauss–Hermite quadratures}},
\newblock \bibinfo{journal}{J. Comput. Phys.} \bibinfo{volume}{316}
  (\bibinfo{year}{2016}) \bibinfo{pages}{760--788}.
\bibitem[{Ambruş et~al.(2019)Ambruş, Sharipov, and Sofonea}]{Ambrus2019}
\bibinfo{author}{V.~E. Ambruş}, \bibinfo{author}{F.~Sharipov},
  \bibinfo{author}{V.~Sofonea},
\newblock \bibinfo{title}{{Lattice Boltzmann approach to rarefied gas flows
  using half-range Gauss-Hermite quadratures: Comparison to DSMC results based
  on ab initio potentials}},
\newblock \bibinfo{journal}{AIP Conference Proceedings} \bibinfo{volume}{2132}
  (\bibinfo{year}{2019}) \bibinfo{pages}{060012}.
\bibitem[{Ambru\ifmmode~\mbox{\c{s}}\else \c{s}\fi{} and
  Sofonea(2012)}]{Ambru2012}
\bibinfo{author}{V.~E. Ambru\ifmmode~\mbox{\c{s}}\else \c{s}\fi{}},
  \bibinfo{author}{V.~Sofonea},
\newblock \bibinfo{title}{{High-order thermal lattice Boltzmann models derived
  by means of Gauss quadrature in the spherical coordinate system}},
\newblock \bibinfo{journal}{Phys. Rev. E} \bibinfo{volume}{86}
  (\bibinfo{year}{2012}) \bibinfo{pages}{016708}.
\bibitem[{Shi(2022)}]{shi2022}
\bibinfo{author}{Y.~Shi},
\newblock \bibinfo{title}{{Velocity discretization for lattice Boltzmann method
  for noncontinuum bounded gas flows at the micro- and nanoscale}},
\newblock \bibinfo{journal}{Phys. Fluids} \bibinfo{volume}{34}
  (\bibinfo{year}{2022}) \bibinfo{pages}{082013}.
\bibitem[{Yong(2023)}]{shi2023}
\bibinfo{author}{S.~Yong},
\newblock \bibinfo{title}{{Comparison of different Gaussian quadrature rules
  for lattice Boltzmann simulations of noncontinuum Couette flows: From the
  slip to free molecular flow regimes}},
\newblock \bibinfo{journal}{Phys. Fluids} \bibinfo{volume}{35}
  (\bibinfo{year}{2023}) \bibinfo{pages}{072015}.
\bibitem[{Hu and Li(2018)}]{HU2018}
\bibinfo{author}{W.~Q. Hu}, \bibinfo{author}{Z.~H. Li},
\newblock \bibinfo{title}{{Investigation on different discrete velocity
  quadrature rules in gas-kinetic unified algorithm solving Boltzmann model
  equation}},
\newblock \bibinfo{journal}{Comput. Math. Appl.} \bibinfo{volume}{75}
  (\bibinfo{year}{2018}) \bibinfo{pages}{4179--4200}.
\bibitem[{Gutnic et~al.(2004)Gutnic, Haefele, Paun, and
  Sonnendrücker}]{GUTNIC2004}
\bibinfo{author}{M.~Gutnic}, \bibinfo{author}{M.~Haefele},
  \bibinfo{author}{I.~Paun}, \bibinfo{author}{E.~Sonnendrücker},
\newblock \bibinfo{title}{{Vlasov simulations on an adaptive phase-space
  grid}},
\newblock \bibinfo{journal}{Comput. Phys. Commun.} \bibinfo{volume}{164}
  (\bibinfo{year}{2004}) \bibinfo{pages}{214--219}.
\bibitem[{Mehrenberger et~al.(2006)Mehrenberger, Violard, Hoenen, {Campos
  Pinto}, and Sonnendrücker}]{MEHRENBERGER2006}
\bibinfo{author}{M.~Mehrenberger}, \bibinfo{author}{E.~Violard},
  \bibinfo{author}{O.~Hoenen}, \bibinfo{author}{M.~{Campos Pinto}},
  \bibinfo{author}{E.~Sonnendrücker},
\newblock \bibinfo{title}{{A parallel adaptive Vlasov solver based on
  hierarchical finite element interpolation}},
\newblock \bibinfo{journal}{Nucl. Instrum. Meth. A} \bibinfo{volume}{558}
  (\bibinfo{year}{2006}) \bibinfo{pages}{188--191}.
\bibitem[{Chen et~al.(2019)Chen, Liu, Wang, and Zhong}]{chen2019}
\bibinfo{author}{J.~Chen}, \bibinfo{author}{S.~Liu}, \bibinfo{author}{Y.~Wang},
  \bibinfo{author}{C.~Zhong},
\newblock \bibinfo{title}{{Conserved discrete unified gas-kinetic scheme with
  unstructured discrete velocity space}},
\newblock \bibinfo{journal}{Phys. Rev. E} \bibinfo{volume}{100}
  (\bibinfo{year}{2019}) \bibinfo{pages}{043305}.
\bibitem[{Zhao et~al.(2020)Zhao, Wu, Chen, Yang, and Shu}]{zhao2020}
\bibinfo{author}{X.~Zhao}, \bibinfo{author}{C.~Wu}, \bibinfo{author}{Z.~Chen},
  \bibinfo{author}{L.~Yang}, \bibinfo{author}{C.~Shu},
\newblock \bibinfo{title}{{Reduced order modeling-based discrete unified gas
  kinetic scheme for rarefied gas flows}},
\newblock \bibinfo{journal}{Phys. Fluids} \bibinfo{volume}{32}
  (\bibinfo{year}{2020}) \bibinfo{pages}{067108}.
\bibitem[{P{\'e}rez and Pi{\~n}ar(1996)}]{perez1996}
\bibinfo{author}{T.~E. P{\'e}rez}, \bibinfo{author}{M.~A. Pi{\~n}ar},
\newblock \bibinfo{title}{{On Sobolev orthogonality for the generalized
  Laguerre polynomials}},
\newblock \bibinfo{journal}{J. Approx. Theory} \bibinfo{volume}{86}
  (\bibinfo{year}{1996}) \bibinfo{pages}{278--285}.
\bibitem[{Cassity(1965)}]{cassity1965}
\bibinfo{author}{C.~Cassity},
\newblock \bibinfo{title}{{Abcissas, coefficients, and error term for the
  generalized Gauss-Laguerre quadrature formula using the zero ordinate}},
\newblock \bibinfo{journal}{Math. Comput.} \bibinfo{volume}{19}
  (\bibinfo{year}{1965}) \bibinfo{pages}{287--296}.
\bibitem[{Gil et~al.(2021)Gil, Segura, and Temme}]{Gil2020}
\bibinfo{author}{A.~Gil}, \bibinfo{author}{J.~Segura}, \bibinfo{author}{N.~M.
  Temme},
\newblock \bibinfo{title}{{Fast and reliable high-accuracy computation of
  Gauss–Jacobi quadrature}},
\newblock \bibinfo{journal}{Numer. Algorithms} \bibinfo{volume}{87}
  (\bibinfo{year}{2021}) \bibinfo{pages}{1391–1419}.
\bibitem[{Sun and Boyd(2002)}]{SUN2002}
\bibinfo{author}{Q.~Sun}, \bibinfo{author}{I.~D. Boyd},
\newblock \bibinfo{title}{{A direct simulation method for subsonic, microscale
  gas flows}},
\newblock \bibinfo{journal}{J. Comput. Phys.} \bibinfo{volume}{179}
  (\bibinfo{year}{2002}) \bibinfo{pages}{400--425}.
\bibitem[{Huang et~al.(2013)Huang, Xu, and Yu}]{Huang2013}
\bibinfo{author}{J.~C. Huang}, \bibinfo{author}{K.~Xu},
  \bibinfo{author}{P.~Yu},
\newblock \bibinfo{title}{{A unified gas-kinetic scheme for continuum and
  rarefied flows III: microflow simulations}},
\newblock \bibinfo{journal}{Commun. Comput. Phys.} \bibinfo{volume}{14}
  (\bibinfo{year}{2013}) \bibinfo{pages}{1147–1173}.
\bibitem[{Wang et~al.(2023)Wang, Liang, and Xu}]{wl2023}
\bibinfo{author}{L.~Wang}, \bibinfo{author}{H.~Liang}, \bibinfo{author}{J.~Xu},
\newblock \bibinfo{title}{{Optimized discrete unified gas kinetic scheme for
  continuum and rarefied flows}},
\newblock \bibinfo{journal}{Phys. Fluids} \bibinfo{volume}{35}
  (\bibinfo{year}{2023}) \bibinfo{pages}{017106}.
\bibitem[{Zhu and Guo(2019)}]{ZHU2019}
\bibinfo{author}{L.~Zhu}, \bibinfo{author}{Z.~Guo},
\newblock \bibinfo{title}{{Application of discrete unified gas kinetic scheme
  to thermally induced nonequilibrium flows}},
\newblock \bibinfo{journal}{Comput. Fluids} \bibinfo{volume}{193}
  (\bibinfo{year}{2019}) \bibinfo{pages}{103613}.
\bibitem[{Shakhov(1968)}]{Shakhov1968}
\bibinfo{author}{E.~Shakhov},
\newblock \bibinfo{title}{{Generalization of the Krook kinetic relaxation
  equation}},
\newblock \bibinfo{journal}{Fluid Dynam.} \bibinfo{volume}{3}
  (\bibinfo{year}{1968}) \bibinfo{pages}{95–96}.
\bibitem[{Li et~al.(2021)Li, Hu, Wu, and Peng}]{Li2021}
\bibinfo{author}{Z.~Li}, \bibinfo{author}{W.~Hu}, \bibinfo{author}{J.~Wu},
  \bibinfo{author}{A.~Peng},
\newblock \bibinfo{title}{{Improved gas-kinetic unified algorithm for high
  rarefied to continuum flows by computable modeling of the Boltzmann
  equation}},
\newblock \bibinfo{journal}{Phys. Fluids} \bibinfo{volume}{33}
  (\bibinfo{year}{2021}) \bibinfo{pages}{126114}.
\bibitem[{Yang and Huang(1995)}]{YANG1995}
\bibinfo{author}{J.~Yang}, \bibinfo{author}{J.~Huang},
\newblock \bibinfo{title}{{Rarefied flow computations using nonlinear model
  Boltzmann equations}},
\newblock \bibinfo{journal}{J. Comput. Phys.} \bibinfo{volume}{120}
  (\bibinfo{year}{1995}) \bibinfo{pages}{323--339}.
\bibitem[{Kovvali(2022)}]{kovvali2022}
\bibinfo{author}{N.~Kovvali}, \bibinfo{title}{{Theory and applications of
  Gaussian quadrature methods}}, \bibinfo{publisher}{Springer Nature},
  \bibinfo{year}{2022}.
\bibitem[{Shizgal(1981)}]{shizgal1981}
\bibinfo{author}{B.~Shizgal},
\newblock \bibinfo{title}{{A Gaussian quadrature procedure for use in the
  solution of the Boltzmann equation and related problems}},
\newblock \bibinfo{journal}{J. Comput. Phys.} \bibinfo{volume}{41}
  (\bibinfo{year}{1981}) \bibinfo{pages}{309--328}.
\bibitem[{Doman(2015)}]{doman2015}
\bibinfo{author}{B.~G.~S. Doman}, \bibinfo{title}{{The classical orthogonal
  polynomials}}, \bibinfo{publisher}{World Scientific}, \bibinfo{year}{2015}.

\end{thebibliography}
\end{document}